\documentclass[11pt,a4paper]{article}

\usepackage[top=12mm,bottom=12mm,left=28mm,right=28mm,head=12mm,includeheadfoot]{geometry}
\usepackage{parskip}

\makeatletter \@clubpenalty=\clubpenalty \makeatother

\usepackage{tocloft}

\makeatletter
\g@addto@macro\bfseries{\boldmath}
\renewcommand\section{\@startsection{section}{1}{\z@}%
    {-3.5ex \@plus -1.3ex \@minus -.7ex}%
    {2.3ex \@plus.4ex \@minus .4ex}%
    {\large\bfseries}}
\renewcommand\subsection{\@startsection{subsection}{2}{\z@}%
    {-2.3ex\@plus -1ex \@minus -.5ex}%
    {1.2ex \@plus .3ex \@minus .3ex}%
    {\normalsize\bfseries}}
\renewcommand\subsubsection{\@startsection{subsubsection}{3}{\z@}%
    {-2.3ex\@plus -1ex \@minus -.5ex}%
    {1ex \@plus .2ex \@minus .2ex}%
    {\normalsize\bfseries}}
\renewcommand\paragraph{\@startsection{paragraph}{4}{\z@}%
    {1.75ex \@plus1ex \@minus.2ex}%
    {-1em}%
    {\normalsize\bfseries}}
\renewcommand\subparagraph{\@startsection{subparagraph}{5}{\parindent}%
    {1.75ex \@plus1ex \@minus .2ex}%
    {-1em}%
    {\normalsize\bfseries}}
\makeatother

\usepackage{mlmodern}              
\usepackage{amssymb}               
\usepackage{mathrsfs}              
\usepackage[new]{old-arrows}       
\usepackage{microtype}

\usepackage[dvipsnames]{xcolor}
\definecolor{maroon}{RGB}{167,74,74}
\definecolor{magreen}{RGB}{59,125,37}
\definecolor{mablue}{RGB}{59,136,195}
\definecolor{mayellow}{RGB}{242,147,24}
\colorlet{mabrown}{maroon!50!magreen}
\colorlet{maorange}{maroon!50!mayellow}
\colorlet{macyan}{magreen!50!mablue}

\colorlet{red}{maroon}
\colorlet{green}{magreen}
\colorlet{blue}{mablue}
\colorlet{yellow}{mayellow}
\colorlet{brown}{mabrown}
\colorlet{orange}{maorange}
\colorlet{cyan}{macyan}

\definecolor{rred}{RGB}{167,33,74}
\definecolor{bblue}{RGB}{29,136,255}
\definecolor{ppurple}{RGB}{113,84,165}
\definecolor{ppink}{RGB}{255,55,219}

\colorlet{tqft}{blue!20}
\colorlet{triv}{gray!20}
\colorlet{Gcs}{red!25}
\colorlet{Hcs}{tqft}

\newcommand{\red}[1]{\textcolor{maroon}{#1}}

\usepackage{physics}               
\usepackage{mathtools}             
\usepackage[thicklines]{cancel}
\usepackage{stackengine}           
\usepackage{interval}
\intervalconfig{soft open fences}

\usepackage{tikz}
\usetikzlibrary{decorations.markings}
\tikzset{
    line/.style={line width=0.25mm},
    curve/.style={line,smooth,tension=1},
    ->-/.style={decoration={markings,
        mark=at position #1 with {\arrow[>=stealth]{>}}},postaction={decorate}},
    -<-/.style={decoration={markings,
        mark=at position #1 with {\arrow[>=stealth]{<}}},postaction={decorate}},
    bg/.style={opacity=.5},
}

\usepackage{booktabs}
\usepackage{tabularx}

\usepackage[inline]{enumitem}
\usepackage{csquotes}              
\usepackage[normalem]{ulem}        
\usepackage{comment}               

\usepackage[style=numeric-comp, eprint=true, natbib, backend=biber, maxbibnames=10,
            giveninits=true, isbn=false, url=false, doi=false, date=year,
            sorting=none, useprefix=true]{biblatex}
\DeclarePrintbibliographyDefaults{heading=bibintoc}   

\DeclareFieldFormat*{title}{\emph{#1}}
\DeclareFieldFormat*{journaltitle}{#1}
\DeclareFieldFormat{pages}{#1}
\renewbibmacro{in:}{}

\renewbibmacro*{date}{%
    \iffieldundef{year}
        {}
        {\printtext[parens]{\printdate}}}
\renewbibmacro*{issue+date}{%
    \printfield{issue}%
    \setunit*{\addspace}%
    \usebibmacro{date}%
    \newunit}
\renewbibmacro*{publisher+location+date}{%
    \printlist{location}%
    \iflistundef{publisher}
        {\setunit*{\addcomma\space}}
        {\setunit*{\addcolon\space}}%
    \printlist{publisher}%
    \setunit*{\addspace}%
    \usebibmacro{date}%
    \newunit}
\renewbibmacro*{institution+location+date}{%
    \printlist{location}%
    \iflistundef{institution}
        {\setunit*{\addcomma\space}}
        {\setunit*{\addcolon\space}}%
    \printlist{institution}%
    \setunit*{\addspace}%
    \usebibmacro{date}%
    \newunit}
\renewbibmacro*{organization+location+date}{%
    \printlist{location}%
    \iflistundef{organization}
        {\setunit*{\addcomma\space}}
        {\setunit*{\addcolon\space}}%
    \printlist{organization}%
    \setunit*{\addspace}%
    \usebibmacro{date}%
    \newunit}
\renewbibmacro*{location+date}{%
    \printlist{location}%
    \setunit*{\addspace}%
    \usebibmacro{date}%
    \newunit}

\AtEveryBibitem{%
    \clearfield{urldate}%
    \clearfield{month}%
    \clearfield{day}%
    \clearfield{issn}%
    \clearfield{version}}

\DeclareFieldFormat{doilink}{%
    \iffieldundef{doi}{%
        \iffieldundef{url}{%
            \iffieldundef{isbn}{%
                \iffieldundef{issn}{%
                    #1%
                }{%
                    \href{https://books.google.com/books?vid=ISSN\thefield{issn}}{#1}%
                }%
            }{%
                \href{https://books.google.com/books?vid=ISBN\thefield{isbn}}{#1}%
            }%
        }{%
            \href{\thefield{url}}{#1}%
        }%
    }{%
        \href{https://doi.org/\thefield{doi}}{#1}%
    }%
}
\DeclareBibliographyDriver{article}{%
    \usebibmacro{bibindex}%
    \usebibmacro{begentry}%
    \usebibmacro{author/translator+others}%
    \setunit{\labelnamepunct}\newblock
    \usebibmacro{title}%
    \iffieldundef{journaltitle}
        {\setunit{\addspace}\usebibmacro{date}\clearfield{year}}
        {\addcomma\space}%
    \newunit
    \printlist{language}%
    \newunit\newblock
    \usebibmacro{byauthor}%
    \newunit\newblock
    \usebibmacro{bytranslator+others}%
    \newunit\newblock
    \printfield{version}%
    \newunit\newblock
    \printtext[doilink]{%
        \usebibmacro{journal+issuetitle}%
        \newunit
        \usebibmacro{byeditor+others}%
        \newunit
        \usebibmacro{note+pages}%
    }%
    \newunit\newblock
    \iftoggle{bbx:isbn}
        {\printfield{issn}}
        {}%
    \newunit\newblock
    \usebibmacro{doi+eprint+url}%
    \newunit\newblock
    \usebibmacro{addendum+pubstate}%
    \setunit{\bibpagerefpunct}\newblock
    \usebibmacro{pageref}%
    \usebibmacro{finentry}}

\usepackage{hyperref}
\hypersetup{colorlinks,
    linkcolor=mablue,
    citecolor=magreen,
    urlcolor=maroon,
    anchorcolor=mablue,
    filecolor=mablue,
    menucolor=mablue}

\usepackage[noabbrev]{cleveref}

\crefname{subsection}{subsection}{subsections}
\crefname{equation}{}{}

\numberwithin{equation}{section}

\newcommand{\nn}{\nonumber \\}
\newcommand{\beq}{\begin{equation}} 
\newcommand{\eeq}{\end{equation}}   
\renewcommand{\t}[1]{{\text{#1}}}

\newcommand{\nspace}[1][1]{\kern-#1em}

\renewcommand{\geq}{\geqslant}
\renewcommand{\leq}{\leqslant}

\renewcommand{\tilde}{\widetilde}

\newcommand{\eqv}{\Longleftrightarrow}

\newcommand{\suchthat}{\;\middle|\;}

\newcommand{\pd}{\partial}  
\newcommand{\pa}{\partial}  
\newcommand{\inv}[1]{{{#1}^{-1}}}

\newcommand{\conj}[1]{{#1^*}}

\newcommand{\ex}[1]{\mathrm{e}^{#1}}
\newcommand{\ii}{i}
\newcommand{\parti}{\mathcal{Z}}                         

\newcommand{\1}{\mathbf{1}}

\newcommand{\w}{\mathbin{\scalebox{0.8}{$\wedge$}}}      
\renewcommand{\H}{\mathrm{H}}                            

\let\C\undefined
\let\U\undefined
\newcommand{\R}{\mathbb{R}}
\newcommand{\C}{\mathbb{C}}

\newcommand{\Z}{\mathbb{Z}}

\newcommand{\T}{\mathbb{T}}

\renewcommand{\S}{S}
\newcommand{\set}[1]{\qty{#1}}

\newcommand{\U}{\mathrm{U}}
\newcommand{\SU}{\mathrm{SU}}

\newcommand{\Spin}{\mathrm{Spin}}
\newcommand{\alg}[1]{\mathfrak{#1}}

\newcommand{\su}{\alg{su}}
\newcommand{\so}{\alg{so}}

\DeclareMathOperator{\homomorphism}{Hom}    \DeclareDocumentCommand\hom{}{\opbraces{\homomorphism}}
\DeclareMathOperator{\representation}{Rep}  \DeclareDocumentCommand\Rep{}{\opbraces{\representation}}
\DeclareMathOperator{\irrep}{Irr}           \DeclareDocumentCommand\irr{}{\opbraces{\irrep}}
\DeclareMathOperator{\conjclass}{Cl}        \DeclareDocumentCommand\cl{}{\opbraces{\conjclass}}

\renewcommand{\vec}{\boldsymbol}

\DeclareDocumentCommand\wedgecommutator{ l m m }{\braces#1{\lbrack}{\rbrack}{#2\w #3}}

\DeclareDocumentCommand\physdiff{ m o g d() }{%
    \IfNoValueTF{#3}{%
        \IfNoValueTF{#4}
        {#1\IfNoValueTF{#2}{}{^{#2}}}
        {\mathinner{#1\IfNoValueTF{#2}{}{^{#2}}\argopen(#4\argclose)}}%
    }
    {\mathinner{#1\IfNoValueTF{#2}{}{^{#2}}#3} \IfNoValueTF{#4}{}{(#4)}}%
}

\newcommand{\kket}[1]{\left.\ket{#1}\!\right\rangle}     

\usepackage[bb=ncmbbk, bfbb]{mathalpha}
\makeatletter
\def\sv@alphabet#1#2#3{%
  \@tfor\sv@letter:=#3\do{%
    \expandafter\edef\csname #1\sv@letter\endcsname{\noexpand#2{\sv@letter}}}}
\sv@alphabet{c}{\mathcal}{ABCDEFGHIJKLMNOPQRSTUVWXYZ}
\sv@alphabet{s}{\mathscr}{ABCDEFGHIJKLMNOPQRSTUVWXYZ}
\sv@alphabet{f}{\mathfrak}{ABCDEFGHIJKLMNOPQRSTUVWXYZabcdefghjklmnopqrstuvwxyz}
\sv@alphabet{bb}{\mathbb}{ABCDEFGHIJKLMNOPQRSTUVWXYZ}
\sv@alphabet{sf}{\mathsf}{ABCDEFGHIJKLMNOPQRSTUVWXYZabcdefghijklmnopqrstuvwxyz}
\makeatother
\def\fri{\mathfrak{i}}

\newcommand{\HH}{\text{HH}}                  
\newcommand{\aent}{A}                        
\newcommand{\coaent}{\bar{\aent}}            
\newcommand{\td}{\tilde{d}}                  
\newcommand{\tD}{\tilde{\cD}}                
\newcommand{\tn}{\tilde{N}}                  
\newcommand{\n}{\sfn}                        
\newcommand{\vp}{{\vphantom{\dagger}}}

\newcommand{\ent}{\cS}                                      
\newcommand{\ee}{\ent}                                      
\newcommand{\entas}{\Delta\ent}                             
\newcommand{\relent}[2]{\cS\qty(#1\,\middle\Vert\, #2)}     

\makeatletter
  \usepackage{twemojis}

\makeatletter
\expandafter\let\csname tikzcache@orig\endcsname\tikzpicture
\expandafter\let\csname endtikzcache@orig\endcsname\endtikzpicture
\let\tikzcache@origbegin\tikzpicture
\def\tikzcache@envname{tikzpicture}
\newbox\tikzcache@box
\newtoks\tikzcache@toks
\newcount\tikzcache@n
\def\tikzpicture{%
  \ifx\@currenvir\tikzcache@envname
    \expandafter\tikzcache@grab
  \else
    \expandafter\tikzcache@origbegin
  \fi}
\def\tikzcache@grab{\tikzcache@toks{}\tikzcache@grab@}
\long\def\tikzcache@grab@#1\end#2{%
  \tikzcache@toks\expandafter{\the\tikzcache@toks#1}%
  \def\tikzcache@arg{#2}%
  \ifx\tikzcache@arg\tikzcache@envname
    \expandafter\tikzcache@process
  \else
    \tikzcache@toks\expandafter{\the\tikzcache@toks\end{#2}}%
    \expandafter\tikzcache@grab@
  \fi}
\long\def\tikzcache@draw#1{\begin{tikzcache@orig}#1\end{tikzcache@orig}}
\def\tikzcache@reg{\csname tikzcache@box@\csname\tikzcache@key\endcsname\endcsname}
\def\tikzcache@process{%
  \edef\tikzcache@key{tikzcache@\detokenize\expandafter{\the\tikzcache@toks}}%
  \ifmeasuring@
    \setbox\tikzcache@box\hbox{\expandafter\tikzcache@draw\expandafter{\the\tikzcache@toks}}%
    \ifcsname\tikzcache@key\endcsname\else
      \global\advance\tikzcache@n\@ne
      \expandafter\newbox\csname tikzcache@box@\the\tikzcache@n\endcsname
      \expandafter\xdef\csname\tikzcache@key\endcsname{\the\tikzcache@n}%
    \fi
    \global\setbox\tikzcache@reg\copy\tikzcache@box
    \box\tikzcache@box
  \else
    \ifcsname\tikzcache@key\endcsname
      \ifvoid\tikzcache@reg
        \expandafter\tikzcache@draw\expandafter{\the\tikzcache@toks}%
      \else
        \copy\tikzcache@reg
        \global\setbox\tikzcache@reg\box\voidb@x
      \fi
    \else
      \expandafter\tikzcache@draw\expandafter{\the\tikzcache@toks}%
    \fi
  \fi
  \let\endtikzpicture\relax
  \end{tikzpicture}}
\makeatother
\begin{document}

\begin{center}
    {\LARGE\textbf{More physics at the entangling surface}\par}
    {\fontsize{13}{16}\selectfont \textit{Boundary conditions and entanglement in 2d TQFT}\par}
    \bigskip
    Jackson R. Fliss,\kern-3pt\textsuperscript{1}
	Alan Rios Fukelman,\kern-3.2pt\textsuperscript{2} and
	Stathis Vitouladitis\textsuperscript{1} \\
	\bigskip
	\footnotesize{
		\textsuperscript{\scriptsize 1}
		{Physique Théorique et Mathématique, Université Libre de Bruxelles \& \\ International Solvay Institutes, CP 231, 1050 Brussels, Belgium \par}		\textsuperscript{\scriptsize 2}
        {School of Mathematical Sciences, Queen Mary University of London, \\  Mile End Road, London E1 4NS, United Kingdom \par} 
    \href{mailto:jackson.fliss@ulb.be}{\footnotesize \sf jackson.fliss@ulb.be}, \quad 
	\href{mailto:a.riosfukelman@qmul.ac.uk}{\footnotesize \sf a.riosfukelman@qmul.ac.uk}, \quad
	\href{mailto:stathis.vitouladitis@ulb.be}{\footnotesize \sf stathis.vitouladitis@ulb.be}
	}
\end{center}
\bigskip

\begin{abstract}
\noindent
    	We present a systematic treatment of entanglement in two-dimensional topological quantum field theories (TQFTs). Our approach clarifies the sensitivity of entanglement measures to the cutting maps that implement Hilbert space factorization. Such maps embed the TQFT Hilbert space into a description with potentially more degrees of freedom, and for entanglement to be well defined they must be isometric, i.e., they must preserve the norms of all states. We show that the na\"ive cutting map based solely upon path integration subject to boundary conditions is not isometric for generic TQFTs. For any 2d TQFT, we give a prescription to construct isometric cutting maps by dressing path integration with suitable edge operators. This allows us to compute entanglement for any state, any bipartition, and any boundary condition. As a consequence, we find a non-integer ``ground-state degeneracy'' à la Affleck--Ludwig, despite all boundary conditions being gapped. We further show how to reconstruct the partition function of the TQFT at arbitrary genus from the multi-interval Rényi entropies of a single state. For theories with (possibly non-invertible) symmetries, we show that the entanglement entropy of the symmetric ground state can distinguish gapped phases with the same symmetry, without any further knowledge of their vacua, while the entanglement asymmetry of the vacua entropically detects symmetry breaking. We illustrate our formalism with a range of applications, from finite gauge theories to Haagerup gapped phases and infrared phases of adjoint QCD$_2$.
\end{abstract}
\pagebreak

\setcounter{tocdepth}{2}
\tableofcontents
\vspace{10pt}

\section{Introduction}\label{sec:intro}

It is hard to overstate the utility of entanglement in characterizing quantum phases of matter. Entanglement entropy provides clear signatures of quantum critical points \cite{Affleck:1991tk,Vidal:2002rm,Latorre:2003kg,Calabrese:2004eu,Calabrese:2009qy,Cardy:2010zs}, as well as the emergence of topological order in gapped systems \cite{Kitaev:2005dm,Levin:2006zz}, while alternative, finer, measures of entanglement such as entanglement asymmetry can detect spontaneous symmetry breaking \cite{Ares:2022koq,Casini:2020rgj,Capizzi:2023yka,Benini:2025lav,Lamas:2025eay,Benini:2025hbj}. These tools can be applied directly at the level of the ground-state wavefunction, which opens the door for diagnosing quantum phase transitions through tensor network \cite{Jiang:2012uea} and variational methods \cite{Hastings:2010zka,Hibat-Allah:2023jpu}, without having to solve for the spectrum of the Hamiltonian.

Entanglement entropy also assumes a special role in gravity. It is hypothesized that the entropy of event horizons, be they black hole \cite{Solodukhin:2011gn,Emparan:2006ni} or cosmological \cite{Gibbons:1976ue,Gibbons:1977mu,Hawking:2000da,Anninos:2020hfj}, can be phrased as an entanglement entropy of some putative microscopic system. This hypothesis is strengthened in the Anti-de Sitter/conformal field theory (AdS/CFT) correspondence where entanglement of the boundary CFT manifests itself geometrically in the bulk \cite{Ryu:2006ef}. 

Fittingly, an increasing amount of attention is being drawn towards how entanglement measures are defined and computed in continuum quantum field theories (QFTs). The central tension lies between entanglement and \emph{locality}: locality in a QFT implies correlations at arbitrarily small scales preventing a well-defined separation of the Hilbert space, i.e. a factorization,%
\footnote{Alternatively we might say that, in a QFT, the von Neumann algebra associated with a subregion is of type III, lacking a trace or an entropic interpretation on density matrices \cite{Araki:1964lyc,Longo:2022lod}. In this paper we will stick to the framework of Hilbert space factorizations.} 
into subregions. Additionally, and equally importantly, there is a tension between entanglement and \emph{non-locality}, in which long-range correlations prevent a strict factorization of a physical Hilbert space into subregions. This is most evident in gauge theories, where gauge constraints prevent this factorization, however is also manifest in low-energy, long-range correlated, systems (where indeed the projection onto the lowest bands can be phrased as a form of an emergent gauge constraint) \cite{Zhang:1988wy,Wen:1990ez,Wen:1991sca,Senthil:1999czm}. Of course, this tension is even starker in gravitational systems where even the coordinate location of a subregion fails to be gauge invariant!\footnote{See \cite{Wong:2025asd} for recent work exploring this direction.}

The lesson is that an entanglement measure must come equipped, explicitly or implicitly, with a \emph{factorization map}:
\begin{equation}\label{eq:introfactor}
    \cH\hookrightarrow\cH_\aent\otimes\cH_{\coaent}~.
\end{equation}
Here $\aent$ denotes a region and $\cH_\aent$ its associated Hilbert space (and similarly for its causal complement, $\coaent$). This map stands as an avatar for an embedding into a putative ultraviolet (UV) description in which this factorization is resolved --- or perhaps replaced entirely by a more meaningful notion of entanglement. Importantly, neither the factorization map nor the UV embedding is unique. It is therefore paramount to delineate which aspects of an entanglement measure depend on the details of \cref{eq:introfactor}, and which are universal and independent of any UV embedding.\footnote{In preparation of this paper, \cite{Sela:2026ipc} appeared which explores questions of this from in 3d Chern-Simons theory and 2d CFTs. The overlap with the work here is minimal.} Similarly paramount is understanding the interplay of this factorization with non-local features of a system, such as those inherited by generalized global symmetries which may be emergent, spontaneously broken, or gauged at low energies. These are the core problems that this paper sets out to address.

Our arena is the class of compact 2d topological quantum field theories (TQFTs). It is an exactly solvable arena and, as such, it allows us to frame and address both problems at once. The reason is that a 2d TQFT can be formulated entirely and precisely by the category of its generalized symmetries (which encodes their fusion rules, associators, and so on), together with the module category on which it acts. Thus, a local factorization cannot avoid confronting how it acts on, or is acted upon by, the generalized symmetries. The approach we follow is to associate a factorization \cref{eq:introfactor} with a set of boundary conditions (BCs) imposed on a fattened entangling surface, i.e. on the boundary of a small neighborhood of $\pd\aent$ (which in 2d is a set of points). This approach is inspired by Ohmori and Tachikawa \cite{Ohmori:2014eia} who highlighted the role of boundary conditions in entanglement in 2d CFTs. Here our focus on TQFTs brings the additional benefit of a rigid mathematical structure, which allows us to develop the story precisely and comprehensively.  

To be specific, in this paper we show how to associate factorization maps with gapped boundary conditions,%
\footnote{For compact TQFTs this set is exhaustive. We briefly speculate on non-compact TQFTs and gapless boundary conditions in the discussion, \cref{sec:disco}.}
$\cB$, labelled by simple objects in a module category, schematically
\begin{equation}
    \iota_{\cB}:\cH\hookrightarrow\cH_{\aent,\cB}\otimes\cH_{\coaent ,\cB}~.
\end{equation}
As such, we illustrate that the entanglement entropy of a state $\ket{\Psi}\in\cH$,
\begin{equation}
    \ent\qty(\rho_{\Psi,\cB})
    =
    -\tr_{\cH_{\aent,\cB}}\!\qty(\rho_{\Psi,\cB}\log\rho_{\Psi,\cB})~,
    \qquad 
    \rho_{\Psi,\cB}
    =\tr_{\cH_{\coaent ,\cB}}\!\qty(\iota_{\cB}\ketbra{\Psi}{\Psi}\iota_{\cB}^\dagger)~,
\end{equation}
depends sensitively on the boundary condition $\cB$. Given the interpretation of $\iota_{\cB}$ as an embedding into a description with potentially more degrees of freedom associated to the boundary, $\pa A$ --- what are familiarly called {\it edge modes} \cite{Li:2008kda,Buividovich:2008gq,Chandran:2011ihz,Swingle:2011hu,Donnelly:2011hn,Qi:2012ngv,Levin:2013gaa,Donnelly:2014fua,Donnelly:2015hxa,Soni:2015yga,Wen:2016snr,Wong:2017pdm,Fliss:2023uiv} --- the contribution of $\cB$ to the entanglement can be thought of as a \emph{boundary entropy}, $\log g$, à la Affleck and Ludwig \cite{Affleck:1991tk}, a notion that we make precise. Given that all simple boundary conditions considered are gapped, one might suspect this count of boundary degrees of freedom is discrete, $g\in\bbN$. While this suspicion is naïvely correct, we illustrate how suitable freedom in the factorization map can lead to \emph{non-integer} \textquote{ground-state degeneracy,} even in 2d TQFTs. Explaining how and when this happens is one of the main results of this paper.

Along the way we isolate the conditions under which a boundary condition leads to an \emph{isometric} map, i.e. one preserving Hilbert space norms
\begin{equation}
    \braket{\Psi}{\Psi}=\mel{\Psi}{\iota_\cB^\dagger\, \iota_\cB^{\vphantom{\dagger}}}{\Psi}~,
\end{equation}
thereby \textquote{zipping up} the entangling cut. This is precisely the role of the \textquote{entanglement brane} of \cite{Donnelly:2016jet,Donnelly:2018ppr}; here, however, we find that path integral preparation with boundary conditions alone does not guarantee its existence. In its absence, we show how to engineer an isometric map nonetheless, and moreover that this construction is not unique. As a byproduct of this formalism, we show that, with a suitable choice of factorization map, the entire TQFT --- in the form of its partition functions on Riemann surfaces of arbitrary genus --- can be reconstructed from the R\'enyi entropies of a single state.%
\footnote{As shown recently in \cite{McNamara:2026isz}, a unitary QFT can be reconstructed fully by its partition function on arbitrary manifolds. For TQFTs in particular, see also \cite{Blanchet:1995tft,Khovanov:2020tft} for earlier works in this direction.}
We regard this as a second main result of the paper. This parallels \cite{DelZotto:2026fpw,DelZotto:2026dkg} which constructs higher-dimensional TQFT data from multipartite entanglement measures. Surprisingly, however, in two dimensions bipartite entanglement alone suffices.

Our focus on 2d TQFTs is more than a technical convenience: they can be viewed as the infrared (IR) effective field theories of 2d gapped systems, and we encounter many examples for which we can make our formal results concrete. These examples include systems of varying physical relevance, from quantum computing to 2d QCD. Importantly, these examples illustrate how entanglement entropy can serve as an \textquote{order parameter} diagnosing distinct gapped phases, even possessing the same generalized symmetries. We further apply our formalism in calculating entanglement asymmetry, through which we quantify the role of factorization in the subregion breaking of a generalized symmetry.

Lastly, our reasons for initiating a broad study into entanglement in 2d TQFTs arise from interests in quantum gravity at low energies. As indicated above, entanglement measures play an interesting role in encoding UV aspects of quantum gravity in IR measurable, and often geometric, quantities. Central puzzles related to this, such as how to interpret horizon entropies as degeneracies should incorporate a suitable notion of a factorization map that resolves the tension between defining local subregions and diffeomorphism invariance. This puzzle is amplified for the quantum gravity sphere path integral, conjectured to reproduce the entropy of a cosmological horizon upon Wick rotation, simply because it admits no obvious partition function interpretation. In low dimensions, such as 2d and 3d, the absence of propagating gravitons allows gravity to admit an expression as a TQFT, at least formally \cite{Achucarro:1986uwr,Witten:1988hc,Blommaert:2018oro,Iliesiu:2019xuh,Mertens:2022ujr,Collier:2023fwi}. The formalisms developed in this paper are first steps towards an exploration of gravitational applications of entanglement in TQFTs that will appear in a future paper.

A summary of the paper follows. We begin with a review of 2d compact TQFTs in \cref{sec:2dTQFT}, providing a definition of the symmetry category composed of their topological operators, their module categories defining the TQFT, and associating states to a set of irreducible boundary conditions in the closed sector and to a set of boundary changing operators in the open sector. This will set up the necessary ingredients for discussing factorization maps in \cref{sec:entanglement} for which we will provide a general construction. We will then illustrate the dependence of different entanglement measures, the entanglement entropy across two and more partitions, and entanglement asymmetry, on the choice of factorization map. In this section we define the analogue of the Affleck--Ludwig boundary entropy as well as indicate how to reconstruct a TQFT through its multi-interval entanglement. In \cref{sec:examp} we apply our formalism to concrete and physically relevant examples. This section is pedagogical and encyclopedic in nature: with each example we intend to present a self-contained treatment of a TQFT in question. As such this section may be optionally skipped for readers that are interested only in the broad mathematical formalism. Lastly, we finish this paper with a discussion of our results, future applications, and open questions in \cref{sec:disco}.

\section{A 2d TQFT primer}\label{sec:2dTQFT}

To set the stage, we first review several standard facts about 2d TQFTs that will be instrumental for the study of entanglement entropy to be done in what follows. Along the way we will also emphasize a few results that, to our knowledge, have not been highlighted before. Two-dimensional TQFTs have been studied since \cite{Atiyah:1989vu,Dijkgraaf:1989,Durhuus:1993cq,Abrams:1996ty}, while their extension to open/closed theories --- which is what allows us to include boundary conditions --- was developed in \cite{Lazaroiu:2000rk,Moore:2006dw,Lauda:2005wn}. For some more modern developments, see also \cite{Thorngren:2019iar,Komargodski:2020mxz,Huang:2021zvu,Huang:2021ytb}.

TQFTs often arise as infrared descriptions of gapped phases of QFTs. From this perspective they are naturally discussed in the language of spontaneous symmetry breaking, understood in the modern sense to include higher-form as well as non-invertible symmetries. Such symmetries are best understood as being organized by a category. Accordingly, we henceforth focus on (semisimple) 2d TQFTs equipped with a fusion-category symmetry $\cC$ \cite{Huang:2021zvu}.

Let us briefly recall the salient features of symmetry categories in two dimensions. We will only expose a minimal set of aspects relevant for our purposes and refer the reader to \cite{Bhardwaj:2017xup} for a more comprehensive account. The topological line operators (topological defect lines, or TDLs) implementing the symmetry are labelled by simple objects of $\cC$, which we denote by \(a,b,c,\ldots\in\irr(\cC)\). We denote the corresponding TDLs as $L_a, L_b, L_c$. The fusion of two line operators produces, in general, a line which needs not be simple, but instead decomposes as a finite direct sum of simple lines:
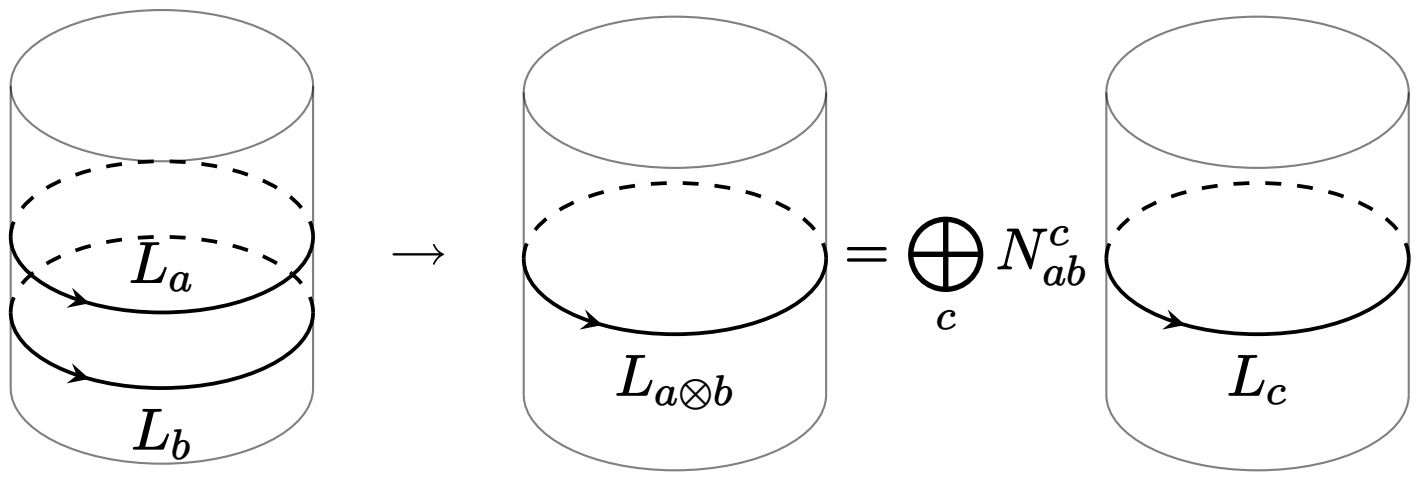
\begin{align}\label{eq:fusion-bulk}
	\begin{gathered}
		\begin{tikzpicture}[scale=.5]
			\draw [bg] (0,4) ++ (0:2 and 1) arc (0:180:2 and 1);
			\draw [bg] (0,4) ++ (180:2 and 1) arc (180:360:2 and 1);
			\draw [bg] (-2,0) -- (-2,4);
			\draw [bg] (2,0) -- (2,4);
			\draw [bg] (0,0) ++ (180:2 and 1) arc (180:360:2 and 1);
			\draw [line,->-=.3] (0,1) ++ (180:2 and 1) arc (180:270:2 and 1) node[below] {$L_b$} arc (270:360:2 and 1);
			\draw [line,dashed] (0,1) ++ (0:2 and 1) arc (0:180:2 and 1);
			\draw [line,->-=.3] (0,2) ++ (180:2 and 1) arc (180:270:2 and 1) node[above] {$L_a$} arc (270:360:2 and 1);
			\draw [line,dashed] (0,2) ++ (0:2 and 1) arc (0:180:2 and 1);
		\end{tikzpicture}
	\end{gathered}
	\quad \to \quad
	\begin{gathered}
		\begin{tikzpicture}[scale=.5]
			\draw [bg] (0,4) ++ (0:2 and 1) arc (0:180:2 and 1);
			\draw [bg] (0,4) ++ (180:2 and 1) arc (180:360:2 and 1);
			\draw [bg] (-2,0) -- (-2,4);
			\draw [bg] (2,0) -- (2,4);
			\draw [bg] (0,0) ++ (180:2 and 1) arc (180:360:2 and 1);
			\draw [line,->-=.3] (0,1.8) ++ (180:2 and 1) arc (180:270:2 and 1) node[below] {$L_{a \otimes b}$} arc (270:360:2 and 1);
			\draw [line,dashed] (0,1.8) ++ (0:2 and 1) arc (0:180:2 and 1);
		\end{tikzpicture}
	\end{gathered}
	=
	\bigoplus_c N^c_{ab} \;
	\begin{gathered}
		\begin{tikzpicture}[scale=.5]
			\draw [bg] (0,4) ++ (0:2 and 1) arc (0:180:2 and 1);
			\draw [bg] (0,4) ++ (180:2 and 1) arc (180:360:2 and 1);
			\draw [bg] (-2,0) -- (-2,4);
			\draw [bg] (2,0) -- (2,4);
			\draw [bg] (0,0) ++ (180:2 and 1) arc (180:360:2 and 1);
			\draw [line,->-=.3] (0,1.8) ++ (180:2 and 1) arc (180:270:2 and 1) node[below] {$L_c$} arc (270:360:2 and 1);
			\draw [line,dashed] (0,1.8) ++ (0:2 and 1) arc (0:180:2 and 1);
		\end{tikzpicture}
	\end{gathered}~,
\end{align}
where \(N^c_{ab}\) are \emph{non-negative integers} known as fusion coefficients.

An important quantity associated with a TDL, \(L_a\), is its quantum dimension.  This is defined as the sphere one-point function of a contractible loop of $a$, normalized by the sphere partition function:
\begin{align}
	d_a =
	\frac{1}{\parti\qty\big[\S^2]}\;
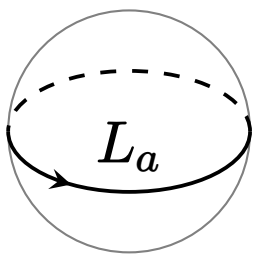
	\begin{gathered}
		\begin{tikzpicture}[scale=.4]
			\draw[bg] (-2,1.8) arc (180:0:2);
			\draw[bg] (-2,1.8) arc (180:360:2);
			\draw [line,->-=.3] (0,1.8) ++ (180:2 and 1) arc (180:270:2 and 1) node[above] {$L_a$} arc (270:360:2 and 1);
			\draw [line,dashed] (0,1.8) ++ (0:2 and 1) arc (0:180:2 and 1);
		\end{tikzpicture}
	\end{gathered}~.
\end{align}
Since the configuration is topologically trivial, the loop labelled by $a$ may be shrunk to a local operator proportional to the identity. The quantity $d_a$ is precisely this proportionality constant.%
\footnote{We assume throughout that all quantum dimensions are positive, or equivalently assume that the underlying fusion category is (pseudo-)unitary \cite{Etingof:2002vpd}.}
For invertible lines this constant is equal to unity; in particular, \(d_{\1} = 1\) for the identity line. Once this normalization is fixed, the remaining quantum dimensions are determined by consistency. Indeed, consider two parallel TDLs labelled by $a$ and $b$ on the sphere. Shrinking them individually yields a factor $d_a d_b$, while fusing first using~\cref{eq:fusion-bulk} and then shrinking gives \(\sum_{c} N^c_{ab}\, d_c\), leading to the constraint:
\begin{align}
	\sum_{c} N^c_{ab}\, d_c = d_a d_b~.
\end{align}
This relation, together with \(d_\1=1\), uniquely determines the set of quantum dimensions. Finally, the quantum dimensions define the total quantum dimension $\cD$ by%
\footnote{
We adopt physics conventions. In the mathematics literature it is more common to define the total (Frobenius--Perron or categorical) dimension of $\cC$ as $\dim\cC = \cD^2$.
\label{foot:qdim-conventions}
}
\begin{align}
	\cD^2 \coloneqq \sum_{a \in \irr(\cC)} d_a^2~.
\end{align}

Once the symmetry category \(\cC\) is fixed, \(\cC\)-symmetric gapped phases are classified by indecomposable \(\cC\)-module categories \(\cM\) on which \(\cC\) acts. In what follows, we adopt this refined description of (semisimple) 2d TQFTs in place of the more familiar formulation in terms of Frobenius algebras. The advantage of this perspective, as we discuss below, is that it clarifies the relation between the Hilbert spaces associated with the circle and the interval, and boundary conditions. Thus, fixing an indecomposable \(\cC\)-module category \(\cM\) fully determines the TQFT.

\subsection{Boundary conditions and closed states}\label{ssec:BC-circle-states}

The natural role of the simple objects \(m,n,k,\ldots\in\irr(\cM)\) is to classify (simple) gapped boundary conditions of the TQFT:
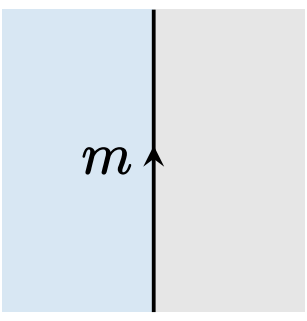
\begin{align}
	\begin{gathered}
		\begin{tikzpicture}[scale=1]
			\fill [fill=tqft] (0,-1) rectangle (-1,1);
			\fill [fill=triv] (0,-1) rectangle (1,1);
			\draw [line,->-=.55] (0,-1) -- (0,0) node [left] {$m$} -- (0,1);
		\end{tikzpicture}
	\end{gathered}~.
\end{align}
In the figure above (and in what follows), the gray shaded region denotes the empty theory, while the non-trivial TQFT under consideration is depicted in blue. In close analogy with TDLs, there is a corresponding notion of boundary quantum dimensions. These are defined by the disc path integral of the TQFT with boundary condition \(m\):
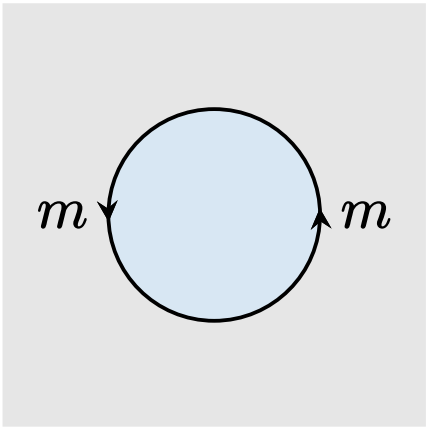
\begin{align}\label{eq:boundary-qdims}
	\td_m = \begin{gathered}
		        \begin{tikzpicture}[scale=0.7]
			\begin{scope}[even odd rule]
				\clip (0,0) circle (1)  (-2,-2) rectangle (2,2);
				\fill [fill=triv] (-2,-2) rectangle (2,2);
			\end{scope}
			\draw [line,->-=.01,->-=.51,fill=tqft] (0,0) circle (1);
			\draw (1,0) node [right] {$m$};
			\draw (-1,0) node [left] {$m$};
		\end{tikzpicture}
	        \end{gathered}
	= 
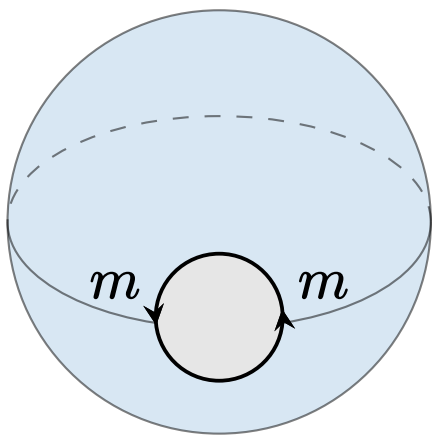
\begin{gathered}
		  \begin{tikzpicture}[scale=.7]
			\fill[fill=tqft] (2,1.8) arc (0:360:2);
			\draw[bg] (-2,1.8) arc (180:0:2);
			\draw[bg] (-2,1.8) arc (180:360:2);
			\draw [bg] (0,1.8) ++ (180:2 and 1) arc (180:270:2 and 1) arc (270:360:2 and 1);
			\draw [bg,dashed] (0,1.8) ++ (0:2 and 1) arc (0:180:2 and 1);
			\draw[line,->-=.02,->-=.52,fill=triv]  (0.6,0.9) arc (0:360:0.6);
			\draw (0.55,1.2) node [right] {$m$};
			\draw (-0.55,1.2) node [left] {$m$};
		\end{tikzpicture}
	  \end{gathered}
\end{align}
From the boundary quantum dimensions one may construct the total boundary quantum dimension,
\begin{align}\label{eq:total-bdy-qdim}
	\tD^2 \coloneqq \sum_{m\in\irr(\cM)} \td_m^{\,2}~,
\end{align}
a quantity that will play a central role in what follows.

A TDL, \(L_a\) with \(a\in\irr(\cC)\), acts on a simple boundary condition by fusing parallel to it. Generically, the result is a direct sum of simple boundary conditions:
\begin{align}\label{eq:action-on-simple-BCs}
	L_a \otimes m = \bigoplus_{n\in\irr(\cM)} \tn^n_{am}\; n~, \qq{with} \tn^{n}_{am}\in \Z_{\geq 0}~.
\end{align}
The coefficients \(\tn_{am}^n\) encode the dimension of the vector space \(\hom(a\otimes m,n)\). The bulk fusion rules, \cref{eq:fusion-bulk}, impose stringent constraints on these coefficients. More precisely, if we regard \(\tn_a\) as a collection of matrices labelled by \(a\), with entries \([\tn_a]^n_{~m} = \tn^n_{am}\), then they obey
\begin{align}
	\sum_{c\in\irr(\cC)} N^c_{ab}\; \tn_c = \tn_a\,\tn_b~.
\end{align}
This structure is known as a \emph{non-negative integer matrix representation} (NIM-rep) of the fusion algebra \cref{eq:fusion-bulk}, and has been studied extensively in the context of 2d CFTs \cite{Verlinde:1988sn,Behrend:1999bn,Cardy:1989ir,Gannon:2001ki,Gaberdiel:2002qa}.

Accordingly, consider the action of a bulk TDL \(L_a\) on a disc with boundary condition \(m\), as in \cref{eq:boundary-qdims}. This immediately yields a constraint on the boundary quantum dimensions. Collecting the boundary quantum dimensions into a vector \(\vec{\td}\) with components \(\td_m\), one finds that \(\vec{\td}\) is a simultaneous eigenvector of the (transposed) NIM-rep matrices \(\tn_a\), with eigenvalue \(d_a\):
\begin{align}\label{eq:FP-equation}
	\tn_a^\sfT\, \vec{\td} = d_a\, \vec{\td}~.
\end{align}
In components, this reads
\begin{align}
	\sum_{n\in\irr(\cM)}\; \tn^n_{am} \td_n = d_a \td_m~.
    \label{bqd_rescaling}
\end{align}
This constraint uniquely determines the boundary quantum dimensions \(\td_m\) up to an overall rescaling. In the mathematics literature, a preferred normalization is \(\tD^2 = \dim(\cM)\), where $\dim(\cM)$ is the Frobenius--Perron dimension of \(\cM\). As will become clear in the rest of the paper, this normalization is natural from the entanglement point-of-view too. For the time being, however, we allow for the possibility that the physically relevant normalization may differ, or even remain unfixed, and proceed while keeping this ambiguity explicit. We return to this issue in the sections that follow.

Physically, one of the primary roles of boundary conditions is to prepare states. To recall, place the TQFT on a cylinder and interpret the non-compact direction as (Euclidean) time. Imposing a boundary condition \(m\) in the infinite Euclidean past prepares a state \(\ket{m}\) on the circle. Since the theory is topological, the precise location of the boundary condition along the non-compact direction is immaterial; it can be brought forward to any finite time. Consequently, \(\ket{m}\) admits the following pictorial path integral representation:
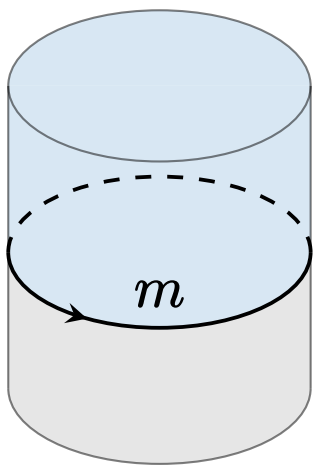
\begin{align}\label{eq:m-state-cylinder}
	\ket{m} =
	\begin{gathered}
		\begin{tikzpicture}[scale=.5]
			\fill [fill=triv] (0,1.8) ++ (0:2 and 1) arc (0:180:2 and 1);
			\fill [fill=triv] (-2,0) rectangle (2,1.8);
			\fill [fill=triv] (0,0) ++ (180:2 and 1) arc (180:360:2 and 1);
			\fill [fill=tqft] (0,4) ++ (0:2 and 1) arc (0:180:2 and 1);
			\fill [fill=tqft] (-2,1.8) rectangle (2,4);
			\fill [fill=tqft] (0,1.8) ++ (180:2 and 1) arc (180:360:2 and 1);
			\draw [bg] (0,4) ++ (0:2 and 1) arc (0:180:2 and 1);
			\draw [bg] (0,4) ++ (180:2 and 1) arc (180:360:2 and 1);
			\draw [bg] (-2,0) -- (-2,4);
			\draw [bg] (2,0) -- (2,4);
			\draw [bg] (0,0) ++ (180:2 and 1) arc (180:360:2 and 1);
			\draw [line,->-=.3] (0,1.8) ++ (180:2 and 1) arc (180:270:2 and 1) node[above] {$m$} arc (270:360:2 and 1);
			\draw [line,dashed] (0,1.8) ++ (0:2 and 1) arc (0:180:2 and 1);
		\end{tikzpicture}
	\end{gathered}~.
\end{align}
The states \(\ket{m}\) furnish a natural basis of the circle Hilbert space and correspond to the vacua of the TQFT.%
\footnote{These states are also known as Cardy branes \cite{Moore:2006dw,Huang:2021zvu,AliAhmad:2025bnd,Popov:2025cha}. Note also that if the TQFT is viewed as the IR description of a gapped model with a UV one-form symmetry, the states $\ket{m}$ can be clustered together in terms of their UV one-form charge into equivalence classes known as \emph{universes} \cite{Hellerman:2006zs,Sharpe:2019ddn,Tanizaki:2019rbk,Komargodski:2020mxz,Aguilera-Damia:2026dbk}.}
Bra states are obtained by a time-reversal of the above picture. The inner product is defined by gluing the corresponding path integrals; under it, the states \(\ket{m}\) are orthonormal. It follows immediately that the dimension of the circle Hilbert space is equal to the number of simple boundary conditions, or equivalently the number of simple objects of \(\cM\), namely
\begin{align}
	\dim\cH_{\S^1} = \abs{\irr(\cM)}~.
\end{align}

Our assumption that the TQFT is semisimple and compact implies that \(m\in\irr(\cM)\) label \emph{gapped} boundary conditions. In particular, these boundary conditions are topological. Consequently, the lower half of the cylinder in \cref{eq:m-state-cylinder} may be deformed into a disc --- precisely as in conformal field theory --- and the disc itself may then be shrunk to a point. In other words, a boundary condition \(m\) imposed on a disc is equivalent to the insertion of a \emph{topological local operator} \(\tilde{\Pi}_m\). Pictorially,
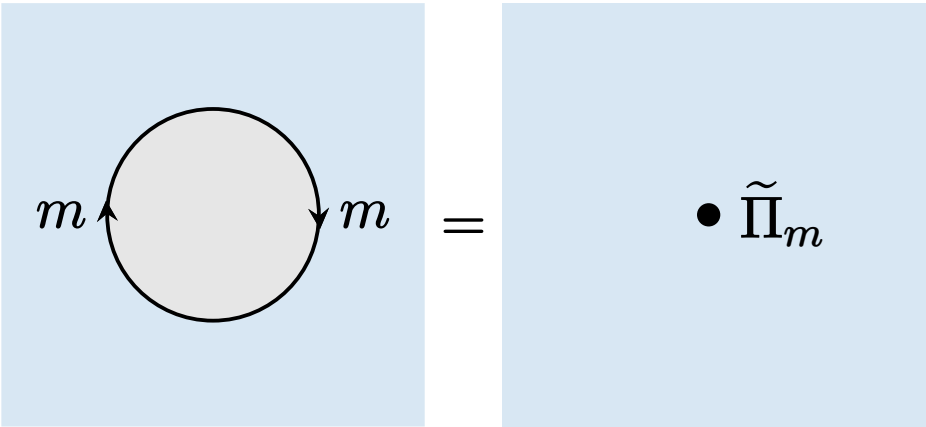
\begin{align}
	\begin{gathered}
		\begin{tikzpicture}[scale=0.7]
			\begin{scope}[even odd rule]
				\clip (0,0) circle (1)  (-2,-2) rectangle (2,2);
				\fill [fill=tqft] (-2,-2) rectangle (2,2);
			\end{scope}
			\draw [line,-<-=.01,-<-=.51,fill=triv] (0,0) circle (1);
			\draw (1,0) node [right] {$m$};
			\draw (-1,0) node [left] {$m$};
		\end{tikzpicture}
	\end{gathered} =
	\begin{gathered}
		\begin{tikzpicture}[scale=0.7]
			\fill [fill=tqft] (-2,-2) rectangle (2,2);
			\draw [fill=black] (-0.05,0) circle (.1);
			\draw (0.05,0.01) node [right] {$\tilde{\Pi}_m$};
		\end{tikzpicture}
	\end{gathered}
    \label{img:Pim}
\end{align}
It follows \cite{Huang:2021zvu} that the topological local operators \(\tilde{\Pi}_m\) satisfy the operator product expansion (OPE)
\begin{align}
	\tilde{\Pi}_m\times \tilde{\Pi}_n = \frac{\delta_{mn}}{\td_m}\, \tilde{\Pi}_m~.
\end{align}
Rescaling the operators according to \(\Pi_m \coloneqq \td_m \tilde{\Pi}_m\), we obtain idempotent operators,
\begin{align}\label{eq:Pim-OPE}
	\Pi_m \times \Pi_n = \delta_{mn}\, \Pi_m~,
\end{align}
so that the \(\Pi_m\) act as mutually orthogonal projectors. From \cref{eq:boundary-qdims} it also follows that the sphere one-point function of \(\Pi_m\) is
\begin{align}\label{eq:Pim-1pt}
	\ev{\Pi_m}_{\S^2} = \td_m^{\, 2}~.
\end{align}
Defined in this way, it is clear that the idempotent operators \(\Pi_m\) form a complete basis of the semisimple Frobenius algebra of the corresponding TQFT \cite{Moore:2006dw}. 

For the moment, let us return to the circle Hilbert space, to emphasize that the above establishes a direct state-operator correspondence for 2d TQFTs.%
\footnote{In the language of Frobenius algebras this perspective goes back at least to \cite{Dijkgraaf:1989,Voronov:1993dz}, and was formulated in the present language in \cite{Huang:2021zvu}; related QFT constructions appear in \cite{Hofman:2024oze,Vitouladitis:2025zoy,Arbalestrier:2026lna,AAPV:2026}.} 
Namely, the states \(\ket{m}\) in \cref{eq:m-state-cylinder} can equivalently be prepared by performing the path integral on a hemisphere with an insertion of the operator \(\Pi_m\):
\begin{align}\label{eq:basis-state-PI}
	\ket{m} = \frac{1}{\td_m}
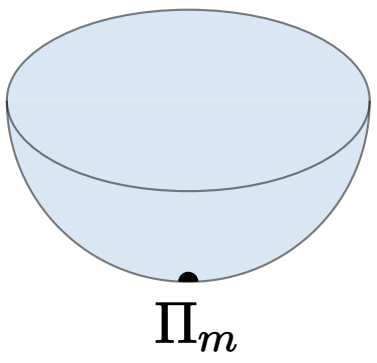
	\begin{gathered}
		\begin{tikzpicture}[scale=0.6]
			\fill [fill=tqft] (0,4) ++ (0:2 and 1) arc (0:180:2 and 1);
			\fill [fill=tqft] (0,4) ++ (180:2 and 1) arc (180:360:2 and 2);
			\draw [bg] (0,4) ++ (0:2 and 1) arc (0:180:2 and 1);
			\draw [bg] (0,4) ++ (180:2 and 1) arc (180:360:2 and 1);
			\draw [bg] (0,4) ++ (180:2 and 1) arc (180:360:2 and 2);
			\begin{scope}[even odd rule]
				\clip (0,4) ++ (180:2 and 1) arc (180:360:2 and 2);
				\draw [fill=black] (0,2) circle (.1);
			\end{scope}
			\draw (0.1,2) node [below] {$\Pi_m$};
		\end{tikzpicture}
	\end{gathered}~.
\end{align}
Bra states correspond again to the time-reversal of the above picture. The states are now orthonormalized \(\braket{n}{m}=\delta_{nm}\).

\subsection{Hartle, Hawking, and the sphere path integral}

Let us also observe the following. Completeness of the states implies that the projectors \(\Pi_m\) furnish a resolution of the identity local operator:
\begin{align}
	\sum_m \Pi_m = \1~.
\end{align}
With this in place, the state prepared by an empty hemisphere path integral can be expressed as a superposition of the basis states \(\ket{m}\), each contributing with weight equal to the boundary quantum dimension \(\td_m\). In the context of theories of gravity, in particular the ones endowed with a positive cosmological constant, such state is referred to as the Hartle--Hawking state \cite{Hartle:1983ai}. In what follows we will use this terminology and denote the Hartle--Hawking state as \(\ket{\HH}\).\footnote{For topological field theorists with no interest in gravity, one may also regard \textquote{HH} as denoting the \textquote{Homogeneous Hemisphere} state.} Explicitly: 
\begin{align}\label{HH}
	\ket{\HH} = \sum_{m\in\irr(\cM)} \td_m \ket{m} =
	\begin{gathered}
		\begin{tikzpicture}[scale=0.5]
			\fill [fill=tqft] (0,4) ++ (0:2 and 1) arc (0:180:2 and 1);
			\fill [fill=tqft] (0,4) ++ (180:2 and 1) arc (180:360:2 and 2);
			\draw [bg] (0,4) ++ (0:2 and 1) arc (0:180:2 and 1);
			\draw [bg] (0,4) ++ (180:2 and 1) arc (180:360:2 and 1);
			\draw [bg] (0,4) ++ (180:2 and 1) arc (180:360:2 and 2);
		\end{tikzpicture}
	\end{gathered}~.
\end{align}
It then follows immediately that the sphere path integral --- namely, the norm of the Hartle--Hawking state --- evaluates to
\begin{align}\label{eq:spherePI}
	\parti\qty\big[\S^2] = \braket{\HH} = \sum_m \td_m^{\,2} = \tD^2~.
\end{align}
Thus the sphere partition function computes the total boundary quantum dimension. We highlight this somewhat surprising fact that, despite being defined on an intrinsically boundary-less manifold, $\parti\qty\big[\S^2]$ depends explicitly on boundary data.

\subsection{Surgery for higher-genus states}
\label{ssec:higher-genus}

Having established that \(\ket{m}\) furnish a basis of states on \(\cH_{\S^1}\), we can use it to determine higher-genus partition functions. In particular, we may insert a resolution of the identity of the form
\begin{align}
	\1 = \sum_m \frac{1}{\td_m^{\,2}}
	
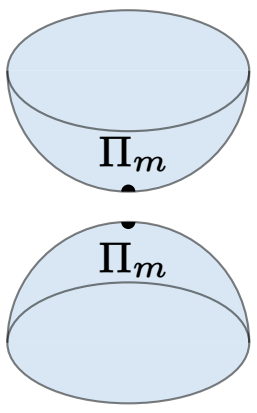
\begin{gathered}
		\begin{tikzpicture}[scale=0.4]
			\fill [fill=tqft] (0,4) ++ (0:2 and 1) arc (0:180:2 and 1);
			\fill [fill=tqft] (0,4) ++ (180:2 and 1) arc (180:360:2 and 2);
			\draw [bg] (0,4) ++ (0:2 and 1) arc (0:180:2 and 1);
			\draw [bg] (0,4) ++ (180:2 and 1) arc (180:360:2 and 1);
			\draw [bg] (0,4) ++ (180:2 and 1) arc (180:360:2 and 2);
			\begin{scope}[even odd rule]
				\clip (0,4) ++ (180:2 and 1) arc (180:360:2 and 2);
				\draw [fill=black] (0,2) circle (.1);
			\end{scope}
			\draw (0.1,2) node [above] {$\scriptstyle  \Pi_m$};
			\fill [fill=tqft] (0,-0.5) ++ (0:2 and 1) arc (360:180:2 and 1);
			\fill [fill=tqft] (0,-0.5) ++ (180:2 and 1) arc (180:0:2 and 2);
			\draw [bg] (0,-0.5) ++ (0:2 and 1) arc (0:180:2 and 1);
			\draw [bg] (0,-0.5) ++ (180:2 and 1) arc (180:360:2 and 1);
			\draw [bg] (0,-0.5) ++ (180:2 and 1) arc (180:0:2 and 2);
			\begin{scope}[even odd rule]
				\clip (0,-0.5) ++ (180:2 and 1) arc (180:0:2 and 2);
				\draw [fill=black] (0,1.5) circle (.1);
			\end{scope}
			\draw (0.1,1.5) node [below] {$\scriptstyle \Pi_m$};
		\end{tikzpicture}
	\end{gathered}~,
\end{align}
along a handle so as to reduce a genus-\(\sfg\) surface to a genus-\((\sfg-1)\) surface with operator insertions. Iterating this procedure, together with the OPE \cref{eq:Pim-OPE}, immediately yields the partition function on a genus-\(\sfg\) Riemann surface, \(\Sigma_\sfg\) \cite{Dijkgraaf:1989,Witten:1991we,Durhuus:1993cq,Fukuma:1993hy,Moore:2006dw}
\begin{align}
	\parti\qty[\Sigma_\sfg] = \sum_{m} \td_m^{\,\chi\qty(\Sigma_\sfg)}~,
    \label{partFgenusg}
\end{align}
where \(\chi(\Sigma_\sfg)=2-2\sfg\) denotes the Euler characteristic of \(\Sigma_\sfg\). Recall that boundary quantum dimensions \(\td_m\) are defined up to an overall rescaling  \cref{bqd_rescaling}. From \cref{partFgenusg} it is manifest that such freedom is equivalent to introducing an Euler counterterm in the path integral formulation of the theory \cite{Witten:1991we}. 

We can use similar surgery arguments to show that the hemisphere states are complete, that is, preparing states on surfaces of higher-genera do not provide an independent basis of states. Specifically, consider the (unnormalized) state, \(\ket{m;\sfg}_\t{PI}\) on \(\S^1\) prepared by performing the path integral on a genus-\(\sfg\) surface \(\Sigma_\sfg\) with a disc removed and with an insertion of the operator \(\Pi_m\). Its norm is
\begin{align}
	_\t{PI}\!\braket{m;\sfg}_\t{PI} = \td_m^{\,\chi(\Sigma_{2\sfg})}~.
\end{align}
Defining the normalized state
\begin{align}
	\ket{m;\sfg} = \frac{1}{\sqrt{\td_m^{\,\chi(\Sigma_{2\sfg})}}}\ket{m;\sfg}_\t{PI}~,
\end{align}
it follows immediately that:
\begin{align}
	\braket{m;\sfg_1}{n;\sfg_2} = \delta_{nm} \frac{\td_m^{\,\chi(\Sigma_{\sfg_1+\sfg_2})}}{\sqrt{\td_m^{\,\chi(\Sigma_{2\sfg_1})}\td_m^{\,\chi(\Sigma_{2\sfg_2})}}} = \delta_{nm}~.
\end{align}
In particular, states prepared by path integrals of different genera, but with the same local operator insertion, are identical.\footnote{In the gravitational context, the Hartle--Hawking state is usually regarded simply as the \textquote{no-boundary} state, and its preparation from the gravitational path integral presumably includes summing over additional topologies sharing the same boundary as the hemisphere. Here in 2d TQFTs we see that such a sum is unnecessary: all higher genus states are the same state.}

\subsection{The interval Hilbert space}\label{sssec:interval-H}

To construct the Hilbert space associated to an interval, it is not sufficient to specify boundary conditions on a spatial slice, as before. One must additionally choose boundary conditions on the two \emph{timelike} boundaries that define the ends of the interval.

To illustrate this, consider first simple boundary conditions \(m,n\in\irr(\cM)\) imposed along the timelike boundaries. This configuration defines a Hilbert space \(\cH_{m,n}\):
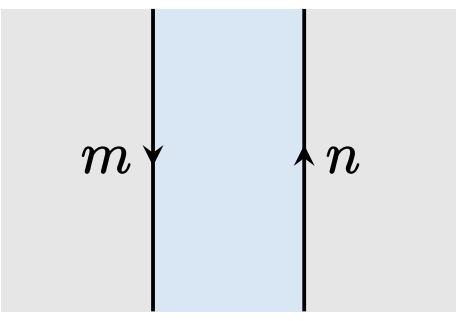
\begin{align}
	\begin{gathered} \begin{tikzpicture}[scale=1]
			\fill [fill=triv] (0,-1) rectangle (1,1);
			\fill [fill=triv] (-2,-1) rectangle (-1,1);
			\fill [fill=tqft] (0,-1) rectangle (-1,1);
			\draw [line,-<-=.55] (-1,-1) -- (-1,0) node [left] {$m$} -- (-1,1);
			\draw [line,->-=.55] (0,-1) -- (0,0) node [right] {$n$} -- (0,1);
		\end{tikzpicture}
	\end{gathered} \rightsquigarrow \cH_{m,n}~.
\end{align}
By arguments entirely analogous to those presented above, one obtains a boundary state-operator correspondence. States \(\ket{\Psi}\in\cH_{m,n}\) are in one-to-one correspondence with \emph{boundary-changing operators} \(\Psi\):
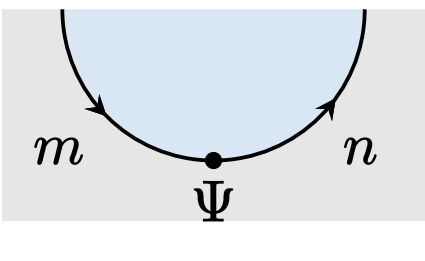
\begin{align}
	\ket{\Psi} =
	\begin{gathered}
		\begin{tikzpicture}
			\begin{scope}[even odd rule]
				\clip (0,0) ++(180:1) arc (180:360:1)  (-1.4,-1.4) rectangle (1.4,0);
				\fill [fill=triv] (-1.4,-2.6) rectangle (1.4,0);
			\end{scope}
			\fill [fill=tqft] (0,0) ++(180:1) arc (180:360:1);
			\draw [line,->-=.25,->-=.8] (0,0) ++(180:1) arc (180:360:1);
			\draw (-.72,-.72) node[below left] {$m$};
			\draw (.72,-.72) node[below right] {$n$};
			\draw [fill=black] (0,-1) circle (.05) node[below] {$\Psi$};
		\end{tikzpicture}
	\end{gathered}~.
\end{align}

In categorical terms, the interval Hilbert space \(\cH_{m,n}\) is identified with the space of morphisms in \(\cM\) from \(m\) to \(n\):
\begin{align}
	\cH_{m,n}\cong \hom_\cM(m,n)~.
\end{align}
In this particular case, the Hilbert space is almost trivial \cite{Huang:2021zvu}:
\begin{equation}\label{eq:Hmnsimple}
    \cH_{m,n}=
    \begin{cases} \bbC~, & m=n \\
    0~, & \text{otherwise.}
    \end{cases}
\end{equation}
This is because the objects \(m\) and \(n\) being considered are simple. As we will see,  this equation will have interesting implications for the entanglement of 2d TQFTs with gapped boundary conditions.

In order to construct more interesting boundary changing operators, and thus states on our interval Hilbert space, we can use semisimplicity to analyze a richer set of boundary conditions. Indeed, the most general topological boundary condition can be written as a finite non-negative integer linear combination of simple boundary conditions:
\begin{align}\label{eq:B-bc}
    \cB = \bigoplus_{m} g_m\, m~, \qq{with} g_m\in\Z_{\geq 0}~.
\end{align}
The interval Hilbert space therefore takes the general form
\begin{align}
	\cH_{\cB_L,\cB_R} \cong \hom_{\cM}\qty(\cB_{L},\cB_{R})~,
\end{align}
where \(\cB_{L,R}\) denote the boundary conditions imposed on the left and right timelike boundaries, respectively, defined as in \cref{eq:B-bc}, with multiplicities \(g_m^{(L,R)}\). Oftentimes we will choose the same BC on the left and the right ends of the interval. In that case we will denote the resulting Hilbert space as \(\cH_\cB = \cH_{\cB,\cB}\), to simplify notation.

To obtain a more concrete description of the open-sector states, note that simplicity of the \(m\) boundary conditions implies \(\cH_{\cB_L,\cB_R}\) admits the decomposition
\begin{align}
	\cH_{\cB_L,\cB_R} = \bigoplus_{m} \cH^{(m)}_{\cB_L,\cB_R}~,
\end{align}
where each block \(\cH^{(m)}_{\cB_L,\cB_R}\) can be regarded as
\begin{align}
    \cH^{(m)}_{\cB_L,\cB_R}=\C^{g_m^{(L)}} \otimes \C^{g_m^{(R)}}
\end{align}
and is spanned by states of the form
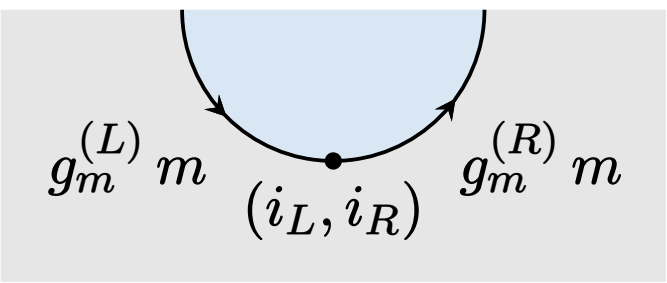
\begin{align}
	\ket{m;\fri} = \ket{m;i_L,i_R} =
	\begin{gathered}
		\begin{tikzpicture}
			\begin{scope}[even odd rule]
				\clip (0,0) ++(180:1) arc (180:360:1)  (-2.2,-1.8) rectangle (2.2,0);
				\fill [fill=triv] (-2.2,-10) rectangle (2.2,0);
			\end{scope}
			\fill [fill=tqft] (0,0) ++(180:1) arc (180:360:1);
			\draw [line,->-=.25,->-=.8] (0,0) ++(180:1) arc (180:360:1);
			\draw (-.7,-.6) node[below left] {$g_m^{(L)}\,m$};
			\draw (.7,-.6) node[below right] {$g_m^{(R)}\,m$};
			\draw [fill=black] (0,-1) circle (.05) node[below] {$\qty(i_L,i_R)$};
		\end{tikzpicture}
	\end{gathered}~.
\end{align}
which are a homomorphism from the \(i_L^\text{th}\) copy of \(m\) on the left to the \(i_R^\text{th}\) copy of \(m\) on the right. We have introduced the shorthand \(\fri=\qty(i_L,i_R)\) for future convenience. The boundary conditions are thus interpreted as stacks of \(g_m^{(L,R)}\) simple \textquote{branes} of type \(m\) on the left and right boundaries, respectively. The vertex \(\qty(i_L,i_R)\) then represents an open string stretching between the \(i_L^\t{th}\) brane on the left and the \(i_R^\t{th}\) brane on the right. In this context we will call \(i_{L,R}=1,2,\ldots,g_m^{(L,R)}\) \textquote{Chan--Paton} indices.  The overlap of the states \(\ket{m;\fri}\) reads:
\begin{align}
	\braket{m;\fri}{n;\fri'} = \td_m \delta_{mn}\delta_{\fri \fri'}~.
\end{align}
Note that while orthogonal, their norm computes the boundary quantum dimension \(\td_m\), as follows from \cref{eq:boundary-qdims}.

Hence a general pure state in the interval Hilbert space can be written as
\begin{align}
	\ket{\Psi} = \sum_{m} \sum_\fri \Psi_\fri^{(m)}\ket{m;\fri} = \sum_{m\in\irr(\cM)}\ \sum_{i_L=1}^{g_m^{(L)}} \sum_{i_R=1}^{g_m^{(R)}} \Psi^{(m)}_{i_L,i_R} \ket{m;i_L,i_R}~.
\end{align}
It follows immediately that the dimension of the total Hilbert space \(\cH_{\cB_L,\cB_R}\) is
\begin{align}
	\dim\cH_{\cB_L,\cB_R} = \sum_m g_m^{(L)} g_m^{(R)}~.
\end{align}
With this, the analysis of the open Hilbert space is essentially complete. Before proceeding, however, let us briefly highlight two special cases of boundary conditions that are of particular physical interest.

\paragraph{Open-to-closed boundary conditions.}

A natural composite boundary condition, which is always available, is obtained by superposing all simple branes with multiplicity one, namely
\begin{align}
	\cB_{1} \coloneqq \bigoplus_{m} m~.
\end{align}
We may then consider the interval Hilbert space obtained by imposing \(\cB_L = \cB_R = \cB_1\) on the two boundaries. In this case,
\begin{align}
	\cH_{\cB_1} = \bigoplus_m \cH_{m,m}~,
\end{align}
and hence
\begin{align}
	\dim\cH_{\cB_1} = \sum_{m\in\irr(\cM)} 1 = \abs{\irr(\cM)}~.
\end{align}
In particular, this coincides with the dimension of the closed Hilbert space. Indeed, it was shown in \cite{Moore:2006dw,Huang:2021zvu} that \(\cH_{\cB_1}\) is isomorphic to the circle Hilbert space \(\cH_{\S^1}\).

\paragraph{Symmetric boundary conditions.}

Whenever all boundary quantum dimensions are integers, \(\td_m\in\Z_{\geq 1}\), a more structured choice of boundary condition becomes available. One may take the multiplicities to equal the boundary quantum dimensions themselves:
\begin{align}
	\cB_\sfS = \bigoplus_{m} \td_m\; m~.
    \label{eq:SBC}
\end{align}
This boundary condition is distinguished by the fact that it is symmetric, in the sense that it is an eigenvector of all TDLs.%
\footnote{In the closed channel, \cref{eq:B_S-eigenvector} states that the boundary state of $\cB_\sfS$ --- i.e., in our notation the state $\ket{\HH}$ --- is an eigenvector of every TDL with eigenvalue $d_a$. When the symmetry category is non-anomalous, i.e. it admits a fiber functor \cite{Thorngren:2019iar}, this is a \emph{strongly symmetric} boundary condition in the language of \cite{Choi:2023xjw}.}
Indeed,
\begin{align}\label{eq:B_S-eigenvector}
	L_a\otimes \cB_\sfS = d_a\, \cB_\sfS~,
\end{align}
where we used \cref{eq:FP-equation} together with the relation \(\tn_a^\sfT = \tn_{\bar{a}}\) for the NIM-rep matrices (see e.g.~\cite{Behrend:1999bn} for a proof). Here \(\bar{a}\) denotes the orientation reversal of the TDL \(a\), and \(d_{\bar{a}} = d_a\). The dimension of the interval Hilbert space with \(\cB_\sfS\) imposed at both ends is
\begin{align}\label{eq:dimHBS}
	\dim\cH_{\cB_\sfS} = \sum_{m} \td_m^{\,2} = \tD^2~,
\end{align}
which coincides with the sphere partition function.

\section{Entanglement in 2d TQFTs}\label{sec:entanglement}

With the formalism we laid out in the previous section we have all the ingredients we need to discuss entanglement in a generic two-dimensional TQFT. As above we fix a fusion category \(\cC\) determining the symmetry of the TQFT and a \(\cC\)-module category \(\cM\) determining the TQFT itself.

\subsection{Cutting maps}\label{ssec:cutting}

We briefly recall the basic setup and further introduce notation convenient to characterize the entanglement in semisimple TQFTs.
We begin by choosing a \emph{cutting map}, that is, a choice of boundary conditions with which to factorize the circle Hilbert space \(\cH_{\S^1}\) into two intervals:
\begin{align}\label{eq:cutting-def}
	\iota_\cB : \cH_{\S^1}\hookrightarrow \cH_\t{edge} \coloneqq \overline{\cH}_{\cB}\otimes\cH_{\cB}~.
\end{align}
This factorization is facilitated by associating degrees of freedom to the boundary. This is the 2d TQFT analogue of ``edge modes'' in higher dimensional TQFTs \cite{Li:2008kda,Chandran:2011ihz,Swingle:2011hu,Qi:2012ngv,Levin:2013gaa,Wen:2016snr,Wong:2017pdm,Fliss:2023uiv} (and gauge theories more generally \cite{Buividovich:2008gq,Donnelly:2011hn,Donnelly:2014fua,Donnelly:2015hxa,Soni:2015yga}). We recall that \(\cH_\cB\) is shorthand for \(\cH_{\cB,\cB}\) and, at this stage, we impose the same boundary condition
\begin{align}\label{eq:cutBC1}
	\cB = \bigoplus_m g_m\; m~,
\end{align}
at both ends of each interval. The generalization to distinct boundary conditions is natural and discussed in the following section. Note also that for the cutting map to be well defined we must require that all \(g_m > 0\): namely if one $g_m=0$ then there would be no image for $\ket{m}$ in $\cH_\text{edge}$. We will assume this henceforth. In \cref{eq:cutting-def}, the overline does not denote complex conjugation; it simply indicates the Hilbert space of a distinct interval that will eventually be traced out. In the Frobenius-algebraic formulation of 2d TQFTs \cite{Lauda:2005wn,Moore:2006dw}, \(\iota_\cB\) corresponds to the open co-multiplication applied after the zipper chosen with boundary conditions \(\cB\). Pictorially, \(\iota_\cB\) is represented by the following Euclidean path integral (or open-closed bordism):
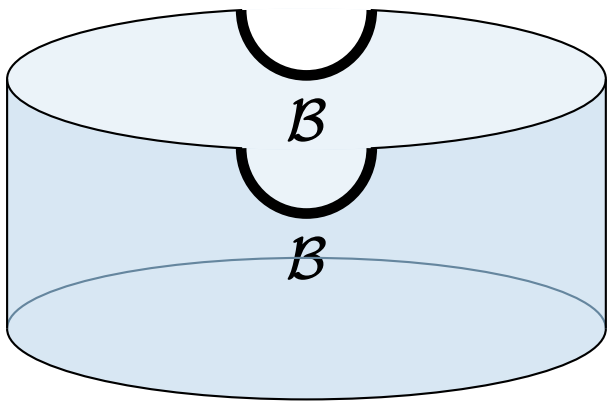
\begin{align}\label{eq:ashtray}
    \iota_\cB =
    \begin{gathered}
        \begin{tikzpicture}[scale=0.6]
			\colorlet{tqftback}{tqft!50}
			\colorlet{backedge}{black!40!blue!70}
			\def\R{3.3}      
			\def\ry{0.78}    
			\def\yT{4.25}    
			\def\yU{1.35}    
			\def\yL{-1.40}   
			\def\yB{-4.20}   
			\def\rtab{0.72}  
			\pgfmathsetmacro{\thetaTab}{acos(\rtab/\R)}
			\pgfmathsetmacro{\aback}{180-\thetaTab}
			\pgfmathsetmacro{\bback}{\thetaTab}
			\pgfmathsetmacro{\afront}{180+\thetaTab}
			\pgfmathsetmacro{\bfront}{360-\thetaTab}
			\fill[blue!20] (-\R,\yL) rectangle (\R,\yU);
			\path[fill=tqft]
				(-\R,\yL)
				arc[start angle=180,end angle=360,x radius=\R,y radius=\ry]
				-- cycle;
			\fill[tqftback]
				(0,\yU) ellipse [x radius=\R, y radius=\ry];
			\path[fill=white]
				(-\R,\yU)
				arc[start angle=180,end angle=\aback,x radius=\R,y radius=\ry]
				arc[start angle=180,end angle=360,x radius=\rtab,y radius=\rtab]
				arc[start angle=\bback,end angle=0,x radius=\R,y radius=\ry]
				arc[start angle=0,end angle=180,x radius=\R,y radius=\ry]
				-- cycle;
			\path[fill=tqftback]
				({\R*cos(\afront)},{\yU+\ry*sin(\afront)})
				arc[start angle=\afront,end angle=\bfront,x radius=\R,y radius=\ry]
				arc[start angle=0,end angle=-180,x radius=\rtab,y radius=\rtab]
				-- cycle;
			\draw
				(-\R,\yU)
				arc[start angle=180,end angle=\afront,
					x radius=\R,y radius=\ry]
				coordinate (lowerarc)
				arc[start angle=180,end angle=360,
					x radius=\rtab,y radius=\rtab]
				arc[start angle=\bfront,end angle=360,
					x radius=\R,y radius=\ry];
			\draw[line width=2pt]
				(lowerarc)
				arc[start angle=180,end angle=360,
					x radius=\rtab,y radius=\rtab] node[midway,below] {$\cB$};
			\draw
				(-\R,\yU)
				arc[start angle=180,end angle=\aback,
					x radius=\R,y radius=\ry]
				coordinate (upperarc)
				arc[start angle=180,end angle=360,
					x radius=\rtab,y radius=\rtab]
				arc[start angle=\bback,end angle=0,
					x radius=\R,y radius=\ry];
				\draw[line width=2pt]
				(upperarc)
				arc[start angle=180,end angle=360,
					x radius=\rtab,y radius=\rtab] node[midway,below] {$\cB$};
			\draw[backedge]
				(-\R,\yL)
				arc[start angle=180,end angle=0,x radius=\R,y radius=\ry];
			\draw
				(-\R,\yL)
				arc[start angle=180,end angle=360,x radius=\R,y radius=\ry];
			\draw (-\R,\yU) -- (-\R,\yL);
			\draw (\R,\yU) -- (\R,\yL);
			\end{tikzpicture}
	\end{gathered}~.
\end{align}
It is straightforward to see that a pure state \(\ket{\Psi}\in\cH_{\S^1}\),
\begin{align}
	\ket{\Psi} = \sum_m \Psi_m \ket{m}~,
    \label{S1PureState}
\end{align}
is embedded via \(\iota_\cB\) into \(\cH_\t{edge}\) as
\begin{align}
	\iota_\cB \ket{\Psi}
	= \sum_m \frac{\Psi_m}{\td_m^{\,2}}\; \sum_{\fri} \overline{\ket{m;\fri}}\otimes\ket{m;\fri}~,
    \label{eq:SimpCutEmbed}
\end{align}
where we denoted \(\overline{\ket{m;\fri}} = \overline{\ket{m;i_L,i_R}} = \ket{m;i_R,i_L}\). Observe that, unless \(g_m=\td_m\) for all $m$, this map does \emph{not} preserve the norm of a generic state. 
To see this, recall that inner products in \(\cH_\t{edge}\) are computed by gluing along the two intervals. Gluing \(\iota_\cB^\dagger\) onto \(\iota_\cB\) in \cref{eq:ashtray}, the two pairs of semicircles carrying \(\cB\) close up into two holes, so that \(\iota_\cB^\dagger\iota_\cB^\vp\) is a cylinder with two holes carrying \(\cB\). Capping the cylinder off with the hemispheres \cref{eq:basis-state-PI} that prepare \(\ket{m}\) and \(\bra{m'}\), we find
\begin{align}
	\mel{m'}{\iota_\cB^\dagger\,\iota_\cB^\vp}{m}
	= \frac{1}{\td_{m'}\td_m}\,
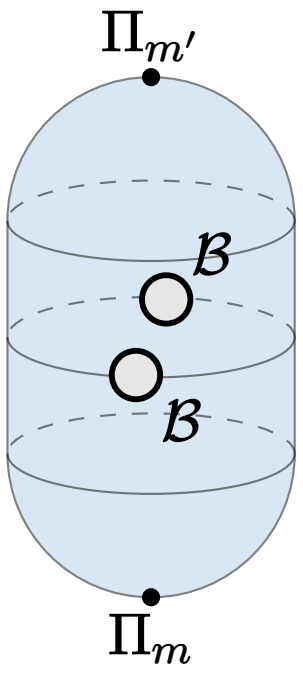
	\begin{gathered}
		\begin{tikzpicture}[
				hole/.style={circle, draw, line width=1.1pt, fill=triv, inner sep=0pt, minimum size=.32cm},
				op/.style={circle, fill=black, inner sep=0pt, minimum size=.12cm}]
			\filldraw[fill=tqft, draw opacity=.5] (-.95,-.77) -- (-.95,.77) arc[start angle=180, end angle=0, radius=.95]
				-- (.95,-.77) arc[start angle=0, end angle=-180, radius=.95] -- cycle;
			\draw[bg] (-.95,-0.77) arc[start angle=180, end angle=360, x radius=.95, y radius=.266];
			\draw[bg, dashed] (.95,0) arc[start angle=0, end angle=180, x radius=.95, y radius=.266];
			\draw[bg] (-.95,0) arc[start angle=180, end angle=360, x radius=.95, y radius=.266];
			\draw[bg, dashed] (.95,-0.77) arc[start angle=0, end angle=180, x radius=.95, y radius=.266];
			\node[hole, label={[inner sep=1pt]below right:$\cB$}] at (-0.1,-0.25) {};
			\node[hole, label={[inner sep=1pt]above right:$\cB$}] at (0.1,0.25) {};
			\draw[bg] (-.95,0.77) arc[start angle=180, end angle=360, x radius=.95, y radius=.266];
			\draw[bg, dashed] (.95,0.77) arc[start angle=0, end angle=180, x radius=.95, y radius=.266];
			\node[op, label={[inner sep=1pt]above:$\Pi_{m'}$}] at (0,1.72) {};
			\node[op, label={[inner sep=1pt]below:$\Pi_m$}] at (0,-1.72) {};
		\end{tikzpicture}
	\end{gathered}
	= \sum_{n,\ell}\frac{g_n\,g_\ell}{\td_{m'}\td_m\td_n\td_\ell}\,
	
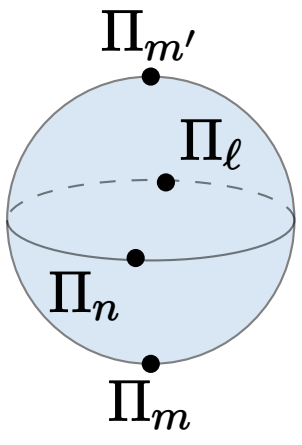
\begin{gathered}
		\begin{tikzpicture}[op/.style={circle, fill=black, inner sep=0pt, minimum size=.12cm}]
			\filldraw[fill=tqft, draw opacity=.5] (0,0) circle (.95);
			\draw[bg] (-.95,0) arc[start angle=180, end angle=360, x radius=.95, y radius=.266];
			\draw[bg, dashed] (.95,0) arc[start angle=0, end angle=180, x radius=.95, y radius=.266];
			\node[op, label={[inner sep=1pt]below left:$\Pi_n$}] at (-.1,-.25) {};
			\node[op, label={[inner sep=1pt]above right:$\Pi_\ell$}] at (.1,.25) {};
			\node[op, label={[inner sep=1pt]above:$\Pi_{m'}$}] at (0,.95) {};
			\node[op, label={[inner sep=1pt]below:$\Pi_m$}] at (0,-.95) {};
		\end{tikzpicture}
	\end{gathered}
	= \delta_{mm'}\,\frac{g_m^2}{\td_m^{\,2}}~,
	\label{eq:iso-main}
\end{align}
where in the last step we used the OPE \cref{eq:Pim-OPE}, which sets \(n=\ell=m=m'\), and the one-point function \cref{eq:Pim-1pt}. Hence
\begin{align}
	\iota_\cB^\dagger\iota_\cB^\vp=\sum_m \qty(\frac{g_m}{\td_m})^2\,\ketbra{m}~,
\end{align}
and the image of \(\ket{m}\) has norm \(g_m/\td_m\). The bare map is therefore isometric if and only if \(g_m=\td_m\) for all \(m\). Since \(g_m\in\Z_{>0}\), this is only possible when all \(\td_m\) are integers, in which case \(\cB=\cB_\sfS\) is the symmetric boundary condition \cref{eq:SBC}.
For \(\cB=\cB_\sfS\), each of the two holes shrinks to the identity operator. In the terminology of \cite{Donnelly:2018ppr,Donnelly:2016jet} it is a \emph{shrinkable} boundary condition, or equivalently, \(\cB_\sfS\) is an \emph{entanglement brane}. In \cite{Brehm:2021wev,Brehm:2024zun} it is an instance of a \emph{cloaking} boundary condition (in a minimal sense).

However, \(\iota_\cB\) need not be taken in this minimal form. One may further postcompose it with arbitrary operators \(\cO\) and \(\overline{\cO}\) acting within \(\cH_\cB\) and \(\overline{\cH}_\cB\), respectively, corresponding to a \emph{decorated bordism}, in category theory terminology. We will explore this possibility below in a later section. For the moment, let us first outline the bare minimum in order to achieve an \emph{isometric cutting map}, that is, one preserving norms. In this case, it is sufficient to take \(\cO\) and \(\overline{\cO}\) to be identical. Following the conventions of \cite{Jafferis:2019wkd}, we denote this operator by \(\sqrt{\bbD}\). The cutting map then becomes
\begin{align}
	\iota_{\bbD,\cB} = \qty(\sqrt{\bbD}\otimes \sqrt{\bbD})\circ \iota_\cB~.
    \label{eq:SimpleCutting}
\end{align}
A direct computation shows that choosing
\begin{align}
	\sqrt{\bbD} = \sum_m \sqrt{\frac{\td_m}{g_m}}\,\1_{m}~,
\end{align}
where \(\1_m\) denotes the identity on \(\cH_\cB^{(m)}\), ensures that \(\iota_{\bbD,\cB}\) is an isometry. With this choice, the embedding of \(\ket{\Psi}\) in the edge mode Hilbert space is exactly
\begin{align}
	\iota_{\bbD,\cB}\ket{\Psi} = \sum_m \Psi_m \kket{\Omega_m}~,
\end{align}
so that each sector \(m\) lifts to a canonically normalized maximally entangled state
\begin{align}
	\kket{\Omega_m} = \frac{1}{\td_m\; g_m} \sum_\fri \overline{\ket{m;\fri}}\otimes\ket{m;\fri}~,
\end{align}
on \(\cH_{\cB}^{(m)}\).

\subsection{Entanglement in pure states}\label{ssec:pure-entanglement}

Let us first compute the entanglement entropy of a generic pure state \(\ket{\Psi}\); mixed states follow directly and are discussed in \cref{app:MixedMult}. After factorizing the Hilbert space, we trace out one factor and retain the complementary subregion. This yields the normalized reduced density matrix
\begin{align}
	\rho_{\Psi,\cB}^\vp \coloneqq \frac{1}{\abs{\Psi}^2}\,
	\tr_{\overline{\cH}_\cB}^\vp\!\qty(\iota^\vp_{\bbD,\cB} \ketbra{\Psi} \iota_{\bbD,\cB}^\dagger)
	= \sum_m \frac{\abs{\Psi_m}^2}{\abs{\Psi}^2}\, \frac{\1_m}{g_m^{2}}~, 
\end{align}
with \(\abs{\Psi}^2 = \sum_m \abs{\Psi_m}^2\). Since the state is block-diagonal in the superselection sectors, the computation of the R\'enyi and entanglement entropies is immediate. The R\'enyi entropies are
\begin{align}
	\ent_\n\qty(\rho_{\Psi,\cB})
	= \frac{1}{1-\n}\log\tr(\rho_{\Psi,\cB}^\n)
	= \frac{1}{1-\n}\log(\sum_m \lambda_m^{\n} g_m^{2(1-\n)})~,
\end{align}
with \(\lambda_m = \abs{\Psi_m}^2/\abs{\Psi}^2\). Taking the limit \(\n\to 1\) gives
\begin{align}\label{eq:Sent-generic-psi}
	\ee\qty(\rho_{\Psi,\cB})
	= -\sum_m \lambda_m
	\log \lambda_m+ {2}\sum_m \lambda_m \log g_m~.
\end{align}
The first term is the Shannon entropy of the probability distribution \(\lambda_m\), while the second encodes the contribution from the boundary multiplicities \(g_m\). Since \(g_m\) are a measure of different possible boundary configurations, this contribution is completely analogous to the Affleck--Ludwig (AL) \emph{boundary entropy} usually described in the context of 2d CFTs \cite{Affleck:1991tk}. Here due to the nature of the boundary condition involved in the cutting map, \cref{eq:cutBC1}, it appears as the log of an integer. We will soon see how to generalize this to non-integer degeneracies, however, before doing so we will briefly generalize the above construction to mixed boundary conditions. We will focus on the case in which we embed the circle Hilbert space in two intervals. The generalization to multiple intervals will follow from there.

\subsection{General cutting maps}
\label{app:mixed-bcs}
So far we have only considered the intervals to have the same boundary condition. Yet, as seen from \cref{eq:Sent-generic-psi}, the entanglement entropy depends sensitively on this specific choice. To address this, we now extend the previous discussion in two directions. First, we allow different boundary conditions at the two ends of the interval. Second, we generalize the cutting map \cref{eq:SimpleCutting} by postcomposing it with more general edge operators after the cut. We call these operators \emph{edge dressings}. 

Let the boundary conditions at the left and right ends of the interval be 
\begin{align}
	\cB_L = \bigoplus_m g_m^{(L)}\,m~,
	\qquad
	\cB_R = \bigoplus_m g_m^{(R)}\,m~.
    \label{LRBC}
\end{align}
An open state in the \(m\)-th sector then carries \(g_m^{(L)}\) Chan--Paton labels at its left end and \(g_m^{(R)}\) labels at its right end. To cut every closed state, no sector may be absent. We therefore assume
\begin{align}
	g_m^{(L)}g_m^{(R)} > 0~,
\end{align}
for every \(m\in\irr(\cM)\). The corresponding bare cutting map is
\begin{align}
	\iota_{\cB_L,\cB_R}:
	\cH_{\S^1}
	\hookrightarrow
	\cH_\t{edge}(\cB_L,\cB_R)
	\coloneqq
	\overline{\cH}_{\cB_R,\cB_L}\otimes \cH_{\cB_L,\cB_R}~.
\end{align}
The interval that is eventually traced out has the reversed ordering of boundary conditions. In a fixed sector, the open-string degeneracy is $g_m^{(L)}g_m^{(R)}.$ The bare map sends
\begin{align}
	\iota_{\cB_L,\cB_R}\ket{\Psi}
	=
	\sum_m\frac{\Psi_m}{\td_m^{\,2}}
	\sum_{i_L=1}^{g_m^{(L)}}\sum_{i_R=1}^{g_m^{(R)}}
	\overline{\ket{m;i_R,i_L}}\otimes \ket{m;i_L,i_R}~,
\end{align}
which generalizes \cref{eq:SimpCutEmbed} and, by the same token, is not an isometry in general. In particular, the image of a normalized basis state \(\ket{m}\) has norm-squared \(g_m^{(L)} g_m^{(R)}/\td_m^{\,2}\).

\paragraph{Scalar edge dressings.}

The simplest way to restore the norm is to dress the two intervals by diagonal edge operators, 
\begin{align}
	\sqrt{\bbD_{\coaent}}
	=
	\sum_m \overline{\delta}_m\,\overline{\1}_m~,
	\qquad
	\sqrt{\bbD_{\aent}}
	=
	\sum_m \delta_m\,\1_m~,
\end{align}
where \(\overline{\1}_m\) and \(\1_m\) are the identities on \(\overline{\cH}_{\cB_R,\cB_L}^{(m)}\) and \(\cH_{\cB_L,\cB_R}^{(m)}\) and $\delta_m$ and $\overline\delta_m$ are constants (we remind the reader that the overline does \emph{not} denote complex conjugation). We call these operators scalar, because they do not mix the Chan--Paton indices of the open-string states; rather, they simply rescale them by \(\overline{\delta}_m\) or \(\delta_m\), respectively.

The dressed cutting map is
\begin{align}
	\iota_\t{iso} \coloneqq \iota_{\bbD_{\coaent},\bbD_\aent;\cB_L,\cB_R}
	=
	\qty(\sqrt{\bbD_{\coaent}}\otimes\sqrt{\bbD_{\aent}})
	\circ
	\iota_{\cB_L,\cB_R}~.
\end{align}
It is isometric precisely when
\begin{align}
	\overline{\delta}_m\,\delta_m = \frac{\td_m}{\sqrt{g_m^{(L)}g_m^{(R)}}}~.
\end{align}
Note that only the product of the two dressing factors is fixed. For any choice obeying this condition, a generic state is embedded to
\begin{align}
	\iota_\t{iso}\ket{\Psi}
	=
	\sum_m \Psi_m\kket{\Omega_m}~,
\end{align}
with unit-norm states $\kket{\Omega_m}$
\begin{align}
	\kket{\Omega_m}
	=
	\frac{1}{\td_m\sqrt{g_m^{(L)}g_m^{(R)}}}
	\sum_{i_L=1}^{g_m^{(L)}}\sum_{i_R=1}^{g_m^{(R)}}
	\overline{\ket{m;i_R,i_L}}\otimes\ket{m;i_L,i_R}~.
\end{align}

For a pure state \(\ket{\Psi}=\sum_m\Psi_m\ket{m}\), we proceed as before and trace out the first factor
\begin{align}
	\rho_{\Psi}
	=
	\sum_m
	\lambda_m\,\frac{\1_m}{g_m^{(L)}g_m^{(R)}}~,
	\qquad
	\lambda_m = \frac{\abs{\Psi_m}^2}{\abs{\Psi}^2}~,
\end{align}
where now \(\1_m\) is the identity on \(\cH_{\cB_L,\cB_R}^{(m)}\), whose dimension is \(g_m^{(L)}g_m^{(R)}\). The state is maximally mixed within each open block, while the closed-sector weights remain \(\lambda_m\). Hence
\begin{align}
	\ent_\n\qty(\rho_{\Psi;\cB_L,\cB_R})
	=
	\frac{1}{1-\n}
	\log\!\qty(
	\sum_m \lambda_m^\n \qty(g_m^{(L)}g_m^{(R)})^{1-\n}
	)~,
\end{align}
and
\begin{align}
	\ee\qty(\rho_{\Psi;\cB_L,\cB_R})
	=
	-\sum_m\lambda_m\log\lambda_m
	+
	\sum_m\lambda_m\log \qty(g_m^{(L)}g_m^{(R)})~.
    \label{EE:MixBoundary}
\end{align}
As in \cref{eq:Sent-generic-psi} we identify the first term as the Shannon entropy of the probability distribution $\lambda_m$ and the second one, encoding the boundary multiplicities, as the AL boundary entropies which again appear as integers. 

\subsection{General edge dressings and non-integer Affleck--Ludwig entropies}\label{sect:genALent}

In order to discover non-integer boundary entropy, we must allow the edge dressing to rotate the Chan--Paton indices. Let \(\cO_{\aent}^{(m)}\) and \(\cO_{\coaent}^{(m)}\) act on \(\cH_{\cB_L,\cB_R}^{(m)}\) and \(\overline{\cH}_{\cB_R,\cB_L}^{(m)}\), respectively. Define
\begin{align}
	\cO_\aent = \bigoplus_m \cO_\aent^{(m)}~,
	\qquad
	\cO_{\coaent} = \bigoplus_m \cO_{\coaent}^{(m)}~,
\end{align}
and the cutting map is given now by
\begin{align}\label{eq:Ocuttingmap}
	\iota_{\cO;\cB_L,\cB_R}
	\coloneqq
	\qty(\cO_{\coaent}\otimes\cO_{\aent})
	\circ
	\iota_{\cB_L,\cB_R}~.
\end{align}
It sends the basis state \(\ket{m}\) to
\begin{align}
	\left.\ket{\Omega_m^{\cO}}\!\!\right>
	=
	\frac{1}{\td_m^{\,2}}
	\sum_{i_L=1}^{g_m^{(L)}}\sum_{i_R=1}^{g_m^{(R)}}
	\cO_{\coaent}^{(m)}\overline{\ket{m;i_R,i_L}}
	\otimes
	\cO_{\aent}^{(m)}\ket{m;i_L,i_R}~.
    \label{genCuttingMap}
\end{align}
It is convenient to introduce the following operators for future convenience
\begin{align}\label{eq:bbOAdef}
	\bbO^{(m)}_{\coaent} \coloneqq \cO_{\coaent}^{(m)\dagger}\cO_{\coaent}^{(m)}~,
	\qquad
	\bbO^{(m)}_{\aent} \coloneqq \cO_{\aent}^{(m)\dagger}\cO_{\aent}^{(m)}~.
\end{align}
It follows that the states \(\ket*{\Omega_m^{\cO}}\!\rangle\) have norm 
\begin{align}
	\ev{\!\!\braket{\Omega_m^{\cO}}{\Omega_m^{\cO}}\!\!}
	=
	\frac{1}{\td_m^{\,4}}
	\sum_{i_L,j_L=1}^{g_m^{(L)}}\sum_{i_R,j_R=1}^{g_m^{(R)}}
	\overline{\bra{m;j_R,j_L}}\,
	\bbO^{(m)}_{\coaent}\,
	\overline{\ket{m;i_R,i_L}}\,
	\bra{m;j_L,j_R}\,
	\bbO^{(m)}_{\aent}\,
	\ket{m;i_L,i_R}~.
\end{align}
Recalling also that the states \(\overline{\ket{m;i_R,i_L}}\) are identified with \(\ket{m;i_L,i_R}\),%
\footnote{The Hilbert spaces \(\cH_{\cB_L,\cB_R}^{(m)}\) and \(\overline{\cH}_{\cB_R,\cB_L}^{(m)}\) are naturally isomorphic, with the isomorphism given precisely by \(I_m \overline{\ket{m;i_R,i_L}} = \ket{m;i_L,i_R}\).}
we recognize the above expression as 
\begin{align}
	\ev{\!\!\braket{\Omega_m^{\cO}}{\Omega_m^{\cO}}\!\!}
	=
	\frac{1}{\td_m^{\,2}}\tr({\bbO_{\coaent}^{(m)}}^\sfT \bbO_{\aent}^{(m)})~,
\end{align}
with the trace taken over \(\cH_{\cB_L,\cB_R}^{(m)}\) 
Thus the dressed cut is isometric if and only if
\begin{align}
	\tr({\bbO_{\coaent}^{(m)}}^\sfT \bbO_{\aent}^{(m)})=\td_m^{\,2}~,
	\qquad
	\t{for all}\ m~.
\end{align}
After tracing out \(\coaent\), the reduced density matrix associated with \(\ket{m}\) is
\begin{align}\label{eq:sigmam}
	\sigma_m
	=
	\frac{1}{\td_m^{\,2}}\;\cO_{\aent}^{(m)}\,{\bbO_{\coaent}^{(m)}}^\sfT\,\cO_{\aent}^{(m)\dagger}~,
	\qquad
	\tr \sigma_m = 1~.
\end{align}
For a general pure state, one obtains
\begin{align}\label{eq:rhosumsigmam}
	\rho_{\Psi}
	=
	\sum_m\lambda_m\, \sigma_m~,
	\qquad
	\lambda_m = \frac{\abs{\Psi_m}^2}{\abs{\Psi}^2}~.
\end{align}
The superselection sectors remain orthogonal, while the edge dressing controls the spectrum within each sector. Consequently,
\begin{align}
	\ent_\n\qty(\rho_\Psi)
	=
	\frac{1}{1-\n}
	\log\!\qty(
	\sum_m \lambda_m^\n \tr \sigma_m^\n
	)~,
\end{align}
and
\begin{align}
	\ee\qty(\rho_\Psi)
	=
	-\sum_m\lambda_m\log\lambda_m
	+
	\sum_m\lambda_m\,\ent\qty(\sigma_m)~.
    \label{genRenyies}
\end{align}
The scalar construction above corresponds to \(\sigma_m=\1_m/g_m^{(L)}g_m^{(R)}\). General edge dressings therefore replace the flat spectrum in each open block by the spectrum of \(\sigma_m\).

\subsection{Multiple intervals and partition function reconstruction}

Let us discuss the generalization to the case in which we dissect the Cauchy slice into multiple disjoint intervals. We will see that, under certain assumptions, the resulting R\'enyi entropies of a single state will give us access to data defining the entire TQFT in question: namely its partition function on Riemann surfaces of arbitrary genus \cite{McNamara:2026isz}. 

Without loss of generality, we will divide the circle into an even number, \(2\sfN\), of intervals. The most general cutting map factorizes the circle Hilbert space as 
\begin{align}\label{eq:MIcutmap}
    \iota_{\{\cO\};\{\cB\}}: \cH_{\S^1}\hookrightarrow \cH_\text{edge} \coloneqq \cH_{\cB_1,\cB_2}\otimes\overline{\cH}_{\cB_2,\cB_3}\otimes \ldots \otimes {\cH}_{\cB_{2\sfN-1},\cB_{2\sfN}}\otimes \overline{\cH}_{\cB_{2\sfN},\cB_1}~.
\end{align}
with an embedding on pure states acting as
\begin{align}\label{eq:MIgenedge}
	\iota_{\{\cO\};\{\cB\}}\ket{\Psi} = \,\sum_m \frac{\Psi_m}{\td_m^{2\sfN}} \sum_{\{\fri\}} \bigotimes_{j=1}^{\sfN}\left( \cO_{2j-1}^{(m)}\ket{m;\fri_{2j-1}}\otimes\cO_{2j}^{(m)}\overline{\ket{m;\fri_{2j}}}\right)~, 
\end{align}
where $\abs{\Psi}^2 = \sum_m \abs{\Psi_m}^2$, $\ket{m;\fri_{2j-1}}\kern-2pt=\kern-2pt\ket{m;i_{2j-1},i_{2j}}$ and $\overline{\ket{m;\fri_{2j}}}\kern-2pt=\kern-2pt\overline{\ket{m,i_{2j+1}, i_{2j}}}\kern-2pt=\kern-2pt\ket{m;i_{2j},i_{2j+1}}$, etc. subject to $\fri_{2\sfN+1}=\fri_1$, and
\begin{equation}
    \sum_{\{\fri\}}\coloneqq\sum_{i_1=1}^{g_m^{(1)}}\ldots\sum_{i_{2\sfN}=1}^{g_m^{(2\sfN)}}~.
\end{equation}
This embedding is an isometry if
\begin{equation}\label{eq:Oisom}
    \frac{1}{{\td_m}^{4\sfN}}\sum_{\{\fri\}}\sum_{\{\fj\}}\qty(\prod_{j=1}^{\sfN}\bra{m;\fj_{2j-1}}\bbO_{2j-1}^{(m)}\ket{m;\fri_{2j-1}}\overline{\bra{m;\fj_{2j}}}\bbO_{2j}^{(m)}\overline{\ket{m;\fri_{2j}}})=1~,
\end{equation}
where we generalize the notation of \cref{eq:bbOAdef} to define $\bbO_j^{(m)}=\cO_j^{(m)\dagger}\cO_j^{(m)}$. We consider this generic situation in \cref{app:MixedMult}; for simplicity of presentation of the results we will specialize the discussion momentarily to the case of scalar dressings:
\begin{equation}\label{eq:MIscalaredege}
    \cO_j^{(m)}=\delta^{(m)}_j\1_{\cH_{\cB_j}}~,
\end{equation}
with isometry condition
\begin{equation}
    \prod_{j=1}^{2\sfN}\delta_j=\prod_{j=1}^{2\sfN}\qty(\frac{\td_m}{g_m^{(j)}})^{1/2}~.
\end{equation}
We trace out the even labelled intervals denoted with overbars (again, without loss of generality). A simple calculation gives the reduced density matrix
\begin{align}
	\rho_{\Psi,\{\cO\};\{\cB\}}^\vp \coloneqq \tr_{\overline{\cH}_\cB}^\vp\!\qty(\iota^\vp_{\{\cO\};\{\cB\}} \ketbra{\Psi} \iota_{\{\cO\};\{\cB\}}^\dagger)
	= \sum_m \frac{\abs{\Psi_m}^2}{\abs{\Psi}^2}\,\1^{\left(\sfN\right)}_m\prod_{j=1}^{2\sfN}\qty(g_m^{(j)})^{-1}~,
\end{align}
where the identity runs over the remaining $\sfN$ interval Hilbert spaces, namely 
\begin{align}
    \1^{\left(\sfN\right)}_m = \frac{1}{\td_m^{\sfN}}\ \sum_{i_1,\ldots,i_{\sfN}}  \ketbra{m;i_1,i_2}{m;i_1,i_2}\otimes\ldots\otimes\ketbra{m;i_{2\sfN-1},i_{2\sfN}}{m;i_{{2\sfN}-1},i_{2\sfN}} ~.
\end{align}
The computation of the R\'enyi and entanglement entropies is again immediate. The R\'enyi entropies are
\begin{align}
	\ent_\n\qty(\rho_{\Psi,\{\cO\};\{\cB\}})
	= \frac{1}{1-\n}\log\tr(\rho_{\Psi,\{\cO\};\{\cB\}}^\n)
	= \frac{1}{1-\n}\log(\sum_m \lambda_m^{\n} \prod_{j=1}^{2\sfN}\qty(g_m^{(j)})^{(1-\n)})~,
\end{align}
with \(\lambda_m = \abs{\Psi_m}^2/\abs{\Psi}^2\), and the limit \(\n\to 1\) gives
\begin{align}\label{eq:Sent-generic-psi-N}
	\ee\qty(\rho_{\Psi,\{\cO\};\{\cB\}})
	= -\sum_m \lambda_m
	\log \lambda_m+ \sum_m \lambda_m \sum_{j=1}^{2\sfN}\log g_m^{(j)}~.
\end{align}
More generally, using the general edge dressings, \cref{eq:MIgenedge}, the entanglement entropy across multiple intervals takes the form
\begin{equation}
    \ee\qty(\rho_{\Psi,\{\cO\},\{\cB\}})
	= -\sum_m \lambda_m
	\log \lambda_m+ \sum_m \lambda_m\,\ee\qty(\boldsymbol{\sigma}_{m;\sfN})~.
\end{equation}
where $\boldsymbol{\sigma}_{m;\sfN}$ is a density matrix over the $m$ sector over $\bigotimes_{j=1}^{\sfN}\cH_{\cB_{2j-1},\cB_{2j}}$ generalizing \cref{eq:sigmam}:
\begin{equation}\label{eq:sigmamN}
    \boldsymbol{\sigma}_{m;\sfN}\coloneqq \frac{1}{\td_m^{\,4\sfN}}\sum_{\{\fri\}}\sum_{\{\fj\}}\bigotimes_{j=1}^{\sfN}\qty(\cO_{2j-1}^{(m)}\ket{m;\fri_{2j-1}}\qty(\bbO_{2j-2}^{(m)})^{\sfT}_{\overline{\fri_{2j-2}};\overline{\fj_{2j-2}}}\bra{m;\fj_{2j-1}}\cO^{(m)\dagger}_{2j-1})~.
\end{equation}
Note that much like the product over $g_m^{(j)}$ in \cref{eq:Sent-generic-psi-N}, $\boldsymbol{\sigma}_{m;\sfN}$ is independent of the state and instead a pure measure of the degrees of freedom introduced by \cref{eq:MIgenedge}, i.e. $\ee\qty(\boldsymbol{\sigma}_{m;\sfN})$ is the Affleck--Ludwig entropy of the cutting map, \cref{eq:MIcutmap}.

Beyond generalizing AL entropies, the R\'enyi entropy of even a single state across multiple intervals contains a wealth of information about the TQFT. Namely, we can show that by choosing the cutting map wisely, the entire TQFT, in the form of its partition function on a Riemann surface of arbitrary genus, $\parti[\Sigma_g]$, can be reconstructed from the R\'enyi entropies of the Hartle--Hawking state, \cref{HH} across multiple intervals.

Let us first illustrate this for the scalar edge dressing, \cref{eq:MIscalaredege}.
The $\sfn^\text{th}$ R\'enyi entropy of the Hartle--Hawking state is
\begin{align}
	\ent_\n\qty(\rho_{\HH,\{\cO\},\{\cB\}})
	= \frac{1}{1-\n}\log\left(\frac{\sum_m \td_m^{\,2\n} \prod_{j=1}^{2\sfN}\qty(g^{(j)}_m)^{(1-\n)}}{\parti\left[S^2 \right]^n}\right)~.
\end{align}
Let us suppose that all of the boundary quantum dimensions are integer, $\td_m\in\bbN$. We can then choose to assign to each edge the symmetric boundary condition $\cB_\sfS$, i.e. $g_m^{(j)}=\td_m$ for each $j$. This yields
\begin{equation}
    \ent_\n\qty(\rho_{\HH,\cB_\sfS})
	= \frac{1}{1-\n}\log\left(\frac{\sum_m \td_m^{\,2\n+ 2\sfN(1-\n)}}{\parti\left[S^2 \right]^n}\right)~,
\end{equation}
We immediately recognize the numerator inside the logarithm as the partition function of the theory on a Riemann surface of genus $\sfg=(\sfn-1)(\sfN-1)$:
\begin{equation}
    \ent_\n\qty(\rho_{\HH,\cB_\sfS})
	= \frac{1}{1-\n}\log\left(\frac{\parti\left[\Sigma_{(\sfn-1)(\sfN-1)} \right]}{\parti\left[S^2 \right]^\sfn}\right)~.
\end{equation}
The geometric interpretation of the above correspondence is natural: $\Sigma_{(\sfn-1)(\sfN-1)}$ is also the $\sfn$-sheeted cover of $S^2$ branched at $2\sfN$ interval points \cite{Calabrese:2009ez}. Since $\ket{\HH}$ is prepared by the hemisphere path integral this is the same surface arrived at through a R\'enyi path integral, by joining consecutively the $\sfn$ $S^2$ sheets at the cuts corresponding to the un-barred intervals. Setting, for instance, $\sfn=2$, we see that every $\parti\qty[\Sigma_\sfg]$ is determined from a 2$^\text{nd}$ R\'enyi entropy of the Hartle--Hawking state (with $\parti[S^2]$ itself determined by the case $\sfN=1$).

When there exists a non-integer boundary quantum dimension then we lose the luxury of imposing the symmetric boundary condition $\cB_\sfS$. However in this case the ability to dress the edges in our cutting map permits enough freedom to repeat the above procedure, although at the expense of losing the geometric interpretation. Namely by choosing the purity of the density matrix $\boldsymbol{\sigma}_{m;\sfN}$ associated with the boundary degrees of freedom, \cref{eq:sigmamN}, to satisfy
\begin{equation}\label{eq:2Rsigcond}
    \tr(\boldsymbol{\sigma}_{m;\sfN}^2)=\frac{1}{\td_m^{\,2\sfN}}~,
\end{equation}
then the 2$^\text{nd}$ R\'enyi entropy of $\ket{\HH}$ again takes the form
\begin{equation}
    \ent_2\qty(\rho_{\HH,\{\cO\},\{\cB\}})=
    -\log(\frac{\sum_{m}\td_m^4\tr(\boldsymbol{\sigma}_{m;\sfN}^2)}{\parti\qty\big[S^2]^2})=-\log\left(\frac{\parti\qty[\Sigma_{\sfN-1}]}{\parti\qty\big[S^2]^2}\right)~.
\end{equation}
Since each boundary quantum dimension is bounded below by one, $\td_m^{-2\sfN}\leq 1$, \cref{eq:2Rsigcond} does not conflict with $\boldsymbol{\sigma}_{m;\sfN}$ as a edge density matrix and herefore it is possible to satisfy \cref{eq:2Rsigcond} by appropriately chosen boundary dressings.

The upshot of this result is that \emph{every partition of a 2d TQFT has an entropic origin}, namely the R\'enyi entropy of the Hartle--Hawking state for some factorization of its multiple interval Hilbert space. This is very much reminiscent of the program set out in \cite{DelZotto:2026fpw,DelZotto:2026dkg} in reconstructing a $d$-dimensional TQFT from the multipartite entanglement of the ground state of a gapped theory which flows to it in the IR. Here we see that the bipartite structure, namely a single configuration of alternating intervals, is enough. This is not surprising and is an accident of two dimensions: because of the topological nature of the theory, alternating intervals is essentially the unique non-trivial configuration to consider the entanglement of multiple parties. Higher dimensions would both admit and require the type of multipartite entropy measures of \cite{DelZotto:2026fpw,DelZotto:2026dkg} in reproducing a TQFT.

\subsection{Entanglement asymmetry}\label{app:entas}

In this section we adapt our formulations towards computing entanglement asymmetry in 2d TQFTs. While most of the formalism presented in this section has been developed elsewhere, e.g. \cite{Ares:2022koq,Benini:2025lav,AliAhmad:2025bnd,Benini:2025hbj}, we present a comprehensive summary both to adapt previous constructions to our formalism and notation, and to provide context for further calculations. Later we will provide novel calculations of entanglement asymmetry in several example systems that, to our knowledge, have not been catalogued yet in the literature. Readers familiar and comfortable with the basics of entanglement asymmetry for ordinary and generalized symmetries may choose to skip this section.

Entanglement asymmetry \cite{Ares:2022koq} is a quantitative entropic measure of symmetry breaking. In the simplest case of invertible zero-form symmetries it is defined as a relative entropy between a given density matrix, defined on the entire system or on a subsystem, and its symmetrized version obtained by averaging over the symmetry group. Explicitly, for a state \(\rho\), we define the symmetrized state as:
\begin{align}\label{eq:rho_S-grp-def}
	\rho_\sfS = \frac{1}{\abs{G}}\sum_{g\in G} L_g\, \rho\, L_g^\dagger~.
\end{align}
Here \(L_g\) are the topological symmetry operators implementing the action of \(G\). For an Abelian symmetry, the symmetrized density matrix can be written equivalently as
\begin{align}\label{eq:rho_S-grp-def-proj}
	\rho_\sfS = \sum_\sfR \Pi_\sfR\, \rho\, \Pi_\sfR~,
\end{align}
where \(\Pi_\sfR\) are projectors onto the irreducible representations of \(G\). For conciseness we have written the above formulas for a finite group; generalizing to continuous groups amounts to replacing the sum with a Haar integral in  \cref{eq:rho_S-grp-def}. The entanglement asymmetry (EA) is then defined as
\begin{align}
	\entas(\rho) = \relent{\rho}{\rho_\sfS}~.
\end{align}
Straightforwardly, by properties of the relative entropy, \(\entas(\rho)\) is non-negative and vanishes if and only if \(\rho=\rho_\sfS\), i.e. whenever \(\rho\) preserves the symmetry. Hence, it is a faithful measure of symmetry breaking. Since \(\Pi_\sfR\) are projectors, one can readily show that \(\entas(\rho)\) can be written as the difference of two entropies:
\begin{align}
	\entas(\rho) = \ent(\rho_\sfS) - \ent(\rho)~.
\end{align}
This rewriting has certain computational advantages; in particular, it enables a replica trick and the introduction of Rényi asymmetries as:
\begin{align}
	\entas_\n(\rho) = \ent_\n(\rho_\sfS) - \ent_\n(\rho)~,
\end{align}
where \(\ent_\n\) is the Rényi entropy of order \(\n\). The Rényi asymmetries are significantly easier to compute than the EA, and are therefore often preferred in practical applications.%
\footnote{Note however that certain properties of EA, such as its monotonicity under renormalization \cite{Benini:2025hbj}, do not necessarily hold for the Rényi asymmetries.}
EA is recovered from the Rényi asymmetry by taking the limit \(\n\to 1\).

EA and its Rényi generalizations have been studied in a variety of contexts, including spin chains, matrix product states and quantum quenches \cite{Ares:2023kcz,Murciano:2023qrv,Rylands:2023yzx,Capizzi:2023xaf} (revealing a quantum Mpemba effect, also observed experimentally \cite{Joshi:2024sup}), quantum and conformal field theory \cite{Fossati:2024ekt,Kusuki:2024gss,Fossati:2024ekt,Benini:2024xjv,Benini:2025lav,Benini:2025hbj}, and toy models of black hole dynamics \cite{Ares:2023ggj,Chen:2024lxe}.

\subsubsection{Entanglement asymmetry for generalized symmetries}\label{app:EA}

The task to generalize EA to non-invertible and higher-form symmetries was recently undertaken in \cite{Benini:2025lav,AliAhmad:2025bnd,Lamas:2025eay,Benini:2025hbj}. For our setup, the construction of \cite{Benini:2025lav} is the most practical, and we will therefore briefly review it here. We refer the reader to the original paper for a more detailed discussion.

In full generality, TDLs acting on a Hilbert space of states%
\footnote{Here we imagine TDLs laying flat on a spatial slice, i.e. acting on the states, but not twisting the Hilbert space.}
form an algebra. In the case of two-dimensional theories, the relevant algebras are:
\begin{enumerate*}[label=(\alph*)]
	\item the fusion algebra \cref{eq:fusion-bulk}, which acts on the untwisted circle Hilbert space \(\cH_{\S^1}\);
    	\item the tube algebra, \(\operatorname{Tube}(\cC)\), \cite{Ocneanu:2016chi} which acts on the direct sum of circle Hilbert spaces, twisted by TDLs:
	      \(\cH_\t{Tube} = \bigoplus_{a} \cH_{\S^1}^a\); and
	\item the strip algebra, \(\operatorname{Strip}_\cC(\cM)\), \cite{Kitaev:2011dxc,Cordova:2024iti} which acts on the interval Hilbert space, associated to a \(\cC\)-module category, \(\cM\), with a given choice of boundary conditions, \(\cH_{\cB_L,\cB_R}\).
\end{enumerate*}
%

\paragraph{Fusion algebra.}

We begin with the simplest case of symmetrizing with respect to the fusion algebra on the circle. Consider a fusion category symmetry \(\cC\) consisting of simple TDLs, \(L_a\), with fusion rules \cref{eq:fusion-bulk} (repeated here for convenience),
\begin{align}
	L_a\otimes L_b = \bigoplus_c N_{ab}^c\, L_c~.
\end{align}
In this case, equation \cref{eq:rho_S-grp-def} is replaced by
\begin{align}\label{eq:symmetrizer-fusion}
	\rho_\sfS = \sum_{a,b\in\irr(\cC)} \tilde{K}^{ab} \, L_a\, \rho\, L_b~.
\end{align}
Here \(\tilde{K}^{ab}\) is the inverse of the matrix
\begin{align}\label{eq:K-pairing}
	K_{ab} = \sum_{c,d\in\irr(\cC)} N_{ac}^d	N_{bd}^c~.
\end{align}
This symmetrizer map, \(\sfS: \rho\mapsto \rho_\sfS\), is the unique quantum channel%
\footnote{One can show that \(\sfS\) is a quantum channel  --- i.e. a completely positive trace-preserving (CPTP) operator --- by noting that the Wedderburn--Artin theorem implies the existence of Kraus operators \(K_\alpha = D^{\phantom{\alpha}a}_{\alpha}\, L_a\), for some matrix \(D\), such that \(\rho_\sfS\) can be written in Kraus form: \(\rho_\sfS = \sum_\alpha K_\alpha\, \rho\, K_\alpha^\dagger\).}
that is \emph{idempotent}, \(\sfS^2 = \sfS\), \emph{commutes with the symmetry} action, \(L_a\, \rho_\sfS = \rho_\sfS\, L_a\), and \emph{preserves symmetric states}, i.e. if \(L_a \rho = \rho L_a\) for all \(a\), then \(\rho_\sfS = \rho\). Hence, it is the natural generalization of the symmetrizer map for ordinary group symmetries.

Furthermore, one can easily show that in the case of invertible symmetries, the above construction reduces to the standard symmetrizer \cref{eq:rho_S-grp-def} for finite group symmetries. In category theoretic terms, an invertible symmetry corresponds to the fusion category of \(G\)-graded vector spaces, potentially twisted by a non-trivial 3-cocycle $\omega\in\H^3(G;\U(1))$ amounting to an 't Hooft anomaly, i.e. \(\cC = \mathsf{Vec}_G^\omega\). The fusion coefficients are then \(N_{g\,h}^k=\delta_{k,gh}\), with \(g,h,k\in G\) and so
\begin{align}
	K_{g\,h} =
	\abs{G}\,\delta_{g,\inv{h}}~
	\quad\implies\quad \tilde{K}^{g\,h} = \frac{1}{\abs{G}}\,\delta_{g,\inv{h}}~,
\end{align}
and the symmetrizer reduces to the standard one for finite group symmetries. Note that the anomaly enters only through the \(F\)-symbols of the fusion category, to which the fusion algebra is blind. Hence the fusion symmetrizer is insensitive to the presence of an 't Hooft anomaly. In contrast, the tube algebra and the corresponding symmetrizer, not discussed in this paper, are sensitive to the anomaly.

\paragraph{Strip algebra.} We will not discuss the tube algebra symmetrizer here, since our main focus lies on the interval Hilbert space. We refer the reader to \cite{Benini:2025lav} for more details. We jump directly to the strip algebra symmetrizer, which is the relevant one for the interval Hilbert space. The key ingredient of the strip algebra is the basis of junction spaces between a TDL and two irreducible boundary conditions, represented graphically as:
\begin{align}
	\begin{gathered}
		\begin{tikzpicture}[scale=1]
			\fill [fill=tqft] (0,-1) rectangle (-1,1);
			\fill [fill=triv] (0,-1) rectangle (1,1);
			\draw [line,->-=.55] (0,-1) -- (0,-0.5) node [shift={(0,-0.2)}, right] {$m$} -- (0,0);
			\draw [line,->-=.55] (0,0) 	-- (0,0.5) node [shift={(0,0.2)}, right] {$n$} -- (0,1);
			\draw [line,->-=.6]  (-1,-0.7) node[shift={(0.44,0.3)}, above] {$a$} -- (0,0);
			\draw [fill=black] (0,0) circle (0.07) node[right] {$\rho$};
		\end{tikzpicture}
	\end{gathered}~.
\end{align}
Bringing \(L_a\) to lie parallel to the boundary condition \(m\) and fusing it, we recover precisely the junction between \(L_a\otimes m\) and \(n\), denoted here by \(\rho\). The strip algebra is generated by a slight modification of the above picture. Its generators are given by:
\begin{align}\label{eq:sfHdef}
	\sfH_I \equiv \sfH_{(a;\rho,\mu;k,\ell,n,m)} = \begin{gathered}
		\begin{tikzpicture}[scale=1]
			\fill [fill=triv] (-0.5,-1) rectangle (0,1);
			\fill [fill=tqft] (0,-1) rectangle (1,1);
			\fill [fill=triv] (1,-1) rectangle (1.5,1);			
			\draw [line,->-=.55] (1,-1) -- (1,-0.5) node [shift={(0,-0.2)}, right] {$m$} -- (1,0);
			\draw [line,->-=.55] (1,0) 	-- (1,0.5) node [shift={(0,0.2)}, right] {$n$} -- (1,1);
			\draw [fill=black] (1,0.1) circle (0.07) node[right] {$\mu$};
			\draw [line,->-=.6]  (0,-0.1) node[shift={(0.5,0.07)}, above] {$a$} -- (1,0.1);
			\draw [line,->-=.55] (0,-1) -- (0,-0.5) node [shift={(0,-0.2)}, left] {$k$} -- (0,0);
			\draw [line,->-=.55] (0,0) 	-- (0,0.5) node [shift={(0,0.2)}, left] {$\ell$} -- (0,1);
			\draw [fill=black] (0,-0.1) circle (0.07) node[left] {$\rho$};
		\end{tikzpicture}
	\end{gathered}~.
\end{align}
Here \(a\in\irr(\cC)\), \(k,\ell,m,n\in\irr(\cM)\), while \(\rho\in\hom(k,a\otimes \ell)\) and \(\mu\in\hom(a\otimes m,n)\) label bases of the junction spaces associated to the two boundary actions, and we use the shorthand notation \(I\) to refer collectively to all of these labels. Thus, the bulk line acts on the two endpoints of the interval and changes the boundary sector from \((k,m)\) to \((\ell,n)\). 

This generates an algebra, with the multiplication obtained by stacking two such strips and reducing the two bulk lines with the associator of \(\cC\) and the module associators of \(\cM\). We denote by \(\tilde{F}\) the module \(F\)-symbols, i.e. the change-of-basis matrices between \((L_b\otimes L_a)\otimes k\) and \(L_b\otimes (L_a\otimes k)\), and similarly for the right boundary label \(m\). In components we write
\begin{align}\label{eq:H-fusion}
	\sfH_I\,\sfH_J
	=
	\sum_K Q^K_{IJ}\,\sfH_K~.
\end{align}
Here \(Q^K_{IJ}\) are the structure constants of the strip algebra. For composable strips,  one finds \cite{Benini:2025lav}
\begin{align}
	Q^K_{IJ}
    =
	\sum_{\alpha}
	\sqrt{\frac{d_a d_b}{d_c}}\,
	\qty[\tilde{F}^{\,t}_{b a k}]_{(c;\alpha\tau)(\ell;\rho\sigma)}
	\qty[\tilde{F}^{\,p}_{b a m}]^{*}_{(c;\alpha\pi)(n;\mu\nu)}~,
\end{align}
where \(I=(a;\rho,\mu;k,\ell,m,n)\), \(J=(b;\sigma,\nu;\ell,t,n,p)\), and \(K=(c;\tau,\pi;k,t,m,p)\). Here \(\alpha\) labels a basis of \(\hom_\cC(L_b\otimes L_a,L_c)\), while \(\rho,\sigma,\tau\) and \(\mu,\nu,\pi\) label the left and right boundary junction spaces, respectively.

The symmetrizer is then the universal one for a finite semisimple symmetry algebra \cite{Benini:2025lav}. Define the regular trace pairing
\begin{align}
	K_{IJ}
	=
	\tr_{\operatorname{Strip}_\cC(\cM)}
	\qty(\sfH_I\sfH_J)
	=
	\sum_{L,M} Q^L_{IM}\,Q^M_{JL}~,
\end{align}
and denote its inverse by \(\tilde{K}^{IJ}\). For any density matrix \(\rho\) on the interval Hilbert space the corresponding symmetrized density matrix is
\begin{align}\label{eq:strip-symmetrizer}
	\rho_\sfS
	=
	\sum_{I,J}\tilde{K}^{IJ}\,\sfH_I\,\rho\,\sfH_J~.
\end{align}
This is the direct analogue of the group average \cref{eq:rho_S-grp-def}. It projects \(\rho\) onto the subspace of operators commuting with the strip algebra. The interval entanglement asymmetry with respect to the categorical symmetry is then defined as
\begin{align}
	\entas_{\operatorname{Strip}_\cC(\cM)}(\rho)
	=
	\relent{\rho}{\rho_\sfS}
	=
	\ent(\rho_\sfS)-\ent(\rho)~,
\end{align}
with the corresponding Rényi version obtained by replacing \(\ent\) with \(\ent_\n\).

The choice of boundary condition now matters more crucially than in the entropy formulas above. If a strongly symmetric boundary exists, the module category has a single simple object, denoted \(*\), and the full strip algebra gives the desired symmetry action. This can happen only when the relevant quantum dimensions are integers, since one needs \(\tn^{*}_{a\, *}=d_a\). Note that this is a necessary, but not sufficient, condition. In fact, we will encounter situations where we can construct the boundary condition \(\cB_\sfS\), but that do not admit an action of the strip algebra. If only a weakly symmetric boundary exists, one restricts to the reduced strip algebra generated by the junctions that preserve that boundary. If no symmetric boundary exists, the boundary condition itself breaks the symmetry. In this case, \cite{Benini:2025lav} proposes an averaging prescription to extract the symmetry breaking associated with the bulk state, rather than the boundary condition. However, this averaging prescription effectively moves the construction back to the circle Hilbert space, rather than the interval Hilbert space. We will therefore restrict the discussion to theories and choices of boundary condition for which a strongly or weakly symmetric boundary exists.

\paragraph{The symmetry of the Hartle--Hawking state.} The above discussions allow us to isolate another special feature distinguishing the Hartle--Hawking state, $\ket{\HH}$. In this case it is a \emph{completely symmetry preserving state}. Entanglement asymmetry makes this statement quantitative: the EA of $\rho_\HH$ vanishes both directly on the circle Hilbert space, and upon reduction to an interval using the symmetric boundary condition, $\cB_\sfS$, when it exists.

For the circle asymmetry this follows from the fact that any symmetry operator, $L_a$, acts trivially on the Hartle--Hawking state by design:
\begin{equation}
    L_a\rho_{\HH}=\rho_{\HH}L_a=d_a\,\rho_{\HH}~,
\end{equation}
and as such it is its own symmetrized state $\rho_{\HH,\sfS}=\rho_{\HH}$ with vanishing asymmetry
\begin{equation}
    \entas_\t{fus}(\rho_{\HH})=0~.
\end{equation}

For the interval asymmetry, it is simple to verify that $\rho_{\HH}$ reduced upon an interval with the symmetric boundary condition, \cref{eq:SBC}, imposed on both ends is the maximally mixed state on the interval Hilbert space:
\begin{equation}
    \rho_{\HH,\cB_\sfS}=\frac{\1_{\cH_{\cB_\sfS,\cB_\sfS}}}{\tilde{\cD}^2}=\bigoplus_m\left(\frac{\td_m^2}{\tilde{\cD}^2
    }\right)\frac{\1_{\cH^{(m)}_{\cB_\sfS,\cB_{\sfS}}}}{\tilde d_m^2}~,
\end{equation}
which commutes with all operators, including the strip operators implementing the symmetry. Thus once again, it is its own symmetrized state
\begin{equation}
    \left(\rho_{\HH,\cB_{\sfS}}\right)_{\sfS}=\rho_{\HH,\cB_{\sfS}}~,
\end{equation}
with vanishing asymmetry
\begin{equation}
    \entas_\t{strip}(\rho_{\HH,\cB_{\sfS}})=0~.
\end{equation}

\section{Examples and applications}\label{sec:examp}

Having constructed the general cutting maps and characterized the structure of entanglement in general semisimple 2d TQFTs we now proceed to discuss some concrete examples and realizations. This section is intended to serve two purposes: firstly, we apply the generic mathematical formalisms we have developed above towards concrete and physically relevant systems, in some cases filling gaps in the literature. Secondly, we provide a pedagogical summary of the systems being examined, indicating their physical relevance. The aim is a self-contained discussion of each system resulting in concrete computations of its entanglement measures. Readers interested only in the formal, constructive aspects of entanglement may choose to move directly onward to the Discussion (\cref{sec:disco}), although they may find interest in some of the applications below.

The key results of this section have been collected below in \cref{table:summary-examples} for ease of reference. 

\begin{table}[tbp]
\centering
\footnotesize
\renewcommand{\arraystretch}{1.2}
\setlength{\extrarowheight}{1pt}  
\setlength{\tabcolsep}{5pt}
\newcolumntype{Y}{>{\raggedright\arraybackslash\setlength{\lineskiplimit}{1pt}\setlength{\lineskip}{1pt}}X}
\newcommand*{\theory}[2]{{\bfseries\boldmath#1}\enspace{\normalfont\textsection#2}}
\newcommand*{\notdone}{\red{\(\circ\)}}
\newcommand*{\nobs}{no \(\cB_\sfS\)}
\renewcommand*{\toprule}{\cmidrule[1.5pt]{1-3}}  
\renewcommand*{\bottomrule}{\cmidrule[1.5pt]{1-3}}  
\renewcommand*{\midrule}{\cmidrule[1pt]{1-3}}  
\newcommand*{\headrule}{\cmidrule[1.3pt]{1-3}}     
\newcommand*{\rowrule}{\cmidrule[0.15pt]{1-3}}   
\setlength{\aboverulesep}{0pt}  
\setlength{\belowrulesep}{0pt}  
\begin{tabularx}{\textwidth}{|c|Y|Y|}
\toprule
	& \theory{\(\Z_N\) gauge theory}{\ref{ssec:ZN}}
	& \theory{Dijkgraaf--Witten \((G,\omega)\)}{\ref{ssec:DWGen}} \\
\headrule
\(\td_m\)
	& \(\td_q=1\), \(q\in\Z_N\)
	& \(\td_\sfR=\dim\sfR\), \(\sfR\in\irr_\omega(G)\) \\
\rowrule
\(\ee(\rho_\HH)\)
	& \(\log N+N^{-1}\sum_q\log(g_q^{(L)}g_q^{(R)})\)
	& \(\log\abs{G}\)\newline
	  \hspace{1em}\(+\abs{G}^{-1}\sum_\sfR\qty(\dim\sfR)^2\log(g_\sfR^{(L)}g_\sfR^{(R)}/(\dim\sfR)^2)\) \\
\rowrule
\(\entas_\t{fus}(\rho_m)\)
	& \(\log N\) \ \cite{Benini:2025lav,Lamas:2025eay}
	& \(-\sum_\sfc\sigma^\omega_\sfR(\sfc)\log\sigma^\omega_\sfR(\sfc)\) \\
\rowrule
\(\entas_\t{strip}(\rho_{m,\cB_\sfS})\)
	& \(\log N\)
	& \(\log(\abs{G}/(\dim\sfR)^2)\) \\
\midrule
	& \theory{Untwisted \(D_4\) gauge theory}{\ref{sec:DWUnt}}
	& \theory{Twisted \(D_4\) gauge theory}{\ref{sec:DWTw}} \\
\headrule
\(\td_m\)
	& \(\td_{\sfA_{1,2}}=\td_{\sfB_{1,2}}=1\), \(\td_\sfE=2\)
	& \(\td_{\sfE_\pm}=2\) \\
\rowrule
\(\ee(\rho_\HH)\)
	& \(\log 4+\frac18\sum_{\sfR=\sfA_i,\sfB_i}\log(g_\sfR^{(L)}g_\sfR^{(R)})\)\newline
	  \hspace*{1em}\(+\frac12\log(g_\sfE^{(L)}g_\sfE^{(R)})\)
	& \(\log 2+\frac12\sum_\pm\log(g_{\sfE_\pm}^{(L)}g_{\sfE_\pm}^{(R)})\) \\
\rowrule
\(\entas_\t{fus}(\rho_m)\)
	& \(\sfA_i,\sfB_i:\ \frac94\log 2\); \ \(\sfE:\ \log 2\)
	& \(\log 2\) \\
\rowrule
\(\entas_\t{strip}(\rho_{m,\cB_\sfS})\)
	& \(\sfA_i,\sfB_i:\ 3\log 2\); \ \(\sfE:\ \log 2\)
	& \(\log 2\) \\
\midrule
	& \theory{Ising}{\ref{ssec:Ising}}
	& \theory{Fibonacci}{\ref{sec:Fib}} \\
\headrule
\(\td_m\)
	& \(\td_\1=\td_\psi=1\), \(\td_\sigma=\sqrt2\)
	& \(\td_\1=1\), \(\td_\tau=\varphi\) \\
\rowrule
\(\ee(\rho_\HH)\)
	& \(\log 4+\frac14\log(g_\1^{(L)}g_\1^{(R)}g_\psi^{(L)}g_\psi^{(R)})\)\newline
	  \hspace*{1em}\(+\frac12\log(g_\sigma^{(L)}g_\sigma^{(R)}/2)\)
	& \(\log(1+\varphi^2)+\frac{1}{1+\varphi^2}\log(g_\1^{(L)}g_\1^{(R)})\)\newline
	  \hspace*{1em}\(+\frac{\varphi^2}{1+\varphi^2}\log(g_\tau^{(L)}g_\tau^{(R)}/\varphi^2)\) \\
\rowrule
\(\entas_\t{fus}(\rho_m)\)
	& \(\1,\psi:\ \frac32\log 2\); \ \(\sigma:\ \log 2\) \ \cite{Benini:2025lav}
	& \(\approx 0.590\) \ \cite{Benini:2025lav} \\
\rowrule
\(\entas_\t{strip}(\rho_{m,\cB_\sfS})\)
	& \nobs
	& \nobs \\
\midrule
	& \theory{Haagerup \(\cH_3\), OO (regular)}{\ref{sec:HaagOO}}
	& \theory{Haagerup \(\cH_3\), DD}{\ref{sec:HaagDD}} \\
\headrule
\(\td_m\)
	& \(\td_{\alpha^i}=1\), \(\td_{\alpha^i\rho}=\xi\), \(i=0,1,2\)
	& \(\td_{m_0}=1\), \(\td_{m_1}=\xi\) \\
\rowrule
\(\ee(\rho_\HH)\)
	& \(\log 3(1+\xi^2)+\frac{1}{3(1+\xi^2)}\sum_i\log(g_{\alpha^i}^{(L)}g_{\alpha^i}^{(R)})\)\newline
	  \hspace*{1em}\(+\frac{\xi^2}{3(1+\xi^2)}\sum_i\log(g_{\alpha^i\rho}^{(L)}g_{\alpha^i\rho}^{(R)}/\xi^2)\)
	& \(\log(1+\xi^2)+\frac{1}{1+\xi^2}\log(g_{m_0}^{(L)}g_{m_0}^{(R)})\)\newline
	  \hspace*{1em}\(+\frac{\xi^2}{1+\xi^2}\log(g_{m_1}^{(L)}g_{m_1}^{(R)}/\xi^2)\) \\
\rowrule
\(\entas_\t{fus}(\rho_m)\)
	& \(\approx 1.657\) \ \cite{Benini:2025lav}
	& \(\approx 0.288\) \\
\rowrule
\(\entas_\t{strip}(\rho_{m,\cB_\sfS})\)
	& \nobs
	& \nobs \\
\midrule
	& \theory{Haagerup \(\cH_3\), OD}{\ref{sec:HaagOD}}
	& \theory{Topological cosets \(G_k/H_{k'}\)}{\ref{sec:TopoCoset}} \\
\headrule
\(\td_m\)
	& \(\td_{m_0}=\td_{m_1}=\td_{m_2}=1\), \(\td_{m_3}=\xi-2\)
	& \(\td_m=\cD_G/\cD_H\sum_\mu\tn^m_{\mu\1}\,d_\mu\) \\
\rowrule
\(\ee(\rho_\HH)\)
	& \(\log(8-\xi)+\frac{1}{8-\xi}\sum_{i=0}^{2}\log(g_{m_i}^{(L)}g_{m_i}^{(R)})\)\newline
	  \hspace*{1em}\(+\frac{5-\xi}{8-\xi}\log(g_{m_3}^{(L)}g_{m_3}^{(R)}/(5-\xi))\)
	& \(\log\cD_G\cD_H\)\newline
	  \hspace*{1em}\(+\qty(\cD_G\cD_H)^{-1}\sum_m\td_m^{\,2}\log(g_m^{(L)}g_m^{(R)}/\td_m^{\,2})\) \\
\rowrule
\(\entas_\t{fus}(\rho_m)\)
	& \(m_{0,1,2}:\;\approx 1.317\); \ \(m_3:\;\approx 0.654\)
	& dependent on fusion rules \\
\rowrule
\(\entas_\t{strip}(\rho_{m,\cB_\sfS})\)
	& \nobs
	& \nobs\ generically \\
\midrule
	& \theory{\(G_k/G_k\)}{\ref{sec:GmodG}}
	& \theory{\(\Spin(N^2-1)_1/\SU(N)_N\)}{\ref{sssec:adjoint-QCD}} \\
\headrule
\(\td_m\)
	& \(\td_\sfR=\bbS_{0\sfR}/\bbS_{00}\)
	& \(3\cdot 2^{N-2}\) vacua; \ \(N\) even: some \(\td_m=\sqrt2\)\newline
	  \(N=3:\ \td_{\1,v,s,c}=1\), \(\td_{\cN,\bar{\cN}}=2\) \\
\rowrule
\(\ee(\rho_\HH)\)
	& \(-2\log\bbS_{00}+\sum_\sfR\bbS_{0\sfR}^{\,2}\log(g_\sfR^{(L)}g_\sfR^{(R)}/\td_\sfR^{\,2})\)
	& \(\log(2/\bbS_{00})+\bbS_{00}/2\sum_m\td_m^{\,2}\log(g_m^{(L)}g_m^{(R)}/\td_m^{\,2})\) \\
\rowrule
\(\entas_\t{fus}(\rho_m)\)
	& \(- \sum_{a} \abs{\bbS_{a\sfR}}^2\log\abs{\bbS_{a\sfR}}^2\)
	& \(N=3:\ m_{\1,v,s,c}:\;\approx 1.661;\ m_{\cN,\bar{\cN}}:\;\approx 1.099\) \\
\rowrule
\(\entas_\t{strip}(\rho_{m,\cB_\sfS})\)
	& \nobs\ generically
	& \nobs\ generically  \\
\bottomrule
\end{tabularx}
\caption{Collection of the results of this section. Boundary quantum dimensions \(\td_m\) of the vacua \(\ket{m}\); Hartle--Hawking entanglement entropy on an interval with generic boundary conditions \cref{LRBC}; entanglement asymmetry of the vacua on the circle (fusion algebra) and on the interval (strip algebra, \(\cB_\sfS\) at both ends). Here \(\varphi=\frac{1+\sqrt5}{2}\), \(\xi=\frac{3+\sqrt{13}}{2}\) and \(\sigma^\omega_\sfR(\sfc)=\abs{\sfc}\abs{\chi^\omega_\sfR(\sfc)}^2/\abs{G}\) with \(\sfc\in\cl_\omega(G)\) (cf.\ \cref{eq:DW-entas-on-R-states}). Entries with a citation were first computed in the cited works. Entries marked ``\nobs'' have no available symmetric boundary condition, hence interval asymmetry is not defined.}
\label{table:summary-examples}
\end{table}

\subsection{\texorpdfstring{\(\Z_N\)}{Z\_N} gauge theory}
\label{ssec:ZN}
We will first consider the simplest instance of a 2d TQFT, \(\Z_N\) gauge theory \cite{seiberg2015, Bergeron:1994ym, blauLectures2gaugeTheories1993,Kapustin:2014gua,Kitaev:1997wr,Wen:2003yv,Levin:2006zz}. The theory consists of topological line operators $W_q$ and local operators $U_p$ respectively, with $p,q \in \Z_N$ satisfying a \emph{clock algebra} 
\begin{equation}
    U_p \, W_q = \ex{\frac{2\pi i}{N} p\,q}\; W_q \, U_p ~. 
    \label{bf:clock}
\end{equation}
It follows from \cref{bf:clock} that the operator algebra consists of $N$ distinct topological lines. One realization of \(\Z_N\) gauge theory in arbitrary dimensions is given by considering an Abelian $p$-form gauge field coupled to a $(d-p-1)$-form gauge field on a $d$-dimensional manifold $\Sigma$ \cite{Bergeron:1994ym}. The algebra of gauge invariant operators is realized by Wilson surface operators constructed from the corresponding forms. The topological lines correspond then to the Wilson surface operators constructed out of the $p$-form. The clock algebra \cref{bf:clock} can be obtained from a canonical formulation of the theory. In the two dimensional case of interest, the topological lines are the usual Wilson loop operator constructed out of the $\U(1)$ one-form field. This simple theory has seen renewed interest in the context of generalized symmetries and the symmetry TFT (SymTFT) paradigm \cite{Gaiotto:2014kfa,Apruzzi:2021nmk,Kaidi:2022cpf,Antinucci:2024zjp,Brennan:2024fgj,Argurio:2024ewp}, entanglement \cite{Fliss:2023uiv,Fliss:2023dze} and spontaneous symmetry breaking in de Sitter QFT \cite{Aguilera-Damia:2026dbk}.

In the categorical language used above, the underlying category \(\cC\) is the fusion category of \(\Z_N\) representations. Since \(\Z_N\) is a finite Abelian group, Pontryagin duality, or a baby version of Tannaka's reconstruction theorem \cite{Tannaka}, identifies it with the category of \(\Z_N\)-graded vector spaces: 
\begin{align}
    \cC\cong \Rep(\Z_N)\cong \operatorname{Vec}_{\Z_N}~.
\end{align}
The line operators $W_q$ can then be viewed as the simple objects of this category:
\begin{equation}
    \irr(\cC) = \set{W_0, W_1, \cdots, W_{N-1}}~.
\end{equation}
Their fusion rules are given by
\begin{equation}\label{eq:W-fusion}
    W_p \otimes W_q = W_{p+q}~,
\end{equation}
with \(p+q\) on the right-hand side only regarded mod \(N\). They yield the trivial fusion coefficients 
\begin{equation}
    N^c_{ab} = \delta_{c,a+b}~.
\end{equation}
All simple lines are invertible and have quantum dimension $d_q =1$. Thus 
\begin{equation}
    \cD^2 = \sum_{q=0}^{N-1} d^2_q = N ~.
\end{equation}
To proceed, we consider the regular module category $\cM = \cC$. The simple boundary conditions are labelled by $q \in \Z_N$ and the corresponding NIM-rep of the fusion algebra is 
\begin{equation}
    \left[ \tilde{N}_p\right]^{a}{}_{q} = \delta_{a,p+q} ~.
\end{equation}
It follows that the boundary quantum dimensions are 
\begin{equation}
    \td_q = 1 ~, \qquad \tilde{\cD}^2 = N ~,
    \label{BF:QuantumD}
\end{equation}
and the circle Hilbert space is spanned by states $\ket{q}$ with $q=0,1,2,\ldots,N-1$ with dimension 
\begin{equation}
    \dim \cH_{S^1} = N ~.
\end{equation}
From a path integral point of view, the states%
\footnote{These states correspond to the puniverses of \cite{Aguilera-Damia:2026dbk}.}
$\ket{q}$ correspond to the hemisphere partition with the insertion of $\Pi_q$. The topological local operators $\Pi_q$ are related to the natural local vertex operators $U_p$ of the theory by a discrete Fourier transform: 
\begin{align}
    \Pi_q = \frac{1}{N}\sum_{p=0}^{N-1} \ex{-\frac{2\pi\ii}{N} p\, q} U_p~.
\end{align}

In this basis, and with the normalization discussed in \cref{sec:2dTQFT}, the Hartle--Hawking state is 
\begin{equation}
    \ket{\HH} = \sum_{q=0}^{N-1} \ket{q} ~,
    \label{BF:HH}
\end{equation} 
and 
\begin{equation}
    \parti\qty\big[\S^2] = \braket{\HH} = N ~.
\end{equation}
Note that the sphere partition function computed here differs from the one obtained through a path integral computation in \cite{Aguilera-Damia:2026dbk} by an overall normalization. 

\subsubsection{Entanglement entropy}
We consider general left and right boundary conditions \cref{LRBC}
\begin{equation}
    	\cB_L = \bigoplus_{q=0}^{N-1} g_q^{(L)}\,W_q~,
	\qquad
	\cB_R = \bigoplus_{q=0}^{N-1} g_q^{(R)}\,W_q~ , 
    \qquad
    g_q^{(L)},g_q^{(R)}\in\Z_{>0}~,
\end{equation}
for which the degeneracy in a fixed sector is given by $g_q^{(L)}\, g_q^{(R)}$. 
Recall that a general pure state is given by 
\begin{equation}
    \ket{\Psi} = \sum_{q=0}^{N-1} \Psi_q \ket{q}~,
\end{equation}
where the normalized probabilities for each sector are
\begin{equation}
    \lambda_q = \frac{\abs{\Psi_q}^2}{\abs{\Psi}^2}~, \qq{with} \abs{\Psi}^2 = \sum_{q=0}^{N-1} \abs{\Psi_q}^2~.
\end{equation}
The entanglement entropy \cref{EE:MixBoundary} is characterized by the state and choice of boundary conditions 
\begin{align}
	\ee\qty(\rho_{\HH;\{\cB_{L,R}\}})
	=
	-\sum_{q=0}^{N-1}\lambda_q\log\lambda_q
	+
	\sum_{q=0}^{N-1}\lambda_q\log \qty(g_q^{(L)}\, g_q^{(R)})~.
    \label{EE:ExamplesMixBoundary}
\end{align}
For the Hartle--Hawking state \cref{BF:HH}, the probabilities are uniform,
\begin{equation}
    \lambda_q = \frac{1}{N} ~, 
\end{equation}
yielding
\begin{equation}
    \ee\qty(\rho_{\HH;\{\cB_{L,R}\}})
	= \log N + \sum_{q=0}^{N-1} \frac{1}{N} \log \qty(g_q^{(L)}\, g_q^{(R)})~.
\end{equation}
In this case the quantum dimensions are integers \cref{BF:QuantumD}, so we can consider the symmetric boundary condition \(\cB_\sfS\) \cref{eq:SBC}, 
\begin{equation}
    g_q^{(L)} = g_q^{(R)} = \td_q = 1 ~. 
\end{equation}
In this case we note that the edge contribution in \cref{EE:ExamplesMixBoundary} vanishes and we have 
\begin{equation}
    \ee\qty(\rho_{\HH;\cB_\sfS})
	= \log N ~,
\end{equation}
which is the log of the sphere partition function as expected.
\subsubsection{Entanglement asymmetry}

For the case of \(\Z_N\) gauge theory entanglement asymmetry, reviewed in \cref{app:entas}, was studied in \cite{Benini:2025lav,Lamas:2025eay}. We present it nonetheless, to set the stage for later examples.  

\paragraph{Circle asymmetry.}

The symmetry operators of \(\Z_N\) theory are the Wilson lines \(W_q\), \(q\in\Z_N\). From their algebra \cref{eq:W-fusion} one can immediately obtain the trace pairing \cref{eq:K-pairing} as
\begin{align}
	K_{q q'}
	=
	N\,\delta_{q+q',0}~,
	\qquad
	\tilde{K}^{q q'}
	=
	\frac{1}{N}\,\delta_{q+q',0}~.
\end{align}
It is immediate to see that the Wilson lines act on the circle basis by translation,
\begin{align}
	W_q\ket{p}=\ket{p+q}~.
\end{align}
Now take a generic pure state $\ket{\Psi}$ as above. The corresponding normalized density matrix is
\begin{align}
	\rho_\Psi
	=
	\frac{1}{\abs{\Psi}^2}
	\sum_{r,s\in\Z_N}
	\Psi_r\conj{\Psi_s}\ketbra{r}{s}~.
\end{align}
Symmetrizing with respect to the Wilson lines gives
\begin{align}
	\rho_{\Psi,\sfS}
	=
	\frac{1}{N}
	\sum_{q=0}^{N-1}
	W_q\,\rho_\Psi\,W_{-q}
	=
	\frac{1}{N\abs{\Psi}^2}
	\sum_{q,r,s=0}^{N-1}
	\Psi_r\conj{\Psi_s}\ketbra{r+q}{s+q}~.
\end{align}
Since the original state is pure, the circle entanglement asymmetry receives contributions only from the symmetrized density matrix
\begin{align}
    \entas_\t{circle}(\rho_\Psi) = \relent{\rho_\Psi}{\rho_{\Psi,\sfS}} = \ent(\rho_{\Psi,\sfS}) = \log N - \frac{1}{N} \sum_{p=0}^{N-1} \eta_p \log \eta_p~, 
\end{align}
with 
\begin{align}
    \eta_p = \frac{1}{\abs{\Psi}^2}\abs{\sum_{r=0}^{N-1} \Psi_r \ex{-\frac{2\pi\ii}{N}p\,r}}^2~.
\end{align}

For a basis state, \(\rho_p = \ketbra{p}\) the corresponding symmetrized density matrix is maximally mixed
\begin{align}
    \rho_{p,\sfS} = \frac{1}{N} \sum_{q=0}^{N-1} W_q \rho_p W_{-q} = \frac{1}{N} \sum_{q=0}^{N-1} \ketbra{q}~,
\end{align}
and so the resulting entanglement asymmetry is
\begin{align}
    \entas_\t{circle}(\rho_p) = \log N = \log\abs{\Z_N}~,
\end{align}
as expected, since \(\ket{p}\) are \(\Z_N\) symmetry breaking vacua. 

\paragraph{Interval asymmetry.}

The symmetric boundary condition is
\begin{align}
	\cB_\sfS=\bigoplus_{q=0}^{N-1} W_q~.
\end{align}
Since \(g_q^{(L)}=g_q^{(R)}=1\), the interval Hilbert space has one state in each diagonal sector. The reduced strip algebra is again the group algebra of \(\Z_N\). Its generators \(\sfH_p\) act by
\begin{align}
	\sfH_p\ket{q;\fri}=\ket{p+q;\fri}~,
\end{align}
where the Chan--Paton label is trivial for \(\cB_\sfS\).
Therefore
\begin{align}
	\sfH_p\otimes\sfH_{q}=\sfH_{p+q}~,
	\qquad
	K_{pq}^{\operatorname{strip}}
	=
	N\,\delta_{p+q,0}~,
	\qquad
	\tilde{K}_{\operatorname{strip}}^{pq}
	=
	\frac{1}{N}\,\delta_{p+q,0}~.
\end{align}
For a basis state, \(\rho_{q,\cB_\sfS}=\ketbra{q;\fri}{q;\fri}\), and hence
\begin{align}
	\rho_{q,\cB_\sfS;\sfS}
	=
	\frac{1}{N}
	\sum_{p=0}^{N-1}
	\sfH_p\,\rho_{q,\cB_\sfS}\,\sfH_{-p}
	=
	\frac{1}{N}
	\sum_{p=0}^{N-1}
	\ketbra{p;\fri}~.
\end{align}
Thus
\begin{align}
	\entas_{\operatorname{strip}}(\rho_{q,\cB_\sfS})=\log N~.
\end{align}
For a generic pure state, \(\rho_{\Psi,\cB_\sfS}\) is already diagonal in the interval basis
\begin{align}
	\rho_{\Psi,\cB_\sfS}
	=
	\sum_{q=0}^{N-1}
	\lambda_q\ketbra{q;\fri}~,
\end{align}
with $\lambda_q$ as above. The strip symmetrizer sends it to the maximally mixed state,
\begin{align}
	\rho_{\Psi,\cB_\sfS;\sfS}
	=
	\frac{1}{N}
	\sum_{q=0}^{N-1}
	\ketbra{q;\fri}~,
\end{align}
yielding an asymmetry as
\begin{align}
	\entas_{\operatorname{strip}}(\rho_{\Psi,\cB_\sfS})
	=
	\log N
	+
	\sum_{q=0}^{N-1}
	\lambda_q\log\lambda_q~.
\end{align}

\subsection{Dijkgraaf--Witten theory}
\label{ssec:DWGen}

An important generalization of the above example is two-dimensional Dijkgraaf--Witten (DW) theory \cite{Dijkgraaf:1989pz}. This is non-Abelian finite gauge theory, twisted by a discrete torsion. We therefore fix $G$ to be a finite non-Abelian group and take $[\omega]\in\H^2(G,\U(1))$ to be a discrete-torsion class. The path integral of the theory sums over flat
\(G\)-bundles weighted by a cocycle representative of the class $[\omega]$.%
\footnote{Of course, different cocycle representatives are related by gauge transformations, so all gauge-invariant observables only depend on the class \([\omega]\).}

A standard presentation of the theory is as a state sum on an oriented triangulated surface
\(\Sigma\) of genus \(\sfg\) \cite{Dijkgraaf:1989pz,Fukuma:1993hy,Turaev:2007}. In this presentation, we have a group element \(g_{ij}\in G\) on every oriented edge
\((ij)\), with \(g_{ji}=g_{ij}^{-1}\), satisfying the flatness condition, 
\(g_{ij}g_{jk}=g_{ik}\), on every triangle \((ijk)\). The exponentiated action
is
\begin{align}
	\ex{\ii S_\omega[g]}
	=
	\prod_{(ijk)}
	\omega(g_{ij},g_{jk})^{\varepsilon(ijk)}~,
\end{align}
where \(\varepsilon(ijk)=+1\) if the triangle orientation agrees with that of
\(\Sigma\), and \(\varepsilon(ijk)=-1\) otherwise. The resulting partition function is
\begin{align}\label{eq:DW-parti-1}
	\parti_\omega(\Sigma)
	=
	\frac{1}{\abs{G}^{\abs{E}-\abs{F}}}
	\sum_{\substack{g_{ij}\in G\\ g_{ij}g_{jk}=g_{ik}}}
	\prod_{(ijk)}
	\omega(g_{ij},g_{jk})^{\varepsilon(ijk)}~,
\end{align}
where \(\abs{E}\) and \(\abs{F}\) is the number of edges and faces, respectively, in the triangulation of $\Sigma$.%
\footnote{
Our normalization of the partition function differs from that in \cite{Dijkgraaf:1989pz,Turaev:2007} by a factor $\abs{G}^{\chi(\Sigma)}$, introduced via the Euler-counterterm discussed in \cref{ssec:higher-genus}.
}
The cocycle condition ensures
invariance under Pachner moves, so the partition function depends only on the topology of
\(\Sigma\). A central result of \cite{Turaev:2007} is that the partition function \cref{eq:DW-parti-1} can be rewritten in the form \cref{partFgenusg} as
\begin{align}
	\parti_\omega(\Sigma)
    =
	\sum_{\sfR\in\irr_\omega(G)}
	(\dim \sfR)^{\chi(\Sigma)}~,
\end{align}
where \(\irr_\omega(G)\) denotes the set of irreducible projective representations of $G$, with the projective phase given by $\omega$. From here, we can immediately read off the projector basis of local operators, 
\(\Pi_\sfR\) with \(\sfR\in\irr_\omega(G)\) and quantum dimensions \(\td_\sfR = \dim \sfR\). The dimension of the circle Hilbert space follows from the partition function on the torus:
\begin{align}
	\dim\cH_{\S^1} = \parti_\omega\qty\big(\T^2)
	=
	\abs{\irr_\omega(G)}~.
\end{align}
The circle Hilbert space is generated by states \(\ket{\sfR}\) as in \cref{eq:basis-state-PI}. A generic pure state, therefore, takes the form
\begin{align}
	\ket{\Psi}
	=
	\sum_{\sfR\in\irr_\omega(G)}
	\Psi_\sfR\ket{\sfR}~.
\end{align}

\subsubsection{Entanglement entropy}

For general left and right boundary conditions
\begin{align}
	\cB_L
	=
	\bigoplus_{\sfR\in\irr_\omega(G)} g_\sfR^{(L)}\sfR~,
	\qquad
	\cB_R
	=
	\bigoplus_{\sfR\in\irr_\omega(G)} g_\sfR^{(R)}\sfR~,
	\qquad
	g_\sfR^{(L)},g_\sfR^{(R)}\in\Z_{>0}~,
\end{align}
the interval Hilbert space decomposes as
\begin{align}
	\cH_{\cB_L,\cB_R}
	=
	\bigoplus_{\sfR\in\irr_\omega(G)} \cH_\sfR~,
	\qquad
	\cH_\sfR
	=
	\C^{g_\sfR^{(L)}}\otimes \C^{g_\sfR^{(R)}}~,
\end{align}
with \(\dim\cH_\sfR=g_\sfR^{(L)}g_\sfR^{(R)}\). From our general formula \cref{EE:MixBoundary} we get directly
the entanglement entropy:
\begin{align}\label{eq:DW-EE}
	\ee\qty(\rho_{\Psi,\cB_L,\cB_R})
	=
	-\sum_{\sfR\in\irr_\omega(G)}\lambda_\sfR\log\lambda_\sfR
	+
	\sum_{\sfR\in\irr_\omega(G)}\lambda_\sfR
	\log\qty[g_\sfR^{(L)}g_\sfR^{(R)}]~,
\end{align}
where \(\lambda_\sfR = \abs{\Psi_\sfR}^2/\abs{\Psi}^2\). 

\paragraph{Vacuum states.}

Taking our state to be one of the clustering vacua \(\ket{\Psi}=\ket{\sfR}\), one finds
\begin{align}
	\ee\qty(\rho_{\sfR,\cB_L,\cB_R})
	=
	\log(g_{\sfR}^{(L)}g_{\sfR}^{(R)}) = \log\dim\cH_\sfR~.
\end{align}
In other words, the vacua isolate the Affleck--Ludwig entropies. 

\paragraph{Hartle--Hawking state.}

For these theories, the Hartle--Hawking state is
\begin{align}\label{eq:DW-HH-state}
	\ket{\HH} = \sum_{\sfR\in\irr_\omega(G)}\dim\sfR\,\ket{\sfR}~.
\end{align}
Its entanglement entropy reads%
\footnote{In the above we have also used that the sum-of-squares formula continues to hold for projective representations, so \(\sum_{\sfR\in\irr_\omega(G)}(\dim\sfR)^2=\abs{G}\). See for instance \cite{Kobayashi:2025pxs} for a proof.}
\begin{align}\label{eq:DW-HH-entropy}
	\ee\qty(\rho_{\HH,\cB_L,\cB_R}) = \log\abs{G}+ \sum_{\sfR\in\irr_\omega(G)} \frac{(\dim\sfR)^2}{\abs{G}}\log(\frac{g_\sfR^{(L)}g_\sfR^{(R)}}{(\dim\sfR)^2})~.
\end{align}

\paragraph{Symmetric boundary condition.}

Since the quantum dimensions \(\td_\sfR = \dim \sfR\) are integer, the symmetric boundary condition always exists and is given by 
\begin{align}
	\cB_\sfS
	=
	\bigoplus_{\sfR\in\irr_\omega(G)}(\dim \sfR)\,\sfR~.
\end{align}
Taking the symmetric boundary condition at both ends of the interval the interval Hilbert space becomes
\begin{align}
	\cH_{\cB_\sfS} = \bigoplus_{\sfR\in\irr_\omega(G)} \sfR\otimes \sfR^* \cong \C_\omega[G]~.
\end{align}
In the above, \(\C_\omega[G]\) denotes the group algebra of \(G\)  twisted by \(\omega\)\footnote{Concisely, the twisted group algebra $\C_\omega[G]$ is the group algebra of the group $G^\omega \coloneqq G\times \U(1)$ with group multiplication $(g,\alpha)\times(h,\beta) = (gh,\alpha+\beta+\omega(g,h))$.} and the equivalence with the left-hand side follows, essentially, by its Wedderburn decomposition; see \cite{Turaev:2007}. %
With this choice of boundary conditions the second summand in the Hartle--Hawking entropy \cref{eq:DW-HH-entropy} vanishes, and we are left simply with
\begin{align}
	\ee\qty(\rho_{\HH,\cB_\sfS}) = \log\abs{G}~.
\end{align}

\paragraph{Untwisted \(D_4\) gauge theory.} \label{sec:DWUnt}

To illustrate our results further, we consider the simplest non-Abelian gauge theory that admits a non-trivial twist. That is DW theory over the dihedral group of order \(8\):
That is DW theory over the dihedral group of order \(8\):
\begin{align}\label{eq:D4-group}
	D_4
	=
	\left\langle \fa,\fb \suchthat \fa^4=\fb^2=1,\ \fb\fa\fb=\fa^{-1} \right\rangle~.
\end{align}
Since \(\H^2(D_4,\U(1))\cong\Z_2\) there is exactly one non-trivial twist. 

We first study the untwisted case. It has four one-dimensional irreps, \(\sfA_1\), \(\sfA_2\), \(\sfB_1\) and \(\sfB_2\), and one two-dimensional irrep, \(\sfE\). The Hartle--Hawking state is, therefore,
\begin{align}
	\ket{\HH} = \ket{\sfA_1}+\ket{\sfA_2}+\ket{\sfB_1}+\ket{\sfB_2}+2\ket{\sfE}~.
\end{align}
Its entanglement entropy with generic boundary conditions is
\begin{align}
	\ee\qty(\rho_{\HH,\cB_L,\cB_R}) = \log 4 + \frac{1}{8} \sum_{\substack{\sfR=\sfA_i,\sfB_i \\ i=1,2}} \log(g_\sfR^{(L)}g_\sfR^{(R)}) + \frac{1}{2} \log(g_\sfE^{(L)}g_\sfE^{(R)})~.
\end{align}
In the symmetric case, \(g_\sfR^{(L)}=g_\sfR^{(R)}=\dim\sfR\) we get
\begin{align}
	\ee\qty(\rho_{\HH,\cB_\sfS}) = \log 4 + \frac{1}{8} \cdot 0 + \frac{1}{2} \log 4 = \log 8 = \log \abs{D_4}~.
\end{align} 

\paragraph{Twisted \(D_4\) gauge theory.}

For the non-trivial class in \(\H^2(D_4,\U(1))\cong\Z_2\), there are two
irreducible projective representations \(\sfE_+\) and \(\sfE_-\), both of dimension
two \cite{Karpilovsky1985}. The twisted Hartle--Hawking state is then
\begin{align}
	\ket{\HH}
	=
	2\ket{\sfE_+}+2\ket{\sfE_-}~.
\end{align}
Its entropy with generic boundary conditions \(\cB_{L/R} = g_+^{(L/R)}\sfE_+ \oplus g_-^{(L/R)} \sfE_-\),
is
\begin{align}
	\ee\qty(\rho_{\HH,\cB_L,\cB_R})
	=
	\log 2
	+
	\frac{1}{2}\log(g_+^{(L)}g_+^{(R)})
	+
	\frac{1}{2}\log(g_-^{(L)}g_-^{(R)})~.
\end{align}
For the symmetric boundary condition \(\cB_\sfS\): \(g_\pm^{(L)}=g_\pm^{(R)} = 2\)
\begin{align}
	\ee\qty(\rho_{\HH,\cB_\sfS})
	&=\log 2+\log 4
	=\log 8~.
\end{align}
Thus the twist removes the zero-entropy one-dimensional sectors, but leaves the
Hartle--Hawking entropy equal to \(\log\abs{D_4}=\log 8\).

\subsubsection{Entanglement asymmetry}

\paragraph{Circle asymmetry.}

Dijkgraaf--Witten theory has a fusion category symmetry \(\Rep(G)\), generated by Wilson lines \cite{Bhardwaj:2017xup}. Their fusion rules follow from the representation ring of \(G\):
\begin{align}
	W_a \otimes W_b
	=
	\bigoplus_{c\in\irr(G)}
	N_{ab}^{c}\,W_c~,
	\qquad
	N_{ab}^{c}
	=
	\dim\hom(c,a\otimes b)~.
\end{align}
For non-Abelian \(G\) the symmetry is non-invertible. On the basis \(\ket{\sfR}\) of circle states, the Wilson lines act as follows:
\begin{align}\label{eq:DW-W-act-on-states}
	W_a\ket{\sfR}
	=
	\sum_{\sfR'\in\irr_\omega(G)}
	\tn_{a \sfR}^{\sfR'}\ket{\sfR'}~,
	\qquad
	\tn_{a \sfR}^{\sfR'}
	=
	\dim\hom(\sfR',a\otimes \sfR)~.
\end{align}
Note that here \(a,b,c\) are ordinary representations, while \(\sfR,\sfR'\) are projective representations. Tensoring a projective representation with an ordinary representation gives another projective representation with the same projective phase \cite{Bartsch:2022mpm}, which explains \cref{eq:DW-W-act-on-states}. 

It will be convenient to introduce a dual set of basis states that diagonalize the Wilson lines.%
\footnote{This serves as the non-Abelian generalization of the story of \textquote{vacua vs puniverses} of \cite{Aguilera-Damia:2026dbk}.}
In the Abelian case discussed above, the dual basis corresponds to a hemisphere path integral with an insertion of a vertex operator at the south pole. In the non-Abelian case, vertex operators are labelled instead by conjugacy classes; see, e.g. \cite{Sharpe:2021srf}. With discrete torsion turned on, they are in fact labelled by so-called \emph{\(\omega\)-regular conjugacy classes} \cite{Ramgoolam:2022xfk}.%
\footnote{A group element \(g\in G\) is said to be \(\omega\)-regular if it satisfies \(\omega(g,h) = \omega(h,g)\), for all \(h\in G\) that commute with \(g\). An \(\omega\)-regular conjugacy class \cite{Karpilovsky1985} is a conjugacy class consisting of only \(\omega\)-regular elements.}
We denote the set of \(\omega\)-regular conjugacy classes as \(\cl_\omega(G)\). The dual basis of circle states is then given as
\begin{align}
	\ket{\sfc;\omega} = \sqrt{\frac{\abs{\sfc}}{\abs{G}}}\; \sum_{\sfR\in\irr_\omega(G)} \overline{\chi_\sfR^\omega(\sfc)}\ket{\sfR}~, \qq{with} \sfc\in\cl_\omega(G)~,
\end{align}
where \(\chi_\sfR^\omega\) denotes the projective character of \(\sfR\); orthogonality of projective characters shows that \(\braket{\sfc;\omega}{\sfc';\omega}=\delta_{\sfc,\sfc'}\). This is precisely the non-Abelian Fourier transform of the original states \cite{Karpilovsky1985,Ramgoolam:2022xfk}. Inverting the Fourier transform gives \(\ket{\sfR}\) in terms of \(\ket{\sfc;\omega}\) as
\begin{align}\label{eq:DW-inverse-fourier}
	\ket{\sfR} = \sum_{\sfc\in\cl_\omega(G)} \sqrt{\frac{\abs{\sfc}}{\abs{G}}} \chi_\sfR^\omega(\sfc)\ket{\sfc;\omega}~.
\end{align}
Note that the Hartle--Hawking state is a particular instance of the dual basis states. Indeed, 
\begin{align}
	\ket{\HH} = \sum \td_\sfR \ket{\sfR}  = \sum_{\sfc\in\cl_\omega(G)} \sqrt{\frac{\abs{\sfc}}{\abs{G}}}\sum_\sfR \td_\sfR \chi_\sfR^\omega(\sfc)\ket{\sfc;\omega}~.
\end{align}
We recognize \(\chi_\t{reg}^\omega(\sfc) \coloneqq \sum_\sfR \td_\sfR \chi_\sfR^\omega(\sfc)\) as the character of the regular projective representation, satisfying \(\chi_\t{reg}^\omega(g) = \abs{G}\delta_{g,\1}\) \cite{Kobayashi:2025pxs}. Hence the sum localizes on the conjugacy class of the identity and gives 
\begin{align}\label{eq:DW-HH-is-ket1}
	\ket{\HH} = \sqrt{\abs{G}}\ket{\1;\omega}~.
\end{align}
To show that the conjugacy-class basis diagonalizes the action of the Wilson lines, we need to utilize the projective character formula%
\footnote{This mirrors the well-known formula for the untwisted case \begin{align}
	N_{ab}^c = \sum_{\sfc\in \cl(G)} \frac{\abs{\sfc}}{\abs{G}}\; \chi_a(\sfc)\chi_b(\sfc)\overline{\chi_c(\sfc)}~.
\end{align} 
See, for instance, \cite{etingof2011introduction}.}
\begin{align}
	\chi_a(\sfc) \chi_{\sfR}^\omega(\sfc) = \sum_{\sfR'\in\irr_\omega(G)} \tilde{N}_{a \sfR}^{\sfR'}\chi_{\sfR'}^\omega(\sfc) \quad\eqv\quad \tilde{N}_{a \sfR}^{\sfR'} = \sum_{\sfc\in\cl_\omega(G)} \frac{\abs{\sfc}}{\abs{G}}\chi_a(\sfc)\chi_{\sfR}^\omega(\sfc)\overline{\chi_{\sfR'}^\omega(\sfc)}~.
\end{align}

From which we can calculate
\begin{align}\label{eq:DW-W-on-c-states}
	W_a\ket{\sfc;\omega} = \chi_a(\sfc) \ket{\sfc;\omega}~.
\end{align}

We have now almost all ingredients to compute the asymmetry. The last ingredient is the bilinear pairing \cref{eq:K-pairing}. It follows directly from the character formula above that
\begin{align}
	K_{ab} = \sum_{c,d\in\irr(G)} N^d_{ac} N^c_{bd} = \sum_{\sfc\in\cl(G)} \chi_a(\sfc)\,\chi_b(\sfc)~.
\end{align}
Its inverse follows from character orthogonality:
\begin{align}
	\tilde{K}^{ab}
	=
	\sum_{\sfc\in\cl(G)}
	\frac{\abs{\sfc}^2}{\abs{G}^2}\,
	\overline{\chi_a(\sfc)} \; \overline{\chi_b(\sfc)}~.
\end{align}

Consider now the pure basis state \(\rho_\sfR=\ketbra{\sfR}{\sfR}\). Its symmetrization with respect to the fusion algebra (cf. \cref{eq:symmetrizer-fusion}) is
\begin{align}
	\rho_{\sfR,\sfS}
	&=
	\sum_{a,b\in\irr(G)}
	\tilde{K}^{ab}
	W_a\ketbra{\sfR}{\sfR}W_b
	=
	\sum_{\sfc\in\cl_\omega(G)}
	\sigma_\sfR^\omega(\sfc)
	\ketbra{\sfc;\omega}{\sfc;\omega}~,
\end{align}
with 
\begin{align}
	\sigma_\sfR^\omega(\sfc) \coloneqq \abs{\braket{\sfc;\omega}{\sfR}}^2 = \frac{\abs{\sfc}}{\abs{G}}
	\abs{\chi_\sfR^\omega(\sfc)}^2~,
\end{align}
where we used the inverse Fourier transform, \cref{eq:DW-inverse-fourier}, and the action of the Wilson lines on the conjugacy classes, \cref{eq:DW-W-on-c-states}. Hence the fusion symmetrizer dephases the state in the
normalized conjugacy-class basis, as expected from the original definition of entanglement asymmetry (for invertible symmetries) \cite{Ares:2022koq}. Hence, the entanglement asymmetry of the basis state \(\ket{\sfR}\), with respect to the fusion algebra is simply
\begin{align}\label{eq:DW-entas-on-R-states}
	\entas_\t{fus}(\rho_\sfR) = - \sum_{\sfc\in\cl_\omega(G)} \sigma_\sfR^\omega(\sfc)
	\log\sigma_\sfR^\omega(\sfc)
	~.
\end{align}
Let us note a few properties of the entanglement asymmetry. First, it is obviously non-negative. Second, it is maximized if all \(\sigma_\sfR^\omega(\sfc)=1/\abs{\cl_\omega(G)}\). Since \(\abs{\cl_\omega(G)}=\abs{\irr_\omega(G)}\) \cite{Kobayashi:2025pxs} we get the bounds:
\begin{align}
	0 \leq \entas_\t{fus}(\rho_\sfR) \leq \log\abs{\irr_\omega(G)}~.
\end{align}
In fact, the upper bound is a global upper bound for the fusion entanglement asymmetry of a generic state in DW theory. On the other hand, the lower bound is rather loose for these states. Note that the conjugacy class of the identity already contributes some asymmetry since \(\chi_\sfR^\omega(\1) = \dim\sfR = \td_\sfR\). Hence we can obtain a tighter lower bound 
\begin{align}
	\entas_\t{fus}(\rho_\sfR) \geq - \frac{\td_\sfR^{\,2}}{\abs{G}}\log\frac{\td_\sfR^{\,2}}{\abs{G}} - \qty(1-\frac{\td_\sfR^{\,2}}{\abs{G}})\log(1-\frac{\td_\sfR^{\,2}}{\abs{G}})~.
\end{align} 
This is strictly positive, unless \(\abs{\irr_\omega(G)}=1\), in which case the entire computation trivializes. Hence, the states \(\ket{\sfR}\) always break the fusion category symmetry. Conversely, a quick calculation mirroring that of the Hartle--Hawking state shows that the conjugacy-class-basis states preserve the entire symmetry.\footnote{On the other hand, all of these states break the dual 1-form symmetry, generated by the conjugacy classes. We will not compute their 1-form asymmetry here.}

\paragraph{Untwisted \(D_4\) gauge theory.}

Let us return to our recurrent example of \(D_4\) gauge theory to spell out our results. We begin with the untwisted theory. In the presentation of \cref{eq:D4-group} the conjugacy classes are
\begin{align}
	\sfc_1=\set{\1}~,\quad
	\sfc_2=\set{\fa^2}~,\quad
	\sfc_3=\set{\fa,\fa^3}~,\quad
	\sfc_4=\set{\fb,\fa^2\,\fb}~,\quad
	\sfc_5=\set{\fa\,\fb,\fa^3\,\fb}~.
\end{align}
The characters of the irreducible representations \(\sfA_i\), \(\sfB_i\), \(\sfE\) are
\begin{align}
\begin{array}{c|ccccc}
	\chi_{\sfR}(\sfc) & \sfc_1 & \sfc_2 & \sfc_3 & \sfc_4 & \sfc_5 \\ \hline
	\sfA_1 & 1 & 1 & 1 & 1 & 1 \\
	\sfA_2 & 1 & 1 & 1 & -1 & -1 \\
	\sfB_1 & 1 & 1 & -1 & 1 & -1 \\
	\sfB_2 & 1 & 1 & -1 & -1 & 1 \\
	\sfE   & 2 & -2 & 0 & 0 & 0
\end{array}
\end{align}
Applying our general formula, \cref{eq:DW-entas-on-R-states} on the representations above we find 
\begin{align}
	\entas_\t{fus}\qty(\rho_{\sfA_i}) = \entas_\t{fus}\qty(\rho_{\sfB_i}) &= \frac{9}{4}\log 2~, \\
	\entas_\t{fus}(\rho_{\sfE}) &= \log 2~.
\end{align}
An interesting feature here is that, unlike the invertible case, in the non-invertible case different vacua within the same gapped phase, break the symmetry by different amounts, as reflected in their entanglement asymmetry. This was first observed in \cite{Benini:2025lav} and is exemplified further by DW theory. We further see here, that neither vacuum reaches the maximal allowed symmetry breaking of \(\log\abs{\irr_\omega(G)}=\log 5 \approx 1.61\), although the one-dimensional representations arrive quite close, \(\frac{9}{4}\log 2 \approx 1.56\). 

\paragraph{Twisted \(D_4\) gauge theory.}\label{sec:DWTw}

Of the conjugacy classes listed above, only \(\sfc_1\) and \(\sfc_3\) are \(\omega\)-regular \cite{Karpilovsky1985,Sharpe:2021srf}. The corresponding projective character table reads
\begin{align}
\begin{array}{c|cc}
	\chi_\sfR^\omega(\sfc)& \sfc_1 & \sfc_3 \\ \hline
	\sfE_+ & 2 & 1+\ii \\
	\sfE_- & 2 & -1-\ii
\end{array}
\end{align}
in the cocycle convention of \cite{Sharpe:2021srf} and choosing as a representative of \(\sfc_3\) the element \(g=a\). The other choice of
representative would simply complex-conjugate the second column. The asymmetry follows:
\begin{align}
	\entas_\t{fus}\qty(\rho_{\sfE_\pm}) = \log 2 = \log\abs{\irr_\omega(G)}~. 
\end{align} 
Here we see that the two vacua break the symmetry maximally, by the same amount. 

\paragraph{Interval asymmetry.}

We now consider the interval Hilbert space with the symmetric boundary condition \(\cB_\sfS\) imposed at both ends. As in the circle case, in order to compute the asymmetry conveniently, our strategy will be to find a basis of states of the interval Hilbert space that diagonalizes the action of the relevant symmetry, here given by the strip algebra.

Recall that with symmetric boundary conditions the interval Hilbert space is the twisted group algebra
\begin{align}\label{eq:projective-peter-weyl}
	\cH_{\cB_\sfS} \cong \C_\omega[G] \cong \bigoplus_{\sfR\in\irr_\omega(G)} \sfR\otimes\sfR^*~,
\end{align}
with dimension
\begin{align}
	\dim\cH_{\cB_\sfS} = \sum_{\sfR\in\irr_\omega(G)} \td_\sfR^{\,2} = \abs{G}~.
\end{align}
The space \(\cH_{\cB_\sfS}\) is spanned by states \(\ket{\sfR;i_L,i_R}\) with \(\sfR\in\irr_\omega(G)\) and \(i_L,i_R=1,\ldots,\td_\sfR\), where \(\td_\sfR=\dim\sfR\). The second isomorphism in \cref{eq:projective-peter-weyl} is best understood as a projective version of the Peter--Weyl theorem \cite{Cheng-characters}. As in the classical Peter--Weyl theorem, this enables a Fourier transform. In our notation, the orthonormal basis states \(\ket{\sfR;i_L,i_R}\) can be expressed as
\begin{align}\label{eq:interval-projective-fourier}
	\ket{\sfR;i_L,i_R}
	&=
	\sqrt{\frac{\td_\sfR}{\abs{G}}}
	\sum_{g\in G}
	\qty[D_\sfR^\omega(g)]_{i_L\,i_R}\ket{g;\omega}~,
\end{align}
or, in the opposite direction,
\begin{align}\label{eq:inverse-interval-fourier}
	\ket{g;\omega}
	=
	\sum_{\sfR\in\irr_\omega(G)}
	\sqrt{\frac{\td_\sfR}{\abs{G}}}
	\sum_{i_L,i_R=1}^{\td_\sfR}
	\overline{\qty[D_\sfR^\omega(g)]_{i_L\,i_R}}
	\ket{\sfR;i_L,i_R}~.
\end{align}
Here, \(D^\omega_\sfR(g)\) are the projective representation matrices (or projective Wigner \(D\)-functions), satisfying \(D_\sfR^\omega(g)\,D_\sfR^\omega(h)=\omega(g,h)\,D_\sfR^\omega(gh)\). Their traces give precisely the projective characters \(\chi_\sfR^\omega(g)\). Applying the Fourier transform to the normalized states \(\ket{\sfR}=\td_\sfR^{-1/2}\sum_i\ket{\sfR;i,i}\) reproduces the Fourier transform of the circle states, \cref{eq:DW-inverse-fourier}, providing a consistency check.

In our setup, the strip algebra is generated by open Wilson lines labelled by \(a\in\irr(G)\) that end on the symmetric boundary condition \(\cB_\sfS\). Pictorially, its generators are
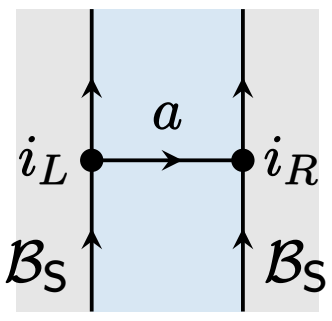
\begin{align}
	\sfH_{a;i_L,i_R} = \sqrt{d_a}
	\begin{gathered}
		\begin{tikzpicture}[scale=1]
			\fill [fill=triv] (-0.5,-1) rectangle (0,1);
			\fill [fill=tqft] (0,-1) rectangle (1,1);
			\fill [fill=triv] (1,-1) rectangle (1.5,1);			
			\draw [line,->-=.55] (1,-1) -- (1,-0.5) node [shift={(0,-0.2)}, right] {$\cB_\sfS$} -- (1,0);
			\draw [line,->-=.55] (1,0) 	-- (1,0.5) -- (1,1);
			\draw [fill=black] (1,0) circle (0.07) node[right] {$i_R$};
			\draw [line,->-=.6]  (0,-0) node[shift={(0.5,0.07)}, above] {$a$} -- (1,0);
			\draw [line,->-=.55] (0,-1) -- (0,-0.5) node [shift={(0,-0.2)}, left] {$\cB_\sfS$} -- (0,0);
			\draw [line,->-=.55] (0,0) 	-- (0,0.5) -- (0,1);
			\draw [fill=black] (0,-0) circle (0.07) node[left] {$i_L$};
		\end{tikzpicture}
	\end{gathered}~.
\end{align}
The normalization factor \(\sqrt{d_a}\) is required for consistency between \cref{eq:H-fusion} and the fusion rules of open Wilson lines, with the module \(F\)-symbols in \cref{eq:H-fusion} given by the Clebsch--Gordan coefficients. The normalized group states \(\ket{g;\omega}\) are given pictorially as
\begin{align}
	\ket{g;\omega} = 
	\frac{1}{\sqrt{\abs{G}}}
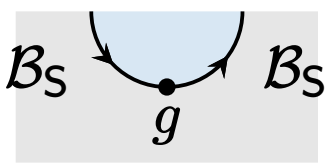
\begin{gathered}
		\begin{tikzpicture}
			\begin{scope}[even odd rule]
				\clip (0,0) ++(180:0.5) arc (180:360:0.5)  (-1,-1) rectangle (1,0);
				\fill [fill=triv] (-1,-1) rectangle (1,0);
			\end{scope}
			\fill [fill=tqft] (0,0) ++(180:0.5) arc (180:360:0.5);
			\draw [line,->-=.25,->-=.8] (0,0) ++(180:0.5) arc (180:360:0.5);
			\draw (-.5,-.1) node[below left] {$\cB_\sfS$};
			\draw (.5,-.1) node[below right] {$\cB_\sfS$};
			\draw [fill=black] (0,-.5) circle (.05) node[below] {$g$};
		\end{tikzpicture}
	\end{gathered}~.
\end{align}
Hence the action of the strip generators on the states is simply
\begin{align}
	\sfH_{a;i_L,i_R}\ket{g;\omega}
	=
	\sqrt{d_a}\,\qty[D_a(g)]_{i_L\,i_R}\ket{g;\omega}~.
\end{align}
From here we could immediately compute the symmetrizer in terms of the generators \(\sfH_{a;i_L,i_R}\). However, it is much easier to switch to a different basis of the strip algebra. Let us define the projectors \(\sfP_g=\ketbra{g;\omega}\), with fusion \(\sfP_g\sfP_h=\delta_{g,h}\sfP_g\). Resolving the identity operator in terms of \(\sfP_g\) gives the map
\begin{align}
	\sfH_{a;i_L,i_R}
	&=
	\sqrt{d_a}
	\sum_{g\in G}
	\qty[D_a(g)]_{i_L\,i_R}\sfP_g~ \qquad \eqv
	\nn
	\sfP_g
	&=
	\sum_{a\in\irr(G)}
	\frac{\sqrt{d_a}}{\abs{G}}
	\sum_{i_L,i_R=1}^{d_a}
	\overline{\qty[D_a(g)]_{i_L\,i_R}}
	\sfH_{a;i_L,i_R}~.
\end{align}
This is nothing but the Fourier transform once again. The upshot is that we can use \(\sfP_g\) as a basis of the strip algebra. In this basis, computing the symmetrizer is trivial. Since \(\sfP_g\ket{h;\omega}=\delta_{g,h}\ket{g;\omega}\), we have
\begin{align}
	K_{gh}^{\t{strip}} = \tr(\sfP_g\sfP_h) = \delta_{gh}
	\qq{and hence}
	\tilde{K}^{gh}_{\t{strip}} = \delta_{gh}~.
\end{align} 
Therefore the strip symmetrizer reads:
\begin{align}\label{eq:DW-interval-strip-symmetrizer}
	\rho_\sfS = \sum_{g\in G} \sfP_g\, \rho\,\sfP_g~.
\end{align}
Notice that neither the algebra nor its symmetrizer depend on \(\omega\). The twist enters only through the state \(\rho\), via the projective representations and its Fourier decomposition \cref{eq:interval-projective-fourier}.

We pause here to note, briefly, that the deeper reason why the symmetrizer is so simple in this basis lies in the fact that the strip algebra is a Hopf algebra \cite{Kitaev:2011dxc,Cordova:2024iti,AliAhmad:2025bnd}. As explained in \cite{Benini:2025lav}, for symmetry algebras given by (weak) Hopf algebras, the symmetrizer simplifies immensely in an idempotent basis. Here, this basis is given by the projectors \(\sfP_g\). 

Continuing, recall that for a generic circle state, the reduced density matrix obtained from the isometric cut is
\begin{align}\label{eq:DW-interval-rho-general}
	\rho_{\Psi,\cB_\sfS}
	=
	\sum_{\sfR\in\irr_\omega(G)}
	\lambda_\sfR\,\frac{\1_\sfR}{\td_\sfR^{\,2}}~,
	\qq{with}
	\1_\sfR
	=
	\sum_{i_L,i_R=1}^{\td_\sfR}
	\ketbra{\sfR;i_L,i_R}{\sfR;i_L,i_R}~,
\end{align}
and \(\lambda_\sfR=\abs{\Psi_\sfR}^2/\abs{\Psi}^2\). Applying the projectors to \(\1_\sfR\) and using \cref{eq:interval-projective-fourier} gives
\begin{align}
	\sfP_g\1_\sfR\sfP_g
	=
	\frac{\td_\sfR^{\,2}}{\abs{G}}\sfP_g~,
\end{align}
where we have also used that \(\sfR\) is unitary, so
\begin{align}
	\sum_{i_L,i_R=1}^{\td_\sfR} \qty[D_\sfR^\omega(g)]_{i_L\,i_R} \overline{\qty[D_\sfR^\omega(g)]_{i_L\,i_R}} = \td_\sfR~.
\end{align}  
Hence, after all this work, we arrive at the conclusion that the symmetrized reduced density matrix is simply the maximally mixed state on the interval Hilbert space:
\begin{align}\label{eq:DW-interval-rho-symmetrized}
	\rho_{\Psi,\cB_\sfS;\sfS}
	=
	\frac{1}{\abs{G}}
	\sum_{g\in G}\sfP_g
	=
	\frac{\1_{\cH_{\cB_\sfS}}}{\abs{G}}~.
\end{align}
This state is manifestly symmetric, but, interestingly, it is completely independent of the seed state \(\ket{\Psi}\), a feature not at all generic. 

The entanglement asymmetry with respect to the strip algebra readily follows for arbitrary states in DW TQFTs:
\begin{align}\label{eq:DW-interval-asymmetry-general}
	\entas_\t{strip}\qty(\rho_{\Psi,\cB_\sfS})
	=
	\log\abs{G} - \ee\qty(\rho_{\Psi,\cB_\sfS})~,
\end{align}
with \(\ee\qty(\rho_{\Psi,\cB_\sfS})\) as in \cref{eq:DW-EE}. Specializing to the vacuum states, \(\ket{\sfR}\), gives
\begin{align}\label{eq:DW-interval-asymmetry-vacuum}
	\entas_\t{strip}\qty(\rho_{\sfR,\cB_\sfS})
	= -	\log\frac{\td_\sfR^{\,2}}{\abs{G}} = -\log\frac{\td_\sfR^{\,2}}{\tD^2}~.
\end{align}

\paragraph{Untwisted \(D_4\) interval asymmetry.}

In our recurring example of \(D_4\) gauge theory, we therefore get, in the untwisted case,
\begin{align}
	\entas_\t{strip}^{D_4}
	\qty(\rho_{\sfA_i,\cB_\sfS})
	=
	\entas_\t{strip}^{D_4}
	\qty(\rho_{\sfB_i,\cB_\sfS})
	&=
	\log 8
	=
	3\log 2~,
	\nn
	\entas_\t{strip}^{D_4}
	\qty(\rho_{\sfE,\cB_\sfS})
	&=
	\log 2~.
\end{align}

\paragraph{Twisted \(D_4\) interval asymmetry.}\label{sec:Ising}

For the non-trivial twist, both irreducible projective representations have dimension two. Consequently,
\begin{align}
	\entas_\t{strip}^{D_4,\omega}
	\qty(\rho_{\sfE_\pm,\cB_\sfS})
	&=
	\log 2~.
\end{align}

\subsection{Ising TQFT}
\label{ssec:Ising}
The Ising TQFT%
\footnote{Not to be confused with the $(2+1)$d Ising anyon theory, which is also
often called Ising TQFT.}
is defined by the regular module category (which is the unique indecomposable module category) of the Ising fusion category, $\cC_\t{Ising}$ \cite{Thorngren:2019iar,Chang:2018iay,Huang:2021zvu}.
Physically it realizes the infrared gapped phase of the tricritical Ising model deformed by \(\varepsilon'\) (of weight $(\frac{3}{5},\frac{3}{5})$) \cite{Huse:1984mn,Chang:2018iay}.
The category $\cC_\t{Ising}$ consists of three simple objects, typically denoted as $\{\1,\eta,\cN\}$, with the following fusion rules:%
\footnote{$\cC_\t{Ising} = \operatorname{TY}(\Z_2,+)$ is the Tambara--Yamagami category over $\Z_2$  \cite{Tambara:1998vmj} with Frobenius--Schur indicator $\epsilon=+$.}
\begin{equation}
    \begin{alignedat}{3}
    \eta \otimes \1 &= \eta ~, \qquad &\cN \otimes \1 &= \cN ~,\qquad \eta \otimes \eta = \1 ~, \\ 
        \eta \otimes \cN &= \cN~, \qquad &\cN \otimes \cN &= \1\oplus\eta ~,
    \end{alignedat}
\end{equation}
and quantum dimensions   
\begin{equation}
    d_\1 = 1~, \quad d_\eta = 1~,\quad d_\cN = \sqrt{2}~, \qquad \cD^2 = 4 ~.
\end{equation}
These may be viewed as the TDLs generating the symmetries of the critical Ising model: $\eta$ is an invertible line generating the $\Z_2$ symmetry, while $\cN$ is a non-invertible line generating the Kramers--Wannier duality. 

The regular module category with $\irr(\cM)=\set{\1,\psi,\sigma}$ --- that is the category of boundary conditions or equivalently topological local operators --- is isomorphic to $\cC_\t{Ising}$ by the replacement $\eta\leftrightarrow\psi$ and $\cN\leftrightarrow \sigma$. The NIM-rep of the corresponding algebra is 
\begin{equation}
    \tilde{N}_{\eta} = \begin{pmatrix}
        0 & 1 & 0 \\ 
        1 & 0 & 0 \\ 
        0 & 0 & 1
    \end{pmatrix} 
    ~, \qquad \tilde{N}_{\cN} = \begin{pmatrix}
        0&0&1\\
        0&0&1 \\ 
        1&1&0
    \end{pmatrix} ~.
\end{equation}
The boundary quantum dimensions can be read off as
\begin{equation}
    \td_{\1} = \td_\psi = 1 ~, \qquad \td_\sigma = \sqrt{2} ~, \qquad \tD^2 = 4 ~.
    \label{Ising:QD}
\end{equation}
We note that the boundary quantum dimensions coincide with the defect quantum dimensions, which is a general feature of regular TQFTs \cite{Huang:2021zvu}. 
The Hartle--Hawking state and sphere partition function are then 
\begin{equation}
    \ket{\HH} = \ket{\1} + \ket{\psi} + \sqrt{2} \ket{\sigma} ~, \qq{and} \parti\qty\big[\S^2] = \braket{\HH} = 4~.
    \label{Ising:HH}
\end{equation}
\subsubsection{Entanglement entropy}
We consider the general pure state
\begin{equation}
    \ket{\Psi} = \Psi_\1 \ket{\1} + \Psi_\psi \ket{\psi} + \Psi_\sigma \ket{\sigma}, \qquad 
\end{equation}
where the normalized probabilities for each sector are now given by 
\begin{equation}
    \lambda_m = \frac{\lvert \Psi_m \lvert^2}{\sum_{n\in \{\1,\psi,\sigma \}}{\lvert \Psi_n\lvert^2}} , \qquad m \in \{\1,\psi,\sigma\} ~.
\end{equation}
For the general mixed boundary condition \cref{LRBC} the R\'enyi entropies are again 
\begin{align}
	\ee\qty(\rho_{\Psi;\cB_L,\cB_R})
	=
	-\sum_m\lambda_m\log\lambda_m
	+
	\sum_m\lambda_m\log(g^{(L)}_m g^{(R)}_m)~. 
    \label{EE:IsingExamplesMixBoundary}
\end{align}
We note that, since the quantum dimensions \cref{Ising:QD} are not integers we cannot consider the symmetric boundary condition \cref{eq:SBC}. We will instead consider 
\begin{equation}
    \cB = g_\1 \1 \oplus g_\psi\, \psi \oplus g_\sigma\, \sigma ~, \qquad g_m \in \mathbb{Z_{>0}} ~,
\end{equation}
where we have fixed $g_m^{(L)}=g_m^{(R)}$. In particular, for the Hartle--Hawking state \cref{Ising:HH} we have the normalized probabilities  
\begin{equation}
    \lambda_\1 = \lambda_{\psi} = \frac{1}{4} ~, \qquad \lambda_\sigma = \frac{1}{2} ~,
\end{equation}
yielding 
\begin{equation}
    \ee\qty(\rho_{\HH,\cB})
	=  \log \parti\qty\big[\S^2] + \frac{1}{2} \log(\frac{g_\1 g_\psi g_\sigma^2}{2}) ~. 
\end{equation}  
Note that, despite not being able to impose the symmetric boundary condition, one can still choose a simple one obeying
\begin{equation}
    {g_\1g_\psi g_\sigma^2} = 2 ~, 
\end{equation}
in which case the edge contribution to the entanglement entropy vanishes. Yet, defined like this, the state is not maximally mixed. Fixing $g_\1=g_\sigma=1$, $g_\psi=2$ we can see this explicitly from the density matrix 
\begin{equation}
    \rho_{\HH,\cB} = \sum_m \lambda_m \frac{\1_m}{g_m^2} = \frac{1}{4} \1_\1 + \frac{1}{16} \1_\psi + \frac{1}{2}\1_\sigma ~. 
\end{equation}
We note that the circle entanglement asymmetry of basis states have already been computed in \cite{Benini:2025lav}; we have included this datum in \cref{table:summary-examples} for completeness.


\subsection{Fibonacci TQFT}\label{sec:Fib}

We now turn to the Fibonacci fusion category, $\cC_\varphi$, which is the simplest fusion category containing a non-invertible line. It has only two simple objects, \(\1\) and \(W\), with fusion rules
\begin{equation}
    \1\otimes \1 = \1~, \qquad \1\otimes W = W ~, \qquad W \otimes W = \1 \oplus W ~,
\end{equation}
and quantum dimensions
\begin{equation}
    d_\1 = 1~, \qquad d_W = \varphi ~, \qquad \cD^2 = 1 + \varphi^2 ~,
\end{equation}
where $\varphi = (1+\sqrt{5})/2$ is the golden ratio. $\cC_\varphi$ admits a unique indecomposable module category, the regular one \cite{Davydov:2010kfz} \(\cM\cong \cC_\varphi\), hence there is a unique gapped phase with Fibonacci symmetry, corresponding to a complete \(\cC_\varphi\)-SSB phase. It has two vacua, which we label by $\irr(\cM)=\set{\1,\tau}$ upon identifying $\tau$ with $W$. 

This phase has two physical origins. One is as the massive phase of the tricritical Ising model, deformed by the subleading magnetic operator $\sigma'$ (of weight $(\tfrac{7}{16},\tfrac{7}{16})$) \cite{Chang:2018iay}. The degenerate but inequivalent vacua of this deformation have been known since the early 1990s \cite{Zamolodchikov:1990xc,Lassig:1990xy}, but the corresponding TQFT was constructed in \cite{Chang:2018iay,Huang:2021zvu} using the non-invertible symmetry the \(\sigma'\)-deformation preserves. A second origin comes from the Fibonacci chain. The uniform golden chain \cite{Feiguin:2006ydp} is itself critical, but turning on specific interactions \cite{Trebst:2008fib,Aasen:2020jwb} opens a gap and leads to two ground states that spontaneously break \(\cC_\varphi\). This is precisely the TQFT described above. 

as well as the NIM-rep 
\begin{equation}
    \tilde{N}_\tau = \begin{pmatrix}
        0 & 1 \\ 
        1 & 1
    \end{pmatrix} ~.
\end{equation}
The Hartle--Hawking state and the sphere partition function are
\begin{equation}
    \ket{\HH} = \ket{\1} + \varphi \ket{\tau} ~, \qquad \parti\qty\big[\S^2] = \braket{\HH   } = 1+ \varphi^2 ~.
    \label{HHFibonacci}
\end{equation}
\subsubsection{Entanglement entropy}
A pure state in the Fibonacci TQFT is given by 
\begin{equation}
    \ket{\Psi} = \Psi_\1 \ket{\1} + \Psi_\tau \ket{\tau} ~, 
\end{equation}
with the corresponding sector probabilities 
\begin{equation}
    \lambda_m = \frac{\lvert \Psi_m \lvert^2}{\sum_{n\in \{\1,\tau \}}{\lvert \Psi_n\lvert^2}} , \qquad m \in \{\1,\tau\} ~.
\end{equation}
For the Hartle--Hawking state \cref{HHFibonacci} they are explicitly given by 
\begin{equation}
    \lambda_\1 = \frac{1}{1+\varphi^2} ~, \qquad \lambda_\tau = \frac{\varphi^2}{1+\varphi^2} ~.
\end{equation}
For the simple boundary conditions 
\begin{equation}
    \cB_L=\cB_R = \cB = g_\1\, \1 \oplus g_\tau\, \tau ~, \qquad g_\1,g_\tau \in \Z_{> 0} ~,  
    \label{ising:SimpleBC}
\end{equation} 
and the Hartle--Hawking state, the entanglement entropy \cref{eq:Sent-generic-psi} is 
\begin{equation}
    \ee\left(\rho_{\HH,\cB} \right) = \log \parti\qty\big[\S^2] - \frac{2\varphi^2}{\parti\qty\big[\S^2]} \log\varphi + \frac{2}{\parti\qty\big[\S^2]} \left(\log g_\1 + \varphi^2 \log g_\tau   \right) ~.
\end{equation}
Since the boundary conditions are given by stacks of simple \textquote{branes} $g_m \in \mathbb{N}$, and $\td_\tau \notin \mathbb{N}$ we again cannot fix the symmetric boundary condition. 

Instead, we can consider a general edge dressing and the corresponding cutting map \cref{genCuttingMap}, the corresponding R\'enyi entropies \cref{genRenyies} can be written as 
\begin{equation}
    \ee\qty(\rho_\HH) = \log \parti\qty\big[\S^2] + \sum_{m\in (\1,\tau)} \frac{\td_m^2}{\parti\qty\big[\S^2]} \left(\cS\left(\sigma_m \right) - \log \td_m^2 \right) ~.
\end{equation}
We explicitly see that the entanglement entropy depends sensitively on the choice of boundary dressing. In particular we note that for a dressing such that 
\begin{equation}
  \ent\left(\sigma_\tau  \right) = \log \varphi^2 ~,
\end{equation}
the non-integer edge sector $\tau$ vanishes. 

We again note that the circle entanglement asymmetry of basis states for this theory have already been computed in \cite{Benini:2025lav} as well; we have included this datum in \cref{table:summary-examples} for completeness.

\subsection{Gapped Haagerup phases}\label{ssec:Haagerup}

We now consider gapped phases enriched by a Haagerup fusion category symmetry. In fact, there are three unitary Haagerup fusion categories, $\cH_1$, $\cH_2$, and $\cH_3$. The first two descend directly from the Haagerup subfactors of \cite{Asaeda:1999exotic,Haagerup:1994principal}, while the latter was constructed in \cite{grossman2012quantum}. In the language of generalized symmetries, $\cH_3$ is connected to $\cH_1$ and $\cH_2$ by a generalized gauging \cite{Bhardwaj:2017xup,Huang:2021zvu,Diatlyk:2023fwf,Putrov:2024uor}, or equivalently the three categories are Morita equivalent. Motivated by recent interest of the community \cite{Huang:2021ytb,Huang:2021zvu,Huang:2021nvb,Vanhove:2021zop,Liu:2022qwn,Putrov:2024uor,Corcoran:2024eeh,Bottini:2025hri,Jia:2024wnu,Hung:2025gcp,Blakeney:2025ext,Albert:2025umy,Huang:2026lqf} we will focus below on the $\cH_3$ fusion category.

From a gapless standpoint, $\cH_3$ (and $\cH_2$) is the simplest example of a fusion category with a non-commutative non-invertible fusion algebra and, as such, it cannot be realized using the Verlinde lines of a diagonal rational CFT \cite{Huang:2021ytb}.
However, assuming a Haagerup CFT,%
\footnote{There is an abundance of evidence for a CFT with Haagerup symmetry \cite{Huang:2021nvb,Vanhove:2021zop,Corcoran:2024eeh,Hung:2025gcp,Albert:2025umy} at various central charges, and a very recent announcement of an analytic construction at $c=8$ \cite{Huang:2026lqf}.
}
it is then interesting to study relevant deformations, preserving the Haagerup symmetry. Unless they hit a strongly coupled fixed point (a second Haagerup CFT), these deformations will necessarily flow to a TQFT with that symmetry, realized linearly or nonlinearly, which is our interest here.

The simple TDLs in the $\cH_3$ fusion category are $\1$, $\alpha$, $\alpha^2$, $\varrho$, $\alpha\varrho$, $\alpha^2\varrho$, and have fusion rules
\begin{equation}\label{eq:H3-fusion-rules}
\renewcommand{\arraystretch}{1.2}
\setlength{\arraycolsep}{7pt}
\begin{array}{c|*{6}{c}}
\otimes
& \1 & \alpha & \alpha^2
& \varrho & \alpha\varrho & \alpha^2\varrho
\\ \hline
\1
& \1 & \alpha & \alpha^2
& \varrho & \alpha\varrho & \alpha^2\varrho
\\
\alpha
& \alpha & \alpha^2 & \1
& \alpha\varrho & \alpha^2\varrho & \varrho
\\
\alpha^2
& \alpha^2 & \1 & \alpha
& \alpha^2\varrho & \varrho & \alpha\varrho
\\
\varrho
& \varrho & \alpha^2\varrho & \alpha\varrho
& \1\oplus H & \alpha^2\oplus H & \alpha\oplus H
\\
\alpha\varrho
& \alpha\varrho & \varrho & \alpha^2\varrho
& \alpha\oplus H & \1\oplus H & \alpha^2\oplus H
\\
\alpha^2\varrho
& \alpha^2\varrho & \alpha\varrho & \varrho
& \alpha^2\oplus H & \alpha\oplus H & \1\oplus H
\end{array}
\end{equation}
with the shorthand $H\coloneqq \varrho \oplus \alpha\varrho \oplus \alpha^2\varrho$.
The simple objects $\1$, $\alpha$ and $\alpha^2$ form an invertible $\Z_3$ subcategory and thus have 
\begin{equation}
    d_\1 = d_\alpha = d_{\alpha^2} = 1 ~,
\end{equation}
while the non-invertible simples, $\varrho$, $\alpha\varrho$, and $\alpha^2 \varrho$ all have the same quantum dimension,
\begin{equation}
    \xi \coloneqq d_\varrho = d_{\alpha \varrho} = d_{\alpha^2 \varrho} = \frac{3 + \sqrt{13}}{2}~.
\end{equation}
The total quantum dimension is then 
\begin{equation}
    \cD^2 = 3\qty(1+\xi^2) = \frac{3}{2}\qty(13+3\sqrt{13}) ~. 
\end{equation}

The possible $\cH_3$-symmetric TQFTs are classified by the module categories of $\cH_3$ of which there are three, and have two, four, and six simple objects, or vacua, respectively. The TQFT with six vacua corresponds to the regular module category, $\cM_6\cong \cH_3$, and was first constructed in \cite{Huang:2021ytb} and further studied in \cite{Huang:2021zvu}. One way to arrive to all gapped Haagerup phases follows  \cite{Bottini:2025hri}. There, the authors constructed a gapless phase with $\cH_3$ symmetry by taking two copies of a $\Z_3$-symmetric CFT
--- in that case, the three-state Potts model ---
that are exchanged by the action of the non-invertible elements of the Haagerup symmetry. A posteriori, this turns out to be equivalent to stacking the original CFT with the two-vacuum $\cH_3$-symmetric TQFT. The gapless theory of \cite{Bottini:2025hri} admits two relevant deformations 
under which each of the seed Potts models flows to either an \emph{ordered} ($\mathbb{Z}_3$-SSB) phase or to a \emph{disordered} phase, depending on the sign of their respective deformation. When both deformations have the same sign, the theory flows to a TQFT with two or six vacua, depending on the overall sign, while both deformations with opposite signs flow to a TQFT with four vacua. Each of these gapped phases is described by an $\cH_3$-module category, $\cM_n$, with $n=2,4,6$ labelling the total number of vacua. Its simple boundary conditions, $m_n$, correspond to each of the vacuum states.  A general pure state in such module category will then be 
\begin{equation}
    \ket{\Psi}_{\cM_n} = \sum_{m\in \irr(\cM_n)} \Psi_m \ket{m} ~.
\end{equation}
The sector probabilities are 
\begin{equation}
    \lambda_{\cM_n,m} = \frac{\abs{\Psi_{m}}_{\cM_n}^2}{\abs{\Psi}_{\cM_n}^2}, \qquad \abs{\Psi}^2_{\cM_n} = \sum_{n\in \irr(\cM_n)}\abs{\Psi_n}_{\cM_n}^2~.
    \label{Haager:Prob}
\end{equation}
For the Hartle--Hawking state they are 
\begin{equation}
    \lambda_{\cM_n,m} = \frac{\td_{m,\cM_n}^{\,2}}{\tilde{\cD}_{\cM_n}^2} ~. 
    \label{Hgr:HHProb}
\end{equation} 
The entanglement entropy for the mixed boundary condition \cref{LRBC} in the Hartle--Hawking state is then 
\begin{equation}
    \ee\qty(\rho_\HH; \cM_n) = \log \tilde{\cD}^2 - \sum_{m\in \irr(\cM_n)} \lambda_{\cM_n,m} \log \frac{\tilde{d}_m^2}{g^{(L)}_{\cM_n,m}\,g^{(R)}_{\cM_n,m}}~,
    \label{Haager:EE}
\end{equation}
as follows from our general formula, \cref{EE:MixBoundary}.

\subsubsection{Disorder-disorder (DD) phase.} \label{sec:HaagDD}
We start by considering the case where both thermal deformations have negative sign. In this case, both copies of the three-state Potts model flow to a disordered, or trivial, gapped phase. Each copy has a single vacuum, on which the $\mathbb{Z}_3$ symmetry acts trivially. In total this phase has two vacua and is described by the module category $\cM_2$ whose simple objects, $m_0$ and  $m_1$ correspond to the vacua. The Haagerup symmetry is encoded in the NIM-rep matrices \cite{Bottini:2025hri} 
\begin{equation}\label{eq:DD-nimrep}
    \tilde{N}_{\alpha} = \begin{pmatrix}
        1 & 0 \\
        0 & 1
    \end{pmatrix} ~, \qquad \tilde{N}_\varrho = \begin{pmatrix}
        0 & 1 \\ 
        1 & 3
    \end{pmatrix} ~.
\end{equation}
As discussed in \cref{eq:FP-equation}, this defines the boundary quantum dimension up to an overall normalization constant which we fix to one in order to obtain
\begin{equation}
    \td_{m_0} = 1 ~, \qquad \td_{m_1} = \xi ~, \qquad \tilde{\cD}^2_{\t{DD}} = 1 + \xi^2 ~.
\end{equation}
In this phase the Hartle--Hawking state and sphere partition function are given by 
\begin{equation}
    \ket{\HH}_{\t{DD}} = \ket{m_0} + \xi \ket{m_1} ~, \qquad \parti_{\t{DD}}\qty\big[\S^2] = 1 + \xi^2 ~. 
\end{equation}
The probabilities \cref{Haager:Prob} for the Hartle--Hawking state are explicitly 
\begin{equation}
    \lambda_{\t{DD},0} = \frac{1}{1+\xi^2} ~, \quad \lambda_{\t{DD},1} = \frac{\xi^2}{1+\xi^2} ~.  
\end{equation}
The entanglement entropy \cref{Haager:EE} is then 
\begin{equation}
    \ee\qty(\rho_\HH)_{\t{DD}} = \log(1+\xi^2) + \frac{1}{1+\xi^2} \log(g^{(L)}_{\t{DD},0}g^{(R)}_{\t{DD},0}) + \frac{\xi^2}{1+\xi^2} \log\frac{g^{(L)}_{\t{DD},1}g^{(R)}_{\t{DD},1}}{\xi^2}~.
\end{equation}

\subsubsection{Order-order (OO) phase.}\label{sec:HaagOO}

We now consider the case where both deformations have instead positive sign, so both Potts models flow to the ordered phase. This is described by six vacuum states $m_i$ $i=0,\ldots,5$. As before, the module category encodes the corresponding boundary conditions. In this case, the corresponding module category is isomorphic to the regular one, $\cM_6 \cong \cC$. As such, the Hartle--Hawking state and sphere partition function are 
\begin{equation}
    \ket{\HH}_\t{OO} = \ket{\1} + \ket{\alpha} + \ket*{\alpha^2} + \xi \ket{\varrho} + \xi \ket{\alpha \varrho} + \xi\ket*{\alpha^2 \varrho} ~, \qquad \parti_\t{OO}\qty\big[\S^2] = 3\qty(1+\xi^2) ~.
\end{equation}
The sector probabilities are now 
\begin{equation}
    \lambda_{\t{OO},i} = \frac{1}{3(1+\xi^2)}~, \quad \lambda_{\t{OO},j} = \frac{\xi^2}{3(1+\xi^2)}~, \qquad i \in \set{\1,\alpha,\alpha^2}~, \quad j \in \set{ \varrho, \alpha\varrho , \alpha^2\varrho}~.
\end{equation}
For which now the entanglement entropy is 
\begin{equation}
\begin{aligned}
    \ee\qty(\rho_\HH)_{\t{OO}} &= 
    \log\left[3\left(1+\xi^2 \right) \right] + 
    \frac{1}{3(1+\xi^2)} \sum_{i=\1,\alpha,\alpha^2}
    \log(g^{(L)}_{\t{OO},i}g^{(R)}_{\t{OO},i}) \\
 &\phantom{=~}
 +\frac{\xi^2}{3(1+\xi^2)}\sum_{j=\varrho,\alpha\varrho,\alpha^2\varrho}\log\frac{g^{(L)}_{\t{OO},j}g^{(R)}_{\t{OO},j}}{\xi^2}~.
 \end{aligned}
\end{equation}

\subsubsection{Order-disorder (OD) phase.}\label{sec:HaagOD}

The remaining phase is described by the last available module category, $\cM_4$, whose simple boundary conditions $m_i$, $i=0,\ldots,3$ correspond to each vacuum state and we have $\irr(\cM_4)=\{m_0,m_1,m_2,m_3 \}$. While the approach of \cite{Bottini:2025hri} cannot conclusively identify this phase with a deformation of the doubled Potts model, the vacuum structure is suggestive of the case where one of the Potts models flows to an ordered phase with three vacua, and the other to the disordered, trivially gapped phase. Accordingly we will call this phase ``order-disorder'' (OD).

The action of the TDLs is encoded in the NIM-rep matrices%
\footnote{We have computed these NIM-rep matrices from the data in \cite{grossman2012quantum} and which differ from those computed in \cite{Bottini:2025hri}. We thank Lea Bottini for discussions on this point.}
\begin{equation}\label{eq:OD-nimrep}
    \tilde{N}_\alpha = 
    \begin{pmatrix}
        0 & 0 & 1& 0  \\ 
        1 & 0 & 0& 0 \\ 
        0 & 1 & 0 & 0 \\ 
        0 & 0 & 0 &1
    \end{pmatrix} ~, \qquad 
    \tilde{N}_\varrho = 
    \mqty(
    0 & 1 & 1 & 1 \\ 
    1 & 1 & 0 & 1 \\ 
    1 & 0 & 1 & 1 \\ 
    1 & 1 & 1 & 1
    )~.
\end{equation}
which again, after fixing the scaling ambiguity defines the boundary quantum dimensions 
\begin{equation}
    \td_{m_0} = \td_{m_1} = \td_{m_2} = 1 ~, \quad \td_{m_3} = \xi-2 ~, \qquad \tilde{\cD}_{\t{OD}}^2 = 3 + (\xi-2)^2=8-\xi~.
\end{equation}
In this phase, the Hartle--Hawking state and sphere partition function are then 
\begin{equation}
    \ket{\HH}_{\t{OD}} = \ket{m_0} + \ket{m_1} + \ket{m_2} + (\xi-2) \ket{m_3} ~, \qquad \parti_{\t{OD}}\qty\big[\S^2] =  8-\xi ~.
\end{equation}
For this state, the sector probabilities are then 
\begin{equation}
    \lambda_{\t{OD},i} = \frac{1}{8-\xi}~,\qquad  i = 0,1,2 ~, \qquad \lambda_{\t{OD},3} = \frac{5-\xi}{8-\xi}~. 
\end{equation} 
In this case, the entanglement entropy is now 
\begin{equation}
    \ee\qty(\rho_\HH)_{\t{OD}} = \log(8-\xi) + \frac{1}{8-\xi} 
    \sum_{i={0,1,2}}\log(
    g^{(L)}_{\t{OD},i}g^{(R)}_{\t{OD},i}) + \frac{5-\xi}{8-\xi} \log(\frac{g^{(L)}_{\t{OD},3}g^{(R)}_{\t{OD},3}}{5-\xi})~.
\end{equation}

\subsubsection{Entropic diagnostic of gapped phases}

Our results suggest an entanglement diagnostic for microscopic Hamiltonians whose fusion category symmetry, here \(\cH_3\), is known but whose infrared gapped phase is not. To that end, consider the lowest-energy state in the trivial sector of the full symmetry category. To prepare it, one can first project an arbitrary initial state onto this sector and then apply an imaginary-time evolution.

The resulting state flows in the infrared to the Hartle--Hawking state, as this is the unique symmetric combination of the TQFT vacua. Note, moreover, that this preparation requires no prior knowledge of the individual vacua or of the gapped phase itself.

The three phases nevertheless have distinct Hartle--Hawking probability distributions, and hence give different vacuum contributions to their entanglement entropies. To isolate these contributions, let us set all boundary degeneracies to unity, \(g^{(L)}_{i}=g^{(R)}_{i}=1\); such a choice is always available within the TQFT. Our previous expressions then give
\begin{align}
\ee\qty(\rho_\HH)_{\t{DD}} &= \log(1+\xi^2) - \frac{\xi^2}{1+\xi^2} \log{\xi^2} \approx 0.288~, \\
\ee\qty(\rho_\HH)_{\t{OD}} &= \log(8-\xi) - \frac{5-\xi}{8-\xi} \log(5-\xi) \approx 1.356~, \\
\ee\qty(\rho_\HH)_{\t{OO}} &= \ee\qty(\rho_\HH)_\t{DD} + \log 3 \approx 1.387~.
\end{align}
Comparing the entropy of the microscopic state prepared above with these three values uniquely identifies the infrared gapped phase, providing a very simple entropic diagnostic.%
\footnote{Note, however, that while simple, this diagnostic is not faithful in general. For instance, it may happen that distinct gapped phases have the same entanglement entropy. In many cases, higher Rényi entropies would still distinguish the phases. In some unfortunate cases, such as phases differing by an SPT, none of these crude diagnostics works. One would then need to probe further symmetry data, for instance through more intricate entropic order parameters \cite{Casini:2019kex,Casini:2020rgj,Ares:2022koq,Benini:2025lav,AliAhmad:2025bnd,Benini:2025hbj}.}
%

\subsubsection{Entanglement asymmetry}

We now compute the entanglement asymmetry of the three Haagerup phases. Since the Haagerup fusion category admits no symmetric boundary condition, we content ourselves with the asymmetry with respect to the fusion algebra acting on the circle Hilbert space. For the OO phase, corresponding to the regular TQFT, the asymmetry was computed in \cite{Benini:2025lav}. The asymmetries of the other two phases follow from a simple extension of their computation.

The coefficients \(\tilde{K}^{ab}\) of the symmetrizer, \cref{eq:symmetrizer-fusion}, are fixed by the fusion ring alone and require no input of the individual phases. The symmetrized density matrix reads \cite{Benini:2025lav}
\begin{align}\label{eq:H3-symmetrizer}
	\rho_{m,\sfS}
	=
	\frac16\sum_{a\in\irr(\cH_3)} L_a\,\rho_m\,L_a^\dagger
	+\frac{1}{26}\qty\big(A\,\rho_m\,A-B\,\rho_m\,B)
	-\frac{1}{39}\qty\big(A\,\rho_m\,B+B\,\rho_m\,A)~,
\end{align}
where 
\begin{align}
	A\coloneqq L_\1+L_\alpha+L_{\alpha^2}~,
	\qquad
	B\coloneqq L_\varrho+L_{\alpha\varrho}+L_{\alpha^2\varrho}~.
\end{align}
The TQFT itself enters through the action of the TDLs on the circle Hilbert space. From \cref{eq:action-on-simple-BCs} it follows that the simple TDLs, \(L_a\), are represented on \(\cH_{\S^1}\) by the NIM-rep matrices \(\tn_a\). This allows us to immediately compute the symmetrized density matrix for each vacuum of each phase.

\paragraph{DD phase.} 

In the DD phase, the NIM-rep matrices are given by \cref{eq:DD-nimrep}. For the two vacua, \(m_0\) and \(m_1\), we find
\begin{alignat}{3}
	\rho_{m_0} &= \ketbra{m_0} = \mqty(\dmat[0]{1,0})~,\qquad &&\rho_{m_0,\sfS} = \frac{1}{13}\mqty(11 & -3 \\ -3 & 2)~, \\
	\rho_{m_1} &= \ketbra{m_1} = \mqty(\dmat[0]{0,1})~,\qquad &&\rho_{m_1,\sfS} = \frac{1}{13}\mqty(2 & 3 \\ 3 & 11)~. 
\end{alignat}
In both cases, the symmetrized density matrix has eigenvalues \(\frac{1}{2}\qty(1\pm 3/\sqrt{13})\), yielding the asymmetry
\begin{align}
	\entas_\t{fus}(\rho_{m_0}) = \entas_\t{fus}(\rho_{m_1}) =
 	\frac{\log 13}{2}+\frac{\log(5 \sqrt{13}-18)}{\sqrt{13}} \approx 0.288~.
\end{align}

\paragraph{OD phase.} 

In the OD phase, the NIM-rep matrices are given by \cref{eq:OD-nimrep}. There are now four vacua, \(m_i\) with \(i=0,1,2,3\), and each density matrix \(\rho_{m_i}\) is again a \(4\times 4\) matrix with a single \(1\) in the \((i+1,i+1)\) entry. The corresponding symmetrized density matrices are
\begin{align}
	\rho_{m_i,\sfS} = \frac{1}{39}\mqty(
 11 & -2 & -2 & 1 \\
 -2 & 11 & -2 & 1 \\
 -2 & -2 & 11 & 1 \\
 1 & 1 & 1 & 6 \\
)~, \ i=0,1,2~, \qquad
\rho_{m_3,\sfS} = \frac{1}{39} \mqty(
 6 & 6 & 6 & -3 \\
 6 & 6 & 6 & -3 \\
 6 & 6 & 6 & -3 \\
 -3 & -3 & -3 & 21 \\
)~,
\end{align}
with eigenvalues
\begin{align}
	\lambda_{m_i} &= \set{\frac{1}{3},\frac{1}{3},\frac{13+\sqrt{13}}{78},\frac{\sqrt{13}-1}{6 \sqrt{13}}}~, \qquad i = 0,1,2~, \\ 
	\lambda_{m_3} &= \set{\frac{13+\sqrt{13}}{26},\frac{\sqrt{13}-1}{2 \sqrt{13}},0,0}~,
\end{align}
respectively. The resulting circle asymmetries are
\begin{align}
	\entas_\t{fus}(\rho_{m_i}) &= 
	\frac{2 \log 3}{3}-\frac{13+\sqrt{13}}{78} \log(\frac{13+\sqrt{13}}{78})-\frac{13-\sqrt{13}}{78}\log(\frac{13-\sqrt{13}}{78}) \nonumber \\ 
	&\approx 1.317~, \qquad i = 0,1,2 \\[1.3ex]
	\entas_\t{fus}(\rho_{m_3})  &= \frac{1}{2} \log(\frac{13}{3})-\frac{1}{2\sqrt{13}}\log(\frac{\sqrt{13}+7}{6}) \approx 0.654~.
\end{align}

\paragraph{OO phase.}

In the regular TQFT, the NIM-reps coincide with the fusion matrices that can be read off from \cref{eq:H3-fusion-rules}. The symmetrized density matrices of the six vacua then read
\begin{alignat}{2}
	\rho_{m,\sfS} &=  
	\frac{1}{78}
	\mqty(
		16 & 3 & 3 & -2 & -2 & -2 \\
		3 & 16 & 3 & -2 & -2 & -2 \\
		3 & 3 & 16 & -2 & -2 & -2 \\
		-2 & -2 & -2 & 10 & -3 & -3 \\
		-2 & -2 & -2 & -3 & 10 & -3 \\
		-2 & -2 & -2 & -3 & -3 & 10 \\
	)~, \qquad &&m = \1,\alpha,\alpha^2~, \\[1.2ex]
	\rho_{m,\sfS} &= 
	\frac{1}{78}\mqty(
		10 & -3 & -3 & 2 & 2 & 2 \\
		-3 & 10 & -3 & 2 & 2 & 2 \\
		-3 & -3 & 10 & 2 & 2 & 2 \\
		2 & 2 & 2 & 16 & 3 & 3 \\
		2 & 2 & 2 & 3 & 16 & 3 \\
		2 & 2 & 2 & 3 & 3 & 16 \\
	)~, \qquad &&m = \varrho, \alpha\varrho,\alpha^2\varrho~.
\end{alignat}
Both matrices share the eigenvalues
\begin{align}
	\lambda_{m} = \set{\frac{\sqrt{13}+3}{6\sqrt{13}},\frac{1}{6},\frac{1}{6},\frac{1}{6},\frac{1}{6},\frac{\sqrt{13}-3}{6\sqrt{13}}}~,
\end{align}
and hence the fusion asymmetry
\begin{align}
	\entas_\t{fus}\qty(\rho_m) &= \frac{2 \log 6}{3}-\frac{13-3 \sqrt{13}}{78}\log(\frac{13-3 \sqrt{13}}{78})-\frac{13+3 \sqrt{13}}{78}\log(\frac{13+3\sqrt{13}}{78}) \nonumber \\ 
	&\approx 1.657~,
\end{align}
in agreement with \cite{Benini:2025lav}.

\subsection{Topological cosets and adjoint \texorpdfstring{QCD$_2$}{QCD2}}\label{sec:TopoCoset}

A large and physically well-motivated class of 2d TQFTs with non-invertible symmetries arises from gauged Wess--Zumino--Witten (WZW) models \cite{Witten:1983ar}. Let $G$ be a compact, semisimple, connected, and simply connected Lie group and consider the diagonal WZW model at level $k$, whose chiral algebra is $\hat{\fg}_k$. A connected subgroup $H\subset G$, acting on the WZW field by conjugation, defines a non-anomalous symmetry, which may be gauged without adding kinetic terms for the gauge field \cite{Gawedzki:1988hq,Karabali:1989dk}. We denote the resulting theory by $G_k/H_{k'}$, where $k'$ is the level induced on $H$ by the embedding. Its central charge is
\begin{equation}
    c(G_k/H_{k'}) = c(G_k)-c\qty(H_{k'})~,
\end{equation}
which coincides with that of the coset construction of Goddard, Kent, and Olive (GKO) \cite{Goddard:1984vk}. The two theories nonetheless differ, as the GKO coset has a unique vacuum, while the gauged WZW model has additional vacua and can be viewed as the GKO coset coupled to a 2d TQFT \cite{Moore:2026lmq}; see also \cite{Komargodski:2020mxz,Cordova:2024goh}.

Here we are interested in the case $c=0$. The GKO coset is then trivial, and the gauged WZW model defines a bona fide TQFT. Following \cite{Sonnenschein:1992cz,Delmastro:2021otj,Cordova:2024goh} we will call such TQFTs \emph{topological cosets}. The condition of vanishing central charge means that $\hat{\fh}_{k'}\subset \hat{\fg}_k$ is a \emph{conformal embedding} \cite{Schellekens:1986mb,Bais:2008ni}. Apart from the case $H=G$, to which we will return shortly, this is only possible for $k=1$. One of the main physical motivations for topological cosets is that they describe the infrared phase of gapped two-dimensional quantum chromodynamics (QCD) with massless fermions \cite{Komargodski:2020mxz,Delmastro:2021otj,Cordova:2024goh,Cordova:2024vsq,Damia:2024kyt}.

The structure of a topological coset is encoded in the branching rules of the conformal embedding,
\begin{align}\label{eq:branching}
	\chi^{G_k}_{\Lambda}(q) = \sum_{\mu} b_{\Lambda\mu}\,\chi^{H_{k'}}_{\mu}(q)~, \qquad b_{\Lambda\mu}\in\Z_{\geq 0}~,
\end{align}
where \(\Lambda\in\irr(\Rep(G_k))\) and \(\mu\in\irr(\Rep(H_{k'}))\) label the integrable representations of the two affine algebras. For a conformal embedding, the branching coefficients \(b_{\Lambda\mu}\) are independent of the nome \(q\) \cite{Kac:1988tf,DiFrancesco:1997nk}. The branching of the vacuum module defines a distinguished object
\begin{align}\label{eq:algebra-A}
	\cA = \bigoplus_{\mu} b_{0\mu}\,\mu \in \Rep(H_{k'})~.
\end{align}
This is a commutative, symmetric special Frobenius (i.e.\ gaugeable) algebra \cite{Moore:2026lmq}, and gauging it in the diagonal \(H_{k'}\) WZW model produces the \(G_k\) WZW model \cite{Fuchs:2002cm,Bhardwaj:2017xup}. The representations of \(G_k\) are then recovered from \cref{eq:branching} \cite{Kirillov:2001ti,Huang:2014ixa,Cordova:2024goh,Moore:2026lmq}. Regarded as a (non-diagonal) \(H_{k'}\) RCFT, the \(G_k\) WZW model has modular invariant
\begin{align}\label{eq:Z-conformal-embedding}
	Z_{\mu\nu} = \sum_{\Lambda} b_{\Lambda\mu}\, b_{\Lambda\nu}~.
\end{align}

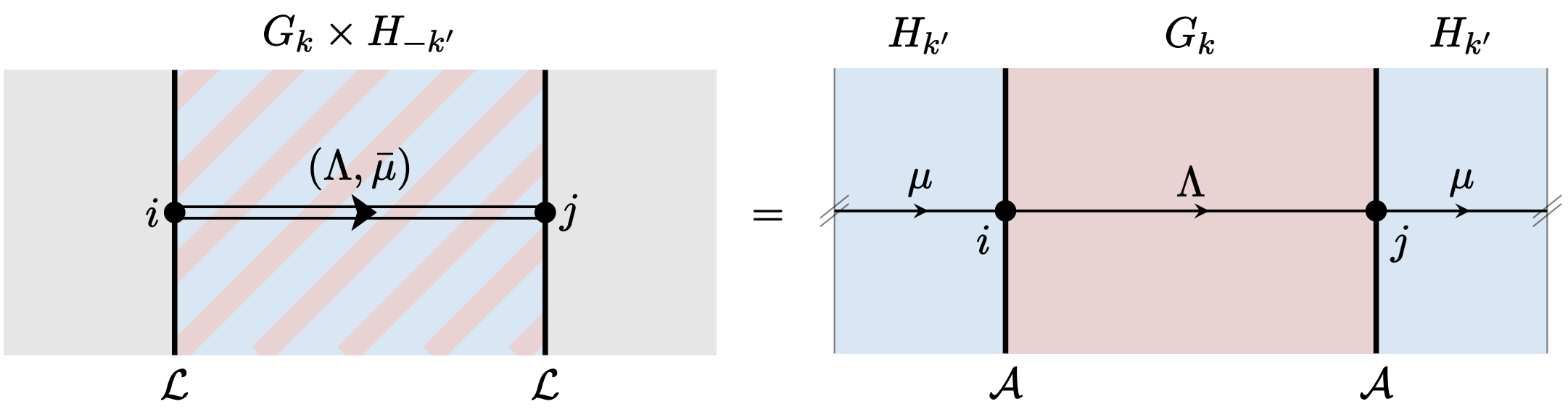
\begin{figure}[!tb]
\begin{align*}
	\begin{gathered}
		\begin{tikzpicture}[scale=1.3]
			\fill [fill=triv] (0,-1) rectangle (1.2,1);
			\fill [fill=triv] (3.8,-1) rectangle (5,1);
			\begin{scope}
				\clip (1.2,-1) rectangle (3.8,1);
				\fill [fill=Hcs] (1.2,-1) rectangle (3.8,1);
				\foreach \x in {-1.2,-0.6,...,4.2}
					\draw [Gcs, line width=2.12mm] (\x,-1) -- ++(2.4,2.4);
			\end{scope}
			\draw [line width=0.5mm] (1.2,-1) -- (1.2,1);
			\draw [line width=0.5mm] (3.8,-1) -- (3.8,1);
			\draw (1.2,-1) node [below] {$\cL$};
			\draw (3.8,-1) node [below] {$\cL$};
			\draw (2.5,1) node [above] {$G_k\times H_{-k'}$};
			\draw [line] (1.2,0.04) -- (3.8,0.04);
			\draw [line] (1.2,-0.04) -- (3.8,-0.04);
			\fill (2.62,0) -- (2.44,0.13) -- (2.48,0) -- (2.44,-0.13) -- cycle;
			\draw (2.5,0) node [above=2pt] {$(\Lambda,\bar\mu)$};
			\draw [fill=black] (1.2,0) circle (0.07) node [left] {$i$};
			\draw [fill=black] (3.8,0) circle (0.07) node [right] {$j$};
		\end{tikzpicture}
	\end{gathered}
	\;\;=\;\;
	\begin{gathered}
		\begin{tikzpicture}[scale=1.3]
			\fill [fill=Hcs] (0,-1) rectangle (1.2,1);
			\fill [fill=Gcs] (1.2,-1) rectangle (3.8,1);
			\fill [fill=Hcs] (3.8,-1) rectangle (5,1);
			\draw [bg] (0,-1) -- (0,1);
			\draw [bg] (5,-1) -- (5,1);
			\draw [bg] (-0.1,-0.11) -- (0.1,0.05);
			\draw [bg] (-0.1,-0.05) -- (0.1,0.11);
			\draw [bg] (4.9,-0.11) -- (5.1,0.05);
			\draw [bg] (4.9,-0.05) -- (5.1,0.11);
			\draw [line width=0.5mm] (1.2,-1) -- (1.2,1);
			\draw [line width=0.5mm] (3.8,-1) -- (3.8,1);
			\draw (1.2,-1) node [below] {$\cA$};
			\draw (3.8,-1) node [below] {$\cA$};
			\draw (0.6,1) node [above] {$H_{k'}$};
			\draw (2.5,1) node [above] {$G_k$};
			\draw (4.4,1) node [above] {$H_{k'}$};
			\draw [line,->-=.55] (0,0) -- (1.2,0);
			\draw [line,->-=.55] (1.2,0) -- (3.8,0);
			\draw [line,->-=.55] (3.8,0) -- (5,0);
			\draw (0.6,0) node [above] {$\mu$};
			\draw (2.5,0) node [above] {$\Lambda$};
			\draw (4.4,0) node [above] {$\mu$};
			\draw [fill=black] (1.2,0) circle (0.07) node [below left] {$i$};
			\draw [fill=black] (3.8,0) circle (0.07) node [below right] {$j$};
		\end{tikzpicture}
	\end{gathered}
\end{align*}
\caption{
The topological coset, folded and unfolded. On the left, the stripes indicate that the folded slab hosts both layers, \(G_k\times H_{-k'}\), while the gray regions are empty. An anyon \(\qty(\Lambda,\bar\mu)\) of \(\cL\), drawn as a double line, stretches between the two boundaries and ends on the junctions \(i\) and \(j\). On the right, the slab has been unfolded and the gray edges identified, so that the \(H_{k'}\) region wraps around a circle. Each gapped boundary has now become a topological interface, labelled by the algebra \(\cA\) that condenses on it. The anyon becomes a single closed line, which runs as \(\Lambda\) inside \(G_k\) and as \(\mu\) inside \(H_{k'}\), crossing the interfaces at \(i\) and \(j\).	
}
\label{fig:unfolding}
\end{figure}

Gauged WZW models are conveniently realized by three-dimensional Chern--Simons (CS) theory with gauge group \(G_k\times H_{-k'}\) on an interval, with \textquote{coset} boundary conditions imposed at both ends \cite{Moore:1989yh,Cordova:2024goh}. For a conformal embedding these boundary conditions are topological, and are specified by the Lagrangian algebra
\begin{equation}\label{eq:coset-Lagrangian}
	\cL = \bigoplus_{\Lambda,\mu} b_{\Lambda\mu}\,\qty(\Lambda,\bar{\mu}) \in \Rep(G_k)\boxtimes\overline{\Rep(H_{k'})}~,
\end{equation}
where the overline denotes orientation reversal, so that \(\overline{\Rep(H_{k'})}\) describes the anyons of \(H_{-k'}\), and \(\bar\mu\) is the conjugate of \(\mu\).%
\footnote{In categorical terms, given a modular tensor category (MTC) \(\cC\), we denote by \(\overline{\cC}\) the category that is isomorphic to \(\cC\) as a monoidal category but with reverse braiding. In the mathematical literature this is usually denoted as \(\cC^\t{rev}\).}
From a modern point of view, this is a symmetry TFT (SymTFT) construction \cite{Gaiotto:2020iye,Apruzzi:2021nmk,Freed:2022qnc,Kaidi:2022cpf,Bhardwaj:2023bbf} for the topological coset. Let us then take one of the two boundaries to be the symmetry boundary, and define \(\cC_{G/H}\) as the fusion category of topological lines living on it, i.e. bulk anyons pushed onto the symmetry boundary. Mathematically, \(\cC_{G/H}\) is the category of \(\cL\)-modules in \(\Rep(G_k)\boxtimes\overline{\Rep(H_{k'})}\) \cite{Kong:2013aya,Cordova:2024goh}. The topological local operators come instead from anyons of \(\cL\) stretched between the two boundaries, as depicted on the left of \cref{fig:unfolding}, and furnish a module category \(\cM\) of \(\cC_{G/H}\). Since both boundaries carry the same Lagrangian algebra, the topological coset is the regular \(\cC_{G/H}\)-symmetric TQFT, i.e.
\begin{align}\label{eq:gWZW-M}
	\cM = \cC_{G/H}~.
\end{align}
In other words, it is the gapped phase in which the \(\cC_{G/H}\) symmetry is completely spontaneously broken \cite{Bhardwaj:2023idu}.

An equivalent picture is obtained by unfolding the \(H_{-k'}\) layer, which flips the sign of its level, as shown on the right of \cref{fig:unfolding}. Each boundary then becomes a topological interface between the \(G_k\) and \(H_{k'}\) CS theories \cite{Fliss:2020cos,Cordova:2024goh}. Such an interface is obtained by condensing, on each side, the anyons of \(\cL\) supported on that factor alone. As we saw above, the \(G_k\) phase is obtained by condensing \(\cA\) in \(H_{k'}\). The topological interface is therefore labelled by \(\cA\), and the lines living on it are the \(\cA\)-modules \cite{Kong:2013aya}. With the orientation chosen above, we then have
\begin{align}\label{eq:CGH}
	\cC_{G/H} \cong \overline{\Rep(H_{k'})}_\cA~.
\end{align}
The simple objects of \(\cM\), which label the simple boundary conditions and hence the vacua, are therefore the simple \(\cA\)-modules in \(\Rep(H_{k'})\). Let us return to the folded picture in order to count them. Each anyon \(\qty(\Lambda,\bar\mu)\) of \(\cL\) can end on either boundary at any one of \(b_{\Lambda\mu}\) junctions, giving in total \(b_{\Lambda\mu}^2\) topological local operators. The number of vacua is then \cite{Komargodski:2020mxz,Cordova:2024goh,Moore:2026lmq}
\begin{align}\label{eq:gWZW-vacua}
	\dim\cH_{\S^1} = \abs{\irr(\cM)} = \sum_{\Lambda,\mu} b_{\Lambda\mu}^{2} = \tr Z~.
\end{align}
Note that, in contrast to the projectors \(\Pi_m\) of \cref{ssec:BC-circle-states}, these operators carry definite generalized charges under \(\cC_{G/H}\) \cite{Bhardwaj:2023ayw}, forming the \textquote{Fourier dual} basis to the projector basis.

Let us now determine the boundary quantum dimensions \(\td_m\), with \(m\in\irr(\cM)\), of the topological coset. We start from the quantum dimension of \(\cA\),
\begin{align}
	d_\cA = \sum_\mu b_{0\mu} d_\mu = \frac{\cD_H}{\cD_G}~,
\end{align}
where \(\cD_{H}\) and \(\cD_{G}\) denote the total quantum dimensions of \(\Rep(H_{k'})\) and \(\Rep(G_k)\), respectively, and we remind the reader of our physics convention for quantum dimensions (cf. \cref{foot:qdim-conventions}). The first equality is simply the definition of \(\cA\) in \cref{eq:algebra-A}, while the second follows from the fact that \(\Rep(G_k)\) is obtained from \(\Rep(H_{k'})\) by condensing \(\cA\) \cite{Kirillov:2001ti,Davydov:2010kfz}. Next, each simple object of \(\cM\) decomposes into simple anyons of \(\Rep(H_{k'})\) as
\begin{align}
	m = \bigoplus_{\mu} \tn_{\mu\1}^m\, \mu~,
\end{align}
where \(\tn_\mu\) is the NIM-rep associated with \(\mu\), and we recall that the identity object of \(\cC_{G/H}\) is \(\cA\) itself. Viewed as an object of \(\Rep(H_{k'})\), \(m\) therefore has quantum dimension
\begin{align}
	d_m^H = \sum_{\mu} \tn_{\mu\1}^m\, d_\mu~.
\end{align}
Dividing by \(d_\cA\) turns this into its quantum dimension in \(\cC_{G/H}\) \cite{Kirillov:2001ti}, which is precisely the boundary quantum dimension we are after:
\begin{align}\label{eq:bqdims-coset}
	\td_m = \frac{d_m^H}{d_\cA} = \frac{\cD_G}{\cD_H} \sum_\mu \tn_{\mu\1}^m\, d_\mu~.
\end{align}
As a check, let us compute the total boundary quantum dimension in two different ways. On the one hand, for a condensable algebra \(\cA\) in an MTC \(\cC\), the fusion category \(\cC_\cA\) of \(\cA\)-modules has total quantum dimension \(\cD_{\cC_\cA}^2 = \cD_\cC^2 / d_\cA\) \cite{Davydov:2010kfz}. Since \(\cM\) is the regular module category, we find
\begin{align}\label{eq:total-bqdim}
	\tilde{\cD}_{G/H}^2 = \cD_{G/H}^2 = \frac{\cD_H^2}{d_\cA} = \frac{\cD_G}{\cD_H}\cD^2_H = \cD_G\,\cD_H~.
\end{align}
On the other hand, using \cref{eq:bqdims-coset} we get
\begin{align}
	\tilde{\cD}_{G/H}^2 &= \sum_m \td_m^{\,2} = \sum_m \qty(\frac{\cD_G}{\cD_H}\sum_\mu \tn_{\mu\1}^m\, d_\mu ) d_m = \frac{\cD_G}{\cD_H} \sum_\mu d_\mu \sum_m \tn_{\mu\1}^m \td_m \\
	&= \frac{\cD_G}{\cD_H} \sum_\mu d_\mu d_\mu \td_\1 = \frac{\cD_G}{\cD_H} \cD_H^2 = \cD_G \cD_H~,
\end{align}
where we replaced one of the two factors of \(\td_m\) by its expression in \cref{eq:bqdims-coset}, exchanged the order of summation, and used the Frobenius--Perron equation \cref{bqd_rescaling}. The two results agree, as they should. In particular, the sphere partition function \cref{eq:spherePI} of a topological coset is
\begin{align}
	\parti_{G/H}\qty\big[\S^2] = \cD_G\,\cD_H~.
\end{align}

\subsubsection{The \texorpdfstring{\(G_k/G_k\)}{G/G} model}\label{sec:GmodG}

For \(H=G\), we obtain the \(G_k/G_k\) model, which has been studied extensively as a 2d TQFT; see for instance \cite{Spiegelglas:1992jg,Blau:1993tv,Witten:1993xi}. The \(G_k/G_k\) model is the regular TQFT for the Verlinde lines of \(G_k\) \cite{Komargodski:2020mxz}:
\begin{align}
	\cC_{G/G} = \cM = \Rep(G_k)~.
\end{align}
The simple TDLs are labelled by the integrable highest-weight representations of \(\hat{\fg}_k\), which we denote \(a,b,c,\cdots\in\irr\qty(\Rep(G_k))\), with \(0\) the vacuum representation. Their fusion coefficients are given by the Verlinde formula \cite{Verlinde:1988sn}
\begin{align}\label{eq:Verlinde-formula}
	N^c_{ab} = \sum_{e} \frac{\bbS_{ae}\,\bbS_{be}\,\bar{\bbS}_{ce}}{\bbS_{0e}}~,
\end{align}
where \(\bbS\) is the modular \(\bbS\)-matrix of \(G_k\), which is symmetric and unitary, with \(\bbS_{0a}>0\). The quantum dimensions and the total quantum dimension are
\begin{align}
	d_a = \frac{\bbS_{0a}}{\bbS_{00}}~, \qquad \cD^2 = \sum_a d_a^2 = \frac{1}{\bbS_{00}^2}~,
\end{align}
where in the last step we also used the unitarity of \(\bbS\). The simple boundary conditions are also labelled by integrable representations, which we denote \(\sfR\in\irr(\cM)\). The NIM-rep is the regular one, \(\tn^{\sfR'}_{a\sfR}=N^{\sfR'}_{a\sfR}\), and, as for any regular TQFT, the boundary quantum dimensions coincide with the bulk ones:
\begin{align}\label{eq:GG-bqd}
	\td_\sfR = d_\sfR = \frac{\bbS_{0\sfR}}{\bbS_{00}}~, \qquad \tD_{G_k/G_k}^2 = \cD_{G_k/G_k}^2 = \frac{1}{\bbS_{00}^2}~.
\end{align}
The circle Hilbert space is spanned by states \(\ket{\sfR}\). Its Hartle--Hawking state and the sphere partition function read
\begin{align}\label{eq:GG-HH}
	\ket{\HH} = \sum_{\sfR} \frac{\bbS_{0\sfR}}{\bbS_{00}}\ket{\sfR}~, \qquad \parti\qty\big[\S^2] = \braket{\HH} = \frac{1}{\bbS_{00}^2}~,
\end{align}
while, more generally, \cref{partFgenusg} gives
\begin{align}\label{eq:GG-genus-g}
	\parti\qty[\Sigma_\sfg] = \sum_{\sfR}\qty(\frac{\bbS_{0\sfR}}{\bbS_{00}})^{2-2\sfg} = \bbS_{00}^{2\sfg-2}\sum_{\sfR} \bbS_{0\sfR}^{\,2-2\sfg}~.
\end{align}
The sum on the right-hand side is the Verlinde formula for the dimension of the space of conformal blocks of \(G_k\) on \(\Sigma_\sfg\) \cite{Verlinde:1988sn},%
\footnote{That is, up to the Euler counterterm \(\bbS_{00}^{2\sfg-2}=\tilde{\cD}^{\chi(\Sigma_\sfg)}\), as discussed in \cref{ssec:higher-genus}.}
which is the standard result for the partition function of the \(G_k/G_k\) model \cite{Blau:1993tv,Witten:1993xi}.

Let us also make the topological local operators explicit. The operators \(\phi_a\) associated with the stretched anyons \((a,\bar a)\) close into the Verlinde (fusion) algebra \cite{Witten:1993xi,Moore:2026lmq},
\begin{align}
	\phi_a\times\phi_b = \sum_c N^c_{ab}\,\phi_c~,
\end{align}
and are related to the projectors \(\Pi_\sfR\) by
\begin{align}\label{eq:GG-Fourier}
	\Pi_\sfR = \bbS_{0\sfR}\sum_{a} \bar{\bbS}_{a\sfR}\,\phi_a~, \qquad \phi_a = \sum_\sfR \frac{\bbS_{a\sfR}}{\bbS_{0\sfR}}\,\Pi_\sfR~.
\end{align}
One can easily show that \(\Pi_\sfR \Pi_{\sfR'}\ = \delta_{\sfR\sfR'} \Pi_\sfR\), using the Verlinde formula \cref{eq:Verlinde-formula} and unitarity of \(\bbS\). This generalizes the discrete Fourier transform of \(\Z_N\) gauge theory, to which it reduces upon setting \(\bbS_{pq}=\ex{2\pi\ii pq/N}/\sqrt{N}\).

For concreteness, consider \(G=\SU(2)\) at level \(k\). The integrable representations are labelled by spins \(j\in\set{0,1/2,1,\ldots,k/2}\) and therefore the theory has \(k+1\) vacua. The modular \(\bbS\)-matrix is given by \cite{Gepner:1986wi}
\begin{align}
	\bbS_{jj'} &= \sqrt{\frac{2}{k+2}}\,\sin(\frac{\pi(2j+1)(2j'+1)}{k+2})~,
\end{align}
leading to quantum dimensions \cite{Witten:1988hf}
\begin{align}
 d_j &= \frac{\sin(\frac{\pi(2j+1)}{k+2})}{\sin(\frac{\pi}{k+2})}~, \qq{and} \cD_{\SU(2)_k/\SU(2)_k}^2 = \frac{k+2}{2\sin[2](\frac{\pi}{k+2})}~.
\end{align}
Let us note here that for \(k=1\) and \(k=2\) the fusion rules of \(\Rep(\SU(2)_k)\) are those of \(\operatorname{Vec}_{\Z_2}\) and of the Ising category, respectively, but the symmetry categories differ from the ones considered there in their associators \cite{Kitaev:2005hzj,Chang:2018iay}.

\paragraph{Entanglement entropy.}

A general pure state reads \(\ket{\Psi}=\sum_\sfR\Psi_\sfR\ket{\sfR}\), with sector probabilities \(\lambda_\sfR=\abs{\Psi_\sfR}^2/\abs{\Psi}^2\), and its entanglement entropy follows from \cref{EE:MixBoundary}. For the Hartle--Hawking state \cref{eq:GG-HH}, the probabilities take the simple form
\begin{align}
	\lambda_\sfR = \frac{d_\sfR^2}{\cD_{G_k/G_k}^2} = \bbS_{0\sfR}^2~,
\end{align}
For the boundary conditions \cref{LRBC}, we then find
\begin{align}
	\ee\qty(\rho_{\HH;\cB_L,\cB_R}) = -\log \bbS_{00}^2 - \sum_\sfR \bbS_{0\sfR}^2 \log\frac{d_\sfR^2}{g_\sfR^{(L)}g_\sfR^{(R)}}~.
\end{align}
Unless all quantum dimensions are integers, as happens for instance for \(\SU(N)_1\), where all Verlinde lines are invertible,\footnote{This fact is related to the existence of a topological interface with the Abelian $\U(1)^{N-1}$ theory \cite{Fliss:2020cos}. At the level of gauged WZW models, it follows from the extension of the affine $\mathfrak{u}(1)^{N-1}$ spectrum generating symmetry algebra to $\mathfrak{su}(N)_1$ by vertex operators \cite{Fuji:2016lgv}, a generalization of the familiar symmetry extension $\mathfrak{u}(1)\rightarrow\mathfrak{su}(2)_1$ of the compact boson at critical radius.} the symmetric boundary condition \cref{eq:SBC} is not available. As we did with the Fibonacci TQFT above, we may instead consider a general edge dressing \cref{genCuttingMap}, for which \cref{genRenyies} gives
\begin{align}
	\ee\qty(\rho_\HH) = \log\cD^2 + \sum_\sfR \bbS_{0\sfR}^2\qty(\ent\qty(\sigma_\sfR) - \log d_\sfR^2)~.
\end{align}
For a dressing with \(\ent\qty(\sigma_\sfR)=\log d_\sfR^2\) the edge contribution cancels and we are left with 
\begin{align}
	\ee\qty(\rho_\HH) = \log\cD_{G_k/G_k}^2=-2\log \bbS_{00}~.
\end{align}
For \(\SU(2)_k\) the Hartle--Hawking entanglement entropy is then
\begin{align}
	\ee\qty(\rho_\HH) = \log\frac{k+2}{2\sin[2](\frac{\pi}{k+2})}~,
\end{align}
agreeing, of course, with \cref{ssec:ZN} and \cref{ssec:Ising} for \(k=1\) and \(k=2\), respectively.  

\paragraph{Entanglement asymmetry.}

With the tools developed above, we can straightforwardly compute the entanglement asymmetry of the \(G_k/G_k\) model. Again, the \(G_k/G_k\) model admits, in general, no symmetric boundary condition, so we compute only the circle asymmetry associated to its fusion category. The key idea is similar to the one employed for the \(\Z_N\) and DW theories above; namely  we will switch to a basis of circle states that diagonalizes the action of the TDLs. This basis is immediately obtained: a well-known consequence of the Verlinde formula is that $\bbS$ diagonalizes the fusion coefficients. In fact, we have already met this basis in \cref{eq:GG-Fourier}. Translated to circle states, it reads
\begin{align}
	\ket*{\hat{a}} = \sum_{\sfR} \bbS_{a\sfR} \ket{\sfR}~, \qquad a \in\irr(\Rep(G_k))~. 
\end{align}
The states \(\ket*{\hat{a}}\) are orthonormal, since \(\bbS\) is unitary. To show that they diagonalize the action of the TDLs, we need, besides unitarity, \(\bbS^\dagger \bbS = \1\), and the Verlinde formula, \cref{eq:Verlinde-formula}, two further standard facts about modular \(\bbS\)-matrices,
\begin{align}
	\bbS_{ab} = \bbS_{ba} = \bbS_{\bar{a}\bar{b}} = \overline{\bbS}_{\bar{a}b}~, \qq{and} \bbS^2 = C~,  
\end{align}
where \(C\) is the charge conjugation matrix with components \(C_{ab} = \delta_{\bar{a}b}\); see, for instance, \cite{Kitaev:2005hzj}. With these, we can immediately check that
\begin{align}
	L_a \ket*{\hat{b}} = \frac{\overline{\bbS}_{ab}}{\bbS_{0b}}\ket*{\hat{b}}~.
\end{align}
Furthermore, the inverse \(\tilde{K}^{ab}\) of the trace pairing, \cref{eq:K-pairing}, follows from \cref{eq:Verlinde-formula} and reads \cite{Benini:2025lav}
\begin{align}
	\tilde{K}^{ab} = \sum_{c} \bbS_{0c}^2\;\overline{\bbS}_{ac}\, \overline{\bbS}_{bc}~.
\end{align}
With these two ingredients, the symmetrized density matrix of any vacuum, \(\rho_\sfR =\ketbra{\sfR}\), takes the very simple form
\begin{align}
	\rho_{\sfR,\sfS} = \sum_a \abs{\bbS_{a\sfR}}^2 \ketbra*{\hat{a}}~.
\end{align} 
The fusion asymmetry is then simply the Shannon entropy of the distribution \(\abs{\bbS_{a\sfR}}^2\),
\begin{align}
	\entas_\t{fus}(\rho_\sfR) = - \sum_{a} \abs{\bbS_{a\sfR}}^2\log\abs{\bbS_{a\sfR}}^2~.
\end{align}

\subsubsection{The \texorpdfstring{\(\Spin(N^2-1)_1/\SU(N)_N\)}{Spin(N\^ 2-1)/SU(N)} model}\label{sssec:adjoint-QCD}

Our final example is the topological coset \(\Spin(N^2-1)_1/\SU(N)_N\), which describes the infrared phase of (bosonized) two-dimensional QCD with gauge group \(\SU(N)\) and a massless Majorana fermion in the adjoint representation \cite{Komargodski:2020mxz,Delmastro:2021otj,Cordova:2024goh}. The branching rules for this conformal embedding were determined in \cite{Kac:1988tf} and analyzed in detail in \cite{Komargodski:2020mxz}, where it was found that the number of vacua is 
\begin{align}\label{eq:adjQCD-vacua}
	\abs{\irr(\cM_\t{IR})} = 3\cdot 2^{N-2}~.
\end{align}
We denote the module category of the topological coset as \(\cM_\t{IR} = \cC_{\Spin(N^2-1)_1/\SU(N)_N}\). The total boundary quantum dimension follows from \cref{eq:total-bqdim},
\begin{align}\label{eq:adjQCD-sphere}
	\parti_\t{IR}\qty\big[\S^2] &= \tD_\t{IR}^2 = \cD_{\Spin(N^2-1)_1}\,\cD_{\SU(N)_N} = \frac{2}{\bbS_{00}}~, \qq{with} \\
	\bbS_{00} &= \frac{1}{\sqrt{N}\,(2N)^{\frac{N-1}{2}}}\prod_{m=1}^{N-1}\qty(2\sin\frac{\pi m}{2N})^{N-m}~.
\end{align}
In the above \(\bbS_{00}\) is the vacuum entry of the modular \(\bbS\)-matrix of \(\SU(N)_N\) \cite{DiFrancesco:1997nk,Ooguri:1999bv} and we have also used the fact that \(\cD_{\Spin(N^2-1)_1}=2\) independently of \(N\). To see this, note that when \(N\) is even, \(\Rep(\Spin(N^2-1)_1)\) has two invertible anyons of quantum dimension one and one anyon of quantum dimension \(\sqrt{2}\) \cite{Kitaev:2005hzj} for a total of \(\cD^2=1+1+2=4\). When \(N\) is odd, it has instead four invertible anyons \cite{Kitaev:2005hzj} and hence \(\cD^2 = 1+1+1+1=4\).

For \(N=2\), since \(\hat{\so}(3)_1\cong\hat{\su}(2)_2\), the model coincides with the \(\SU(2)_2/\SU(2)_2\) model discussed above. For \(N=3\), the symmetry category \(\cM_\t{IR}\) is the \(\Z_2\times\Z_2\) triality category with 6 simple objects \cite{Cordova:2024goh}. It consists of four invertible lines \(\1,v,s,c\), which descend from the \(\Spin(8)_1\) Verlinde lines and fuse according to \(\Z_2\times\Z_2\), together with two lines \(\cN\) and \(\bar{\cN}\) of quantum dimension two, with
\begin{align}\label{eq:fusion-adjqcd-n3}
	x\otimes\cN = \cN~, \qquad \cN\otimes\cN = 2\,\bar{\cN}~, \qquad \cN\otimes\bar{\cN} = \1\oplus v\oplus s\oplus c~,
\end{align}
and similarly for \(\bar{\cN}\), with \(x\) standing for any of the invertible lines \(\1,v,s,c\). The Hartle--Hawking state and the sphere partition function are
\begin{align}\label{eq:adjQCD-N3}
	\ket{\HH} = \ket{\1}+\ket{v}+\ket{s}+\ket{c}+2\ket{\cN}+2\ket*{\bar{\cN}}~, \qquad \parti\qty\big[\S^2]=\braket{\HH}=12~.
\end{align}

\paragraph{Entanglement entropy.}

For a general state, the entanglement entropy follows from \cref{EE:MixBoundary}. For the Hartle--Hawking state, \(\lambda_m=\td_m^{\,2}/\tD^2 = \bbS_{00}\,\td_m^{\,2}/2\), so that
\begin{align}
	\ee\qty(\rho_{\HH;\cB_L,\cB_R}) = \log\frac{2}{\bbS_{00}} - \frac{\bbS_{00}}{2}\sum_{m\in\irr(\cM)}\td_m^{\,2}\log\frac{\td_m^{\,2}}{g_m^{(L)}g_m^{(R)}}~.
\end{align}
For \(N=3\) this reads
\begin{align}
	\ee\qty(\rho_{\HH;\cB_L,\cB_R}) = \log 12 + \frac{1}{12}\sum_{x\in\{\1,v,s,c\}}\log\qty(g_x^{(L)}g_x^{(R)}) + \frac{1}{3}\sum_{\cX\in\{\cN,\bar{\cN}\}}\log\frac{g_\cX^{(L)}g_\cX^{(R)}}{4}~.
\end{align}
Since all boundary quantum dimensions are integers, the symmetric boundary condition \(\cB_\sfS = \1\oplus v\oplus s\oplus c\oplus 2\,\cN\oplus 2\,\bar{\cN}\) is available, and yields
\begin{align}
	\ee\qty(\rho_{\HH;\cB_\sfS}) = \log 12~,
\end{align}
For even \(N\), instead, \(\cM_\t{IR}\) contains a vacuum with \(\td=\sqrt{2}\) and the symmetric boundary condition is not available. As in the previous example, taking a general edge dressing with \(\ent\qty(\sigma_m)=\log\td_m^{\,2}\) gives \(\ee\qty(\rho_\HH)=\log\tD_\t{IR}^2=\log\qty(2/\bbS_{00})\).

\paragraph{Entanglement asymmetry.}

The fusion rules of \(\cM_\t{IR}\) are not known for general \(N\), and strongly symmetric boundaries are not available either.%
\footnote{For $N=3$, while \(\cB_\sfS\) exists, it is not strongly symmetric. In fact, it is not weakly symmetric either, since \(\tilde{N}^x_{\cN x} = 0\), for all invertible objects \(x\), thus failing the criterion of \cite{Choi:2023xjw}.}
However, for the values of \(N\) for which the fusion rules are known, we can at least compute the circle entanglement asymmetry using the same technology as in the previous examples. Our approach follows closely the treatment of the Haagerup phases in \cref{ssec:Haagerup}.

The simplest non-trivial case is \(N=3\), with adjoint quarks of three colors. The fusion matrices, and hence the NIM-rep matrices, can be read off from \cref{eq:fusion-adjqcd-n3}. Inverting the pairing \cref{eq:K-pairing} then gives the coefficients \(\tilde{K}^{ab}\) of the symmetrizer \cref{eq:symmetrizer-fusion},
\begin{align}
	\tilde{K}^{xy} = \frac{1}{4}\delta^{xy}-\frac{1}{24}~, \qquad \tilde{K}^{\cN\bar{\cN}} = \tilde{K}^{\cN\bar{\cN}} = \frac{1}{12}~, 
\end{align}
where \(x,y\in\{\1,v,s,c\}\) and all other entries of \(\tilde{K}^{ab}\) vanish. The fusion matrices also show that \(L_x^\dagger = L_x\) for all invertible lines, while \(L_\cN^\dagger = L_{\bar{\cN}}\) for the two non-invertible ones. The symmetrized density matrix therefore reads
\begin{align}
	\rho_\sfS = \frac{1}{4}\sum_{x} L_x\, \rho\, L_x^\dagger - \frac{1}{24} A\, \rho\, A + \frac{1}{12}\qty(L_\cN\, \rho\, L_\cN^\dagger + L_{\bar{\cN}}\, \rho\, L_{\bar{\cN}}^\dagger)~,
\end{align}
where now \(A \coloneqq L_\1\oplus L_v\oplus L_s\oplus L_c\). The symmetrized density matrices of the six vacua have eigenvalues
\begin{align}
	\lambda_x = \set{\frac{1}{4},\frac{1}{4},\frac{1}{4},\frac{1}{12},\frac{1}{12},\frac{1}{12}}~, \qquad
	\lambda_\cN = \lambda_{\bar{\cN}} = \set{\frac{1}{3},\frac{1}{3},\frac{1}{3}, 0, 0, 0}~,
\end{align}
again with \(x\) labeling the invertible vacua \(\1,v,s,c\). The resulting circle asymmetries are then
\begin{align}
	\entas_\t{fus}(\rho_x) &= \frac{3}{4}\log 4 + \frac{1}{4}\log 12 \approx 1.661~,\qquad \qquad x = \1,v,s,c~, \\ 
	\entas_\t{fus}(\rho_\cN) &= \entas_\t{fus}(\rho_{\bar{\cN}}) = \log 3 \approx 1.099~.
\end{align}
The invertible vacua are thus more asymmetric than the non-invertible ones, but neither attains the upper bound \(\log\abs{\cM_\t{IR}} = \log 6 \approx 1.792\). 

\section{Discussion}\label{sec:disco}

In this paper we scrutinized the definition of entanglement in quantum field theory and, more specifically, its sensitivity to the choice of a \emph{factorization map}, \(\cH\hookrightarrow \cH_{\aent}\otimes\cH_{\coaent}\), which, following \cite{Ohmori:2014eia}, we implemented by imposing boundary conditions on a fattened entangling surface. To keep our investigation under full control, we restricted ourselves to the arena of compact, semisimple two-dimensional TQFTs. Their mathematical classification allowed us to develop a complete and systematic framework for computing entanglement in such theories, for any state (pure or mixed), any bipartition (any number of intervals), and any boundary condition. In particular, we showed how to construct \emph{à la carte} an isometric cutting map even when one naïvely seems unavailable through path-integration alone, i.e., in the absence of an \emph{entanglement brane} \cite{Donnelly:2016jet,Donnelly:2018ppr}.

This construction has several implications, which we explored in the main text and summarize here.
\vspace{-1ex}
\begin{itemize}[topsep=0pt, itemsep=4pt]
	\item We showed that suitably dressed cutting maps lead to a non-integer \textquote{ground-state degeneracy,} analogous to the celebrated Affleck--Ludwig boundary entropy of critical systems \cite{Affleck:1991tk}, even though all boundary conditions of a compact 2d TQFT are gapped.
	\item We demonstrated an entropic reconstruction of the TQFT. For a suitable choice of cutting map, its partition function on an arbitrary closed Riemann surface can be read off from the multi-interval R\'enyi entropies of the Hartle--Hawking state. This reconstruction is a purely infrared avatar of the conjecture of \cite{DelZotto:2026fpw,DelZotto:2026dkg}, with the simplification that in two dimensions, bipartite entanglement suffices.
	\item We established a simple entropic diagnostic of gapped phases with the same symmetry. Without any knowledge of the vacua of the infrared TQFT, the entanglement entropy of the Hartle--Hawking state, i.e., the unique symmetric combination of all vacua, can tell such phases apart, as we demonstrated for the three gapped Haagerup phases in \cref{ssec:Haagerup}. This is morally analogous to the topological entanglement entropy of \((2+1)\)d gapped phases \cite{Kitaev:2005dm,Levin:2006zz}.
	\item We connected our formalism to more refined entropic diagnostics, such as entanglement asymmetry \cite{Ares:2022koq,Benini:2025lav}, which we computed for several TQFTs with non-invertible symmetries.
\end{itemize}
We close by collecting some further directions that we find worth pursuing.

\paragraph{Symmetry-resolved entanglement.}
Another useful refinement of entanglement entropy is its resolution into symmetry sectors \cite{Goldstein:2017bua}. For non-invertible symmetries \cite{Saura-Bastida:2024yye,Heymann:2024vvf,Choi:2024wfm}, this is achieved by decomposing the interval Hilbert space into irreducible representations of the strip algebra \cite{Cordova:2024iti,Choi:2024tri}. In CFTs, Ref. \cite{Choi:2024wfm} found that projecting onto an irreducible representation shifts the boundary entropy by a universal amount, the logarithm of a ratio of quantum dimensions, while leaving the leading term \(\frac{c}{3}\log(\ell/\varepsilon)\) unchanged. In a compact TQFT there is no such leading term, and one might hope for an exact counterpart of their formula, with the Chan--Paton multiplicities, or more generally the edge dressings, playing the role of the \(g\)-functions. This is currently under investigation.

\paragraph{Cloaking asymmetry?}
To define entanglement asymmetry directly on an interval, the symmetry must act on the interval Hilbert space. This requires all topological lines to be able to end topologically on the boundary conditions at the entangling points; in other words, it requires a strongly symmetric boundary condition. In the absence of a symmetric boundary condition, one could imagine dressing a non-symmetric boundary condition by an operator, as we did here to construct isometric cutting maps. A closed loop of such a dressed boundary condition is then clearly symmetric: an isometric cutting map is shrinkable -- it can be shrunk away and reopened on the other side of any topological line, effectively passing right through it. This is a TQFT avatar of what was called a \emph{cloaking} boundary condition in \cite{Brehm:2021wev,Brehm:2024zun}. In our construction, the dressing acts directly on the interval Hilbert space, i.e., on open boundaries, where the shrinking argument is no longer available. It would be interesting to understand whether such a dressed boundary condition can nevertheless support an action of the strip algebra, or of a suitable modification thereof, on the interval. Defining entanglement asymmetry in this way would dispense with the need for a symmetric boundary condition when studying the breaking of non-invertible symmetries.%
\footnote{We note here a related idea in \cite{Benini:2025lav}, which constructs an \emph{averaged} entanglement asymmetry in such cases.}
%

\paragraph{Non-compact TQFTs.}
Much of our construction relied on the compactness and semisimplicity of the TQFTs we analyzed; in particular the finite number of vacua that these theories host.  This in turn implies a finite set of simple boundary conditions, all of them gapped, and hence finite-dimensional edge Hilbert spaces. For many physical reasons, one might want to relax these assumptions and consider instead \emph{non-compact} TQFTs. For instance, such theories appear as SymTFTs for continuous symmetries \cite{Antinucci:2024zjp,Brennan:2024fgj,Bonetti:2024cjk,Argurio:2024ewp}, and in low-dimensional solvable models of quantum gravity \cite{Achucarro:1986uwr,Witten:1988hc,Blommaert:2018oro,Iliesiu:2019xuh,Mertens:2022ujr,Collier:2023fwi}. Interestingly, non-compact TQFTs can host a continuum of edge modes, i.e., gapless boundary conditions. This feature is, in a sense, both good and bad. On the one hand, isometric cutting maps are easier to construct, since they no longer hinge on an arithmetic comparison between the (total) boundary quantum dimensions and the dimensions of the edge Hilbert spaces. On the other hand, both quantities are now infinite, which calls for a UV regulator and costs us some of the power of topological field theory. It would be interesting to understand how our results translate to this arena.

\paragraph{Higher dimensions.}
Naturally, now that we have a complete framework for entanglement in 2d TQFTs, one would like to extend these ideas to higher dimensions. Unfortunately, higher-dimensional TQFTs can famously host gapless edge modes \cite{Witten:1988hf,Elitzur:1989nr,Witten:1996hc,Belov:2006jd,Fliss:2023uiv}, bringing us back to the problem above. Nonetheless, the idea of defining entanglement by fattening the entangling surface and imposing boundary conditions extends to these cases as well; see, for instance, \cite{Wong:2017pdm,Fliss:2023uiv,Benini:2025hbj,Sela:2026ipc}. It would then be interesting to engineer isometric cutting maps for higher-dimensional TQFTs, \emph{assuming} gapped boundary conditions. One could, for instance, hope to isolate in this way the topological entanglement entropy \cite{Kitaev:2005dm,Levin:2006zz} and related quantities \cite{Fliss:2017wop,Lou:2019heg,Fliss:2023dze,Lamas:2025eay}, without the need to subtract a divergent leading contribution.

\paragraph{Gravity.} Lastly, reiterating a point raised in the introduction, there is much interest in what degree the tools developed in this paper apply to the TQFT descriptions of low-dimensional quantum gravity. For instance, one key result of this paper --- that the TQFT partition function can be recovered from an entropic calculation with an appropriate cutting map --- might lend deeper insight into the Gibbons--Hawking interpretation of the sphere path integral as a Euclidean counting of states. To be specific, in instances where there is no natural isometric factorization map, the path integral had to be dressed with a boundary operator in order to reproduce the norm of the Hartle--Hawking state, i.e. $\cZ[S^2]$. If we maintain that the edge modes introduced in a factorization map are avatars of UV degrees of freedom not directly accessed in the TQFT, the equality of $\cZ[S^2]$ with an entanglement entropy is a non-trivial IR datum on the state of these UV degrees of freedom. Investigations in this direction are currently being pursued by the authors; however in order to port these insights into JT gravity, one would need to overcome the difficulties associated to non-compact TQFTs (as noted above). Further investigations into 3d gravity would need to address the additional subtleties associated with higher-dimensional TQFTs.

\paragraph{Acknowledgements.}
It is a pleasure to acknowledge helpful discussions with Dio Anninos, Riccardo Argurio, Francesco Benini, Panos Betzios, Lea Bottini, Michele Fossati, Giovanni Galati, Matthieu Marchand, Ronak Soni, Yuji Tachikawa, and Gabriel Wong. 

JRF is supported by FNRS MISU grant 40024018 \textquote{Pushing horizons in Black hole Physics.} ARF is supported by the UKRI Future Leaders Fellowship ``The materials approach to quantum spacetime'' under reference MR/X034453/1. SV is supported by a Marina Solvay fellowship and by FNRS under grant no. 4.4503.15.  SV would also like to acknowledge the contribution of COST (European Cooperation in Science and Technology) through the action CA22113.

\paragraph{Statement on LLM use.} LLMs were utilized for the purposes of literature search, computational aid and error checking, and copy-editing. No generative content was used in this paper. All ideas originate with the authors and all computations have been performed and checked by the authors.

\appendix
\crefalias{section}{appendix}
\crefalias{subsection}{appendix}
\addtocontents{toc}{\protect\setcounter{tocdepth}{2}}

\section{Entanglement of mixed states and multiple intervals}\label{app:MixedMult}
The discussion of \cref{ssec:cutting,ssec:pure-entanglement} readily generalizes to mixed states and multiple intervals.

\subsection{Mixed states}

A generic mixed state in \(\cH_{\S^1}\) takes the form
\begin{align}
	\rho = \sum_{m,n} \rho_{mn}\ketbra{m}{n}~, \qquad \rho_{mn} = \rho_{nm}^*~,\quad \sum_m \rho_{mm} = 1~,
\end{align}
Under a general edge dressing, the embedding of \(\rho\) into \(\cH_\t{ext}\) by \(\iota_{\cO;\cB_L,\cB_R}\) is, therefore
\begin{align}
	\iota_{\cO;\cB_L,\cB_R}^\vp\, \rho\, \iota_{\cO;\cB_L,\cB_R}^\dagger = \sum_{m,n} \rho_{mn} \left.\ket{\Omega_m^{\cO}}\!\!\right>\!\!\left<\!\!\bra{\Omega_n^{\cO}}\right.~.
\end{align}
Due to $m$ and $n$ being simple boundary conditions, Eq. \cref{eq:Hmnsimple} makes it clear that tracing out \(\overline{\cH}_\cB\) from \(\cH_\t{ext}\) kills all off-diagonal elements of \(\rho\), and the reduced state is simply:
\begin{align}
	\rho_{R} = \tr^\vp_{\overline{\cH}_{\cB_L,\cB_R}}\qty(\iota_{\cO;\cB_L,\cB_R}^\vp\, \rho\, \iota_{\cO;\cB_L,\cB_R}^\dagger) = \sum_m \rho_{mm}\,\sigma_m~.
\end{align}
where $\sigma_m$ is defined as above, \cref{eq:sigmam}. From here the computation of R\'enyi and entanglement entropies of \(\rho\) is identical to the previous subsection and gives:
\begin{align}
	\ent_\n\qty(\rho_{R}) & = \frac{1}{1-\n}\log(\sum_m \rho_{mm}^{\n} \tr\sigma_m^\n)~,         \\
	\ee\qty(\rho)         & = -\sum_m \rho_{mm} \log \rho_{mm} + \sum_m \rho_{mm}\ent\qty(\sigma_m)~.
\end{align}

\subsection{Multiple intervals}

The generalization to the case in which we dissect the Cauchy slice into multiple disjoint intervals is also straightforward. Without loss of generality, we take number of cuts to be even, \(2\sfN\), such that the cutting map factorizes the circle Hilbert space as 
\begin{align}
    \iota_{\{\cO\};\{\cB\}}: \cH_{\S^1}\to \cH_\text{edge} \coloneqq \cH_{\cB_1,\cB_2}\otimes\overline{\cH}_{\cB_2,\cB_3}\otimes \ldots \otimes {\cH}_{\cB_{2\sfN-1},\cB_{2\sfN}}\otimes \overline{\cH}_{\cB_{2\sfN},\cB_1}~.
\end{align}
We will consider a general edge dressing along each entangling cut such that a pure state \cref{S1PureState} is embedded into \(\cH_{\textnormal{edge}}\) as 
\begin{align}
	\iota_{\{\cO\};\{\cB\}}\ket{\Psi} = \sum_m \frac{\Psi_m}{{\td_m}^{2\sfN}} \sum_{\{\fri\}} \bigotimes_{j=1}^{\sfN}\left( \cO_{2j-1}^{(m)}\ket{m;\fri_{2j-1}}\otimes\cO_{2j}^{(m)}\overline{\ket{m;\fri_{2j}}}\right)~, 
\end{align}
where $\ket{m;\fri_{2j-1}}=\ket{m;i_{2j-1},i_{2j}}$ and $\overline{\ket{m;\fri_{2j}}}=\overline{\ket{m,i_{2j+1}, i_{2j}}}=\ket{m;i_{2j},i_{2j+1}}$, etc. subject to $\fri_{\sfN+1}=\fri_1$, and
\begin{equation}
    \sum_{\{\fri\}}\coloneqq\sum_{i_1=1}^{g_m^{(1)}}\ldots\sum_{i_{2\sfN}=1}^{g_m^{(2\sfN)}}~.
\end{equation}
This embedding is an isometry if
\begin{equation}\label{eq:appOisom}
    \frac{1}{{\td_m}^{4\sfN}}\sum_{\{\fri\}}\sum_{\{\fj\}}\qty(\prod_{j=1}^{\sfN}\bra{m;\fj_{2j-1}}\bbO_{2j-1}^{(m)}\ket{m;\fri_{2j-1}}\overline{\bra{m;\fj_{2j}}}\bbO_{2j}^{(m)}\overline{\ket{m;\fri_{2j}}})=1~,
\end{equation}
where we remind the reader the notation, $\bbO_j^{(m)}=\cO_j^{(m)\dagger}\cO_j^{(m)}$. From here we will keep this condition as implicitly imposed.

We now reduce the density matrix by tracing over the even intervals (denoted above by overbars)
\begin{align}
	\rho_{\Psi,\{\cO\},\{\cB\}}^\vp \coloneqq&\frac{1}{\abs{\Psi_m}^2}
	\tr_{\overline{\cH}_\cB}^\vp\!\qty(\iota^\vp_{\{\cO\},\{\cB\}} \ketbra{\Psi} \iota_{\{\cO\},\{\cB\}}^\dagger)~,\nonumber\\
	=& \sum_m \frac{\abs{\Psi_m}^2}{\abs{\Psi}^2{\td_m}^{4\sfN}}\,\sum_{\{\fri\}}\sum_{\{\fj\}}\bigotimes_{j=1}^{\sfN}\qty(\cO_{2j-1}^{(m)}\ket{m;\fri_{2j-1}}\qty(\bbO_{2j-2}^{(m)})^{\sfT}_{\overline{\fri_{2j-2}};\overline{\fj_{2j-2}}}\bra{m;\fj_{2j-1}}\cO^{(m)\dagger}_{2j-1})~,
\end{align}
where $\abs{\Psi}^2 = \sum_m \abs{\Psi_m}^2$. We can write this in a compact notation by denoting 
\begin{equation}
    \boldsymbol{\sigma}_{m;\sfN}\coloneqq \frac{1}{{\td_m}^{4\sfN}}\sum_{\{\fri\}}\sum_{\{\fj\}}\bigotimes_{j=1}^{\sfN}\qty(\cO_{2j-1}^{(m)}\ket{m;\fri_{2j-1}}\qty(\bbO_{2j-2}^{(m)})^{\sfT}_{\overline{\fri_{2j-2}};\overline{\fj_{2j-2}}}\bra{m;\fj_{2j-1}}\cO^{(m)\dagger}_{2j-1}) 
\end{equation}
which we can think of as reduced density matrix over the $m$ sector of $\otimes_j\cH_{\cB_{2j-1},\cB_{2j}}$ (which is normalized due to the condition \cref{eq:appOisom}):
\begin{equation}
    \rho_{\Psi,\{\cO\},\{\cB\}}^\vp=\sum_m \lambda_m\,\boldsymbol{\sigma}_{m;\sfN}~,\qquad\text{with }\lambda_m\coloneqq\frac{\abs{\Psi_m}^2}{\abs{\Psi}^2}.
\end{equation}
From this we read off the $\n^\text{th}$ R\'enyi entropy as
\begin{equation}
    \ent_\n\qty(\rho_{\Psi,\{\cO\},\{\cB\}})
	= \frac{1}{1-\n}\log\tr(\rho_{\Psi,\{\cO\},\{\cB\}}^\n)
	= \frac{1}{1-\n}\log(\sum_m \lambda_m^{\n} \tr\boldsymbol{\sigma}_{m;\sfN}^\n)
\end{equation}
Depackaging the notation (slightly), this is equivalent to
\begin{equation}
    \ent_\n\qty(\rho_{\Psi,\{\cO\},\{\cB\}})
	=\frac{1}{1-\sfn}\log\sum_m\frac{\abs{\Psi_m}^{2\n}}{\abs{\Psi}^{2\n}{\td_m}^{4\sfN\n}}\sum_{\{\fri^{(q)},\fj^{(q)}\}}\prod_{q=1}^{\n}\prod_{j=1}^{\sfN}\qty(\bbO^{(m)}_{2j-2})^{\sfT}_{\overline{\fri_{2j-2}^{(q)}};\overline{\fj_{2j-2}^{(q)}}}\qty(\bbO^{(m)}_{2j-1})_{\fj_{2j-1}^{(q)};\fri_{2j-1}^{(q+1)}}~.
\end{equation}
Taking an $\n\rightarrow1$ limit yields an entanglement entropy
\begin{equation}
    \ee\qty(\rho_{\Psi,\{\cO\},\{\cB\}})
	= -\sum_m \lambda_m
	\log \lambda_m+ \sum_m \lambda_m\,\ee\qty(\boldsymbol{\sigma}_{m;\sfN})~.
\end{equation}

\printbibliography

\end{document}